\documentclass[final]{siamltex}
\usepackage{framed,algorithm,algorithmic}
\usepackage{amsfonts,amsmath,bm,url,color}

\long\def\symbolfootnote[#1]#2{\begingroup%
\def\thefootnote{\fnsymbol{footnote}}\footnote[#1]{#2}\endgroup}

\newcommand{\Expect}[1]{\mbox{}{\mathbb{E}}\left[#1\right]}

\newcommand{\FNorm }[1]{\mbox{}\|#1\|_\mathrm{F}  }
\newcommand{\FNormS}[1]{\mbox{}\|#1\|_\mathrm{F}^2}

\newcommand{\TNorm }[1]{\mbox{}\|#1\|_2  }
\newcommand{\TNormS}[1]{\mbox{}\|#1\|_2^2}

\newcommand{\pinv}[1]{ {#1}^\dagger}

\newcommand{\transp}{^{\textsc{T}}}
\newcommand{\trace}{\text{\rm Tr}}
\newcommand{\mat}[1]{{\ensuremath{\bm{\mathrm{#1}}}}}

\def\rank{\hbox{\rm rank}}

\def\b{{\mathbf b}}
\def\e{{\mathbf e}}

\def\v{{\mathbf v}}

\def\matA{\mat{A}}
\def\matB{\mat{B}}
\def\matC{\mat{C}}
\def\matD{\mat{D}}
\def\matE{\mat{E}}

\def\matI{\mat{I}}

\def\matL{\mat{L}}

\def\matM{\mat{M}}

\def\matQ{\mat{Q}}
\def\matR{\mat{R}}
\def\matS{\mat{S}}

\def\matU{\mat{U}}
\def\matV{\mat{V}}
\def\matW{\mat{W}}
\def\matX{\mat{X}}
\def\matY{\mat{Y}}
\def\matZ{\mat{Z}}
\def\matOmega{\mat{\Omega}}
\def\matSig{\mat{\Sigma}}
\def\matTh{\mat{\Theta}}

\def\matOmega{\mat{\Omega}}
\def\matSig{\mat{\Sigma}}
\def\matTh{\mat{\Theta}}

\def\matPsi{\mat{\Psi}}
\def\matDelta{\mat{\Delta}}
\def\matXi{\mat{\Xi}}

\DeclareMathSymbol{\Prob}{\mathbin}{AMSb}{"50}
\newcommand\remove[1]{}

\def\nnz{{ \rm nnz }}

\def\math#1{$#1$}

\def\frac#1#2{{#1\over #2}}

\def\eqan#1{\begin{eqnarray*}
#1
\end{eqnarray*}}

\DeclareMathSymbol{\R}{\mathbin}{AMSb}{"52}

\def\cl#1{{\cal #1}}

\def\argmin{\mathop{\hbox{argmin}}\limits}

\def\x{{\mathbf x}}
\def\y{{\mathbf y}}

\def\a{{\mathbf a}}
\def\b{{\mathbf b}}

\begin{document} 

\title{Optimal CUR Matrix Decompositions \thanks{An extended abstract appeared at the 46th ACM Symposium on Theory of Computing (STOC).}}

\author{
Christos Boutsidis\thanks{
Yahoo! Labs, New York, NY. Email: boutsidis@yahoo-inc.com}
\and
David P. Woodruff\thanks{
IBM Research, Almaden, CA. Email: dpwoodru@us.ibm.com}
}

\maketitle

\begin{abstract}
\noindent
The CUR decomposition of an $m \times n$ matrix $\matA$ finds an $m \times c$ matrix $\matC$ with a subset of $c < n$ columns of $\matA,$
together with an $r \times n$ matrix $\matR$ with a subset of $r < m$ rows of $\matA,$ as well as a $c \times r$ low-rank matrix $\matU$ such that the matrix $\matC \matU \matR$
approximates the matrix $\matA,$ that is,
$\FNormS{\matA - \matC \matU \matR} \leq (1+\varepsilon) \FNormS{\matA - \matA_k}$, where $\FNorm{.}$ denotes the Frobenius norm and
$\matA_k$ is the best $m \times n$ matrix of rank $k$ constructed via the SVD. We present input-sparsity-time and deterministic algorithms for constructing
such a CUR decomposition where $c=O(k/\varepsilon)$ \emph{and}  $r=O(k/\varepsilon)$ \emph{and} rank$(\matU) = k$. Up to constant factors, our algorithms are simultaneously
\emph{optimal} in $c, r,$ and rank$(\matU)$.
\end{abstract}



\pagestyle{myheadings}
\thispagestyle{plain}
\markboth{BOUTSIDIS AND WOODRUFF}{OPTIMAL CUR MATRIX DECOMPOSITIONS}

\section{Introduction}\label{sec:intro}
Given as inputs a matrix $\matA \in \R^{m \times n}$ and integers $c < n$ and $r < m,$
the CUR factorization of $\matA$ finds $\matC \in \R^{m \times c}$ with $c$ columns of $\matA$,
$\matR \in \R^{r \times n}$ with $r$ rows of $\matA,$ and $\matU \in \R^{c \times r}$ such that
$ \matA = \matC \matU \matR + \matE. $
Here, $\matE = \matA - \matC \matU \matR $ is the residual error matrix. Compare this to the SVD factorization (let $k < \rank(\matA)$),
$\matA = \matU_k \matSig_k \matV_k\transp + \matA_{\rho-k}.$
The SVD residual error $ \matA_{\rho-k}$ is the best possible, under some rank constraints (see Section~\ref{sec:svd}). The matrices
$\matU_k \in \R^{m \times k}$ and $\matV_k \in \R^{n \times k}$ contain the top $k$ left and right singular vectors of $\matA$, while 
$\matSig_k\in \R^{k \times k}$ contains the top $k$ largest singular values of $\matA$.
In CUR,  $\matC$ and $\matR$ contain actual columns and rows of $\matA$, a property which is desirable  for feature selection and data interpretation~\cite{DM09}. This last property
makes CUR attractive in a wide range of applications~\cite{DM09}.

From an algorithmic perspective, the challenge is to construct $\matC, \matU$, and $\matR$ quickly
to  minimize the approximation error $\FNormS{\matA - \matC \matU \matR}$.
Definition~\ref{def:cur} states the precise optimization problem:
\begin{definition}[The CUR Problem]\label{def:cur}
Given $\matA \in \R^{m \times n}$ of rank $\rho = \rank(\matA),$ rank parameter $k < \rho,$ and accuracy parameter $0< \varepsilon < 1,$
construct $\matC \in \R^{m \times c}$ with $c$ columns from $\matA$, $\matR \in \R^{r \times n}$ with $r$ rows from $\matA$,
and $\matU \in \R^{c \times r}$, with $c, r,$ and $\rank(\matU)$ being as small as possible, in order to reconstruct $\matA$ within relative-error:
$$\FNormS{\matA - \matC \matU \matR} \le (1+\varepsilon) \FNormS{\matA -\matA_k}.$$
Here, $\matA_k = \matU_k \matSig_k \matV_k\transp  \in \R^{m \times n}$ is the best rank $k$ matrix obtained via the SVD of $\matA$.
\end{definition}

Despite the significant amount of work and progress on CUR, spanning both the numerical linear algebra community~\cite{Tyr96, Tyr00, GTZ97a, GTZ97b,HP97, Pan03, MG03, GM04}
and the theoretical computer science community~\cite{DK03,DM05,DKM06c,DMM06b,DMM08,DM09,BYM10,GM13,WZ13}, there are several important questions which still remain unanswered:
\begin{enumerate}
\item {\bf Optimal CUR:} Are there any $(1+\varepsilon$)-error CUR algorithms selecting  the optimal number of columns and rows?
\item {\bf Rank $k$ CUR:} Are there any $(1+\varepsilon$)-error CUR algorithms constructing $\matU$ with optimal rank?
\item  {\bf Input-sparsity-time CUR:} Are there any $(1+\varepsilon$)-error algorithms for CUR
running in time proportional to the number of the non-zero entries of $\matA$?
\item {\bf Deterministic CUR:}  Are there any deterministic, polynomial-time, $(1+\varepsilon$)-error CUR algorithms?
\end{enumerate}
Questions 1 and 4 were, for example, asked in Question 12 in the IITK list of  ``Open Problems in Data Streams and Related Topics''~\cite{mcgregor2006open}.
This article settles all of these questions via presenting two novel CUR algorithms. 
Our first algorithm is randomized and runs in ``input-sparsity-time''~(see Section~\ref{sec:alg3}); 
our second algorithm is deterministic and runs in polynomial time~(see Section~\ref{sec:alg2}).
Both algorithms achieve a relative-error bound with $c = O(k/\varepsilon)$ columns, $r = O(k/\varepsilon)$ rows, and $\rank(\matU)=k$.
Additionally, a matching lower bound is proven~(see Section~\ref{sec:lower}), indicating that, up to constant factors, both algorithms select the \emph{optimal} number of columns and rows and construct $\matU$ with
the \emph{optimal} $\rank(\matU)$.

We also present an optimal randomized CUR algorithm in Section~\ref{sec:alg1}. 
This algorithm might be slower than the algorithm in Section~\ref{sec:alg2}, especially on sparse matrices,  
but it is simpler to describe and analyze, hence we use it as a stepping stone for the algorithms in Section~\ref{sec:alg3} and Section~\ref{sec:alg2}. However, this algorithm might be faster when the input matrix is dense.

\subsection{Outline}
We summarize the main contributions of this paper in Subsection~\ref{sec:con}. 
In this subsection, we provide only a high level description of our results, hiding low level details and ideas. 
We give a summary of our techniques on how to obtain optimal CUR algorithms in Subsection~\ref{sec:over}. In Algorithm~\ref{alg1},
we present a so-called proto-algorithm for an optimal, relative-error, rank $k$ CUR. 
Our algorithms in later sections 
(Algorithm~\ref{algcur1} in Section~\ref{sec:alg1}, 
Algorithm~\ref{algcur3} in Section~\ref{sec:alg3}, and 
Algorithm~\ref{algcur2} in Section~\ref{sec:alg2}) are
specific instances of this proto-algorithm. As we explain in Subsection~\ref{sec:over},
each instance corresponds to a specific implementation of the various steps in the proto-algorithm 
in order to obtain the desired properties; for example, to design a deterministic algorithm all steps in the proto-algorithm
should be implemented in a deterministic way, etc. 
Section~\ref{sec:prior} summarizes results from prior literature and puts our CUR
algorithms in context (see Table I). To design our CUR algorithms in Sections~\ref{sec:alg1},~\ref{sec:alg3}, and~\ref{sec:alg2} we need
several ``subset selection tools'' from prior literature, which we summarize in Section~\ref{sec:back1}, as well as new tools,
which we present in Section~\ref{sec:back2}. Finally, we give a lower bound for a CUR algorithm in Section~\ref{sec:lower}. 

\subsection{Summary of contributions}\label{sec:con}
This work has two main technical contributions answering all four open problems related to CUR that we described in Section~\ref{sec:intro}. 

The first  contribution is an input-sparsity-time,  relative-error CUR algorithm selecting an optimal number of columns and rows 
and constructing $\matU$ with optimal rank. The theorem below presents our main result.
\begin{theorem}\label{thm:main1r} [Restatement of Theorem~\ref{thm:main3} in Section~\ref{sec:alg3}]
Given \math{\matA\in\R^{m\times n}} of rank $\rho$, a target rank $1 \leq k < \rho$, and $0 < \epsilon < 1$,
there exists a randomized algorithm to select at most
$$
c = O(k/\varepsilon)
$$
columns
and at most
$$
r = O(k/\varepsilon)
$$
rows from \math{\matA}
to form matrices
$\matC\in\R^{m \times c},$
$\matR\in\R^{r \times n}$
and $\matU \in \R^{c \times r}$ with
$\rank\left(\matU\right) = k$
such that, with constant probability of success,
$$\FNormS{\matA - \matC \matU \matR} \le (1+O(\varepsilon)) \FNormS{\matA -\matA_k}.$$
The matrices $\matC, \matU$, and $\matR$ can be computed in time
$$O\left( nnz\left(\matA\right)\log n + \left(m+n\right) \cdot poly\left(\log n, k,1/\varepsilon \right) \right).$$
\end{theorem}

To the best of our knowledge, this is the \emph{first} algorithm to achieve a relative-error CUR decomposition with $O(k/\varepsilon)$ number of rows \emph{and} columns. This number of rows and columns is asymptotically optimal:
Theorem~\ref{thm:lower2} proves a matching lower bound.
The previous best such relative-error CUR algorithm~\cite{WZ13}
selects $c=O(k/\varepsilon)$ columns and $r = O(k/\varepsilon^2)$ rows from $\matA$. The first ever relative-error CUR algorithm selects  $c=O(k \log k / \varepsilon^2)$
columns and $r=O(c \log c / \varepsilon^2)$ rows~\cite{DMM08}. Additionally, the algorithms in~\cite{DMM08,WZ13} construct $\matU$ with $\rank(\matU)=\omega(k)$, while we obtain $\rank(\matU) = k$,
which is also optimal, up to a constant $2$, according to the lower bound in Theorem~\ref{thm:lower2}.

The running time of the algorithm in Theorem~\ref{thm:main1r} 
is proportional to the number of the non-zero entries of $\matA$.
Recent progress on sketching-based numerical linear algebra~\cite{CW13,MM13,JH13} has provided very accurate approximation algorithms that run in time
proportional to the number of the non-zero entries of the input matrix (input-sparsity-time).
Although~\cite{CW13,MM13,JH13} study several important problems, including low-rank matrix approximation and 
least-squares regression,
the question of whether  an input-sparsity-time CUR
algorithm exists remained unexplored.
To the best of our knowledge, the only prior work addressing CUR  on sparse matrices is in
\cite{BPSS05,Ste99}. Although high quality implementations are provided, the theoretical results are lacking.

An important open problem in~\cite{DMM08,Sublinear12,mcgregor2006open} is whether there exists a polynomial-time \emph{deterministic} relative-error CUR algorithm.
We address this in Theorem~\ref{thm:main2}. The second main contribution of this work, 
our deterministic, relative-error CUR algorithm runs in time $O( mn^3 k/\varepsilon)$.
Additionally, it constructs $\matC$ with $c=O(k/\varepsilon)$ columns, $\matR$ with $r=O(k/\varepsilon)$ rows, and
$\matU$ of rank $k$. The work in~\cite{GS12}
provides a relative-error, deterministic algorithm, based on volume sampling, constructing $\matC$ with  $O(k / \varepsilon)$ columns of $\matA$ such that
$\FNormS{\matA - \matC \pinv{\matC} \matA} \le (1+\varepsilon) \FNormS{\matA -\matA_k}.$ It is not obvious how to
extend this to a rank $k$, column-based, relative-error decomposition, which is a harder instance of the problem.
The best deterministic algorithm for a rank $k,$ column-based, decomposition achieves 
$\FNormS{\matA - \matC \pinv{\matC} \matA} \le (2+\varepsilon) \FNormS{\matA -\matA_k},$
when $\matC$ has $O(k / \varepsilon)$ columns of $\matA$~\cite{BDM11a}.

\subsection{Overview of techniques}\label{sec:over}
In this section, we give a high level overview of our approach and explain
the key points of our CUR algorithms. Our starting point is the following 
tool from prior work which connects matrix factorizations and column subset selection.
\begin{lemma}[Lemma 3.1 in~\cite{BDM11a}]
\label{lem:structural0}
Let $\matA = \matA \matZ \matZ\transp + \matE \in \R^{m \times n}$ be a low-rank matrix factorization of $\matA$, with $\matZ \in \R^{n \times k},$
and $\matZ\transp\matZ=\matI_{k}$.
Let $\matS\in\R^{n\times c}$ ($c \ge k$) be any matrix such that $rank(\matZ\transp \matS) =
rank(\matZ)=k.$
Let $\matC = \matA \matS \in \R^{m \times c}$. Then,
$$
\FNormS{\matA - \matC \pinv{\matC} \matA} \le
  \FNormS{\matA - \Pi^{\mathrm{F}}_{\matC,k}(\matA)} \le
\FNormS{\matE} + \FNormS{\matE\matS (\matZ\transp \matS)^{\dagger}}.
$$
Here, $\Pi^{\mathrm{F}}_{\matC,k}(\matA) = \matC \matX_{opt} \in \R^{m \times n},$ 
where $\matX_{opt}\in \R^{c \times n}$ has rank at most $k$ and $\matC \matX_{opt}$ is the
best rank $k$ approximation to $\matA$ in the column span of $\matC$.
\end{lemma}

If $\matS$ samples columns from $\matA,$ i.e., $\matC = \matA \matS$ consists of columns of $\matA$, then, using this lemma, one obtains a bound for a column-based,
low-rank matrix approximation of $\matA$ (this bound depends on the specific choice of $\matZ$).
The crux of our approach in obtaining an optimal, relative-error, rank $k$ CUR is to use this lemma \emph{twice} and \emph{adaptively}, one time to sample columns from $\matA$
and the other time to sample rows. Here, by adaptively we mean that the two sampling processes are not independent from each other: the rows sampled in the second application
of the lemma depend on the columns sampled in the first application.
Next, we present precisely an overview of this approach.

Assume that for an appropriate matrix $\matZ_1 \in \R^{n \times k}$ we can find $\matS_1 \in \R^{n \times c_1}$ that samples $c_1 = O(k)$ columns from $\matA$
such that after applying Lemma~\ref{lem:structural0} with $\matA$ and $\matZ_1$ gives ($\matC_1 = \matA \matS_1; \matE_1 = \matA - \matA \matZ_1 \matZ_1\transp$):
\eqan{
\FNormS{ \matA - \matC_1\pinv{\matC}_1 \matA } &\le& \FNormS{\matE_1} + \FNormS{\matE_1\matS_1 (\matZ_1\transp \matS_1)^{\dagger}} \\
&\le& O(1) \FNormS{\matA -\matA \matZ_1 \matZ_1\transp} \\
&\le&  O(1) \FNormS{\matA - \matA_k},
}
where in the third inequality we further assumed that
$$
\FNormS{\matE_1} = \FNormS{\matA - \matA \matZ_1 \matZ_1\transp} \le O(1) \FNormS{\matA - \matA_k},
$$
i.e., $\matZ_1$ approximates, within a constant factor, the best rank $k$ SVD of $\matA$ (one obvious choice for $\matZ_1$ is the matrix $\matV_k$ from the rank $k$ SVD of $\matA$; however,
since this choice is costly we will use methods which approximate the SVD - see Section~\ref{sec:approxSVD}).
The bound in the second inequality also requires that the term $\FNormS{\matE_1\matS_1 \pinv{(\matZ_1\transp \matS_1)}}$ is sufficiently small, specifically
$
\FNormS{\matE_1\matS_1 \pinv{(\matZ_1\transp \matS_1)}} \le O(1) \FNormS{\matE_1}.
$
This can be achieved by choosing the sampling matrix $\matS_1$ carefully (we discuss below what choices of the matrix $\matS_1$ are appropriate).

Once we have this matrix $\matC_1$ with $O(k)$ columns of $\matA,$ that approximates the best rank $k$ matrix within a constant factor,
we can use the adaptive sampling method of~\cite{DRVW06} (see Lemma~\ref{lem:adaptivecolumns} in our paper) and additionally sample
$O(k/\varepsilon)$ columns to obtain a matrix $\matC$ with $c=O(k)+ O(k/\varepsilon)$ columns for which
\begin{equation}\label{eqn:crux15}
\FNormS{\matA - \matC \pinv{\matC} \matA} \le
\FNormS{ \matA - \matC \matX_{opt} } \le  (1 + \varepsilon) \FNormS{\matA - \matA_k}.
\end{equation}
Here, the adaptive sampling step turns a constant factor approximation to a relative-error one by sampling an additional of $O(k/\varepsilon)$ columns.
$\matX_{opt}\in \R^{c \times n}$ has rank at most $k$ and $\matC \matX_{opt}$ is the best rank $k$ approximation to $\matA$ in $span(\matC)$.
(See Section~\ref{sec:bestrankk} for the exact computation of $\matX_{opt}$). 
This adaptive sampling step to turn a constant factor approximation to a relative-error one by increasing the number of columns slightly has been previously used in the near-optimal algorithms in~\cite{BDM11a}. 
This $\matC$ would be the matrix with columns of $\matA$ in the  CUR factorization that we aim.

The main idea now is to use again Lemma~\ref{lem:structural0} and sample rows from $\matA$, i.e., apply the lemma to $\matA\transp$.
If $\matC$ had orthonormal columns, then it could immediately play the role of $\matZ$ in Lemma~\ref{lem:structural0}. In that case, $\matE = \matA\transp - \matA\transp \matC \matC\transp,$ and
we already have a bound for that term from the discussion above. 
However, $\matC$ is not orthonormal. But we could hope that we can find such an orthonormal matrix with a similar bound as in Eqn.~\ref{eqn:crux15}.
An obvious choice is to take $\matZ$ to be an orthonormal basis of $\matC$. However, this choice is not desirable because in that case $\matZ$ would have dimensions $m \times c;$
hence, in order to satisfy the rank assumption in Lemma~\ref{lem:structural0}, we would need to sample at least $c$ rows from $\matZ \in \R^{m \times c}$. 
Recall that $c=O(k/\varepsilon)$, hence $O(k/\varepsilon)$ rows would be needed to be sampled in this step. 
This is not desirable because we aim to obtain a CUR decomposition with only $O(k/\varepsilon)$ rows,
and to achieve that, we cannot afford to select $O(k/\varepsilon)$ rows at this step. We can only afford $O(k)$ rows
(because otherwise the adaptive sampling step in the next step, which seems unavoidable, 
would sample $O(k/\varepsilon^2)$ rows from $\matA$).
So, what we really want is some matrix $\matZ$ in the column span of $\matC$ such that  $\matZ \matZ\transp\matA$ approximates 
$\matA$ as good as $\matC \matX_{opt}$ and at the same time the number of columns of $\matZ$ is $O(k)$.
Towards this end, we construct (see Lemma~\ref{lem:KVW})
$\matZ_2\in \R^{m \times k}$ ($\matZ_2 \in span(\matC)$) with
\begin{equation}\label{eqn:crux2}
\FNormS{\matA\transp - \matA\transp \matZ_2 \matZ_2\transp} \le
\left(1+O(\varepsilon) \right) \FNormS{\matA - \matC \matX_{opt}}.
\end{equation}

Now, we want to apply Lemma~\ref{lem:structural0} to $\matA\transp$ and $\matZ_2\in \R^{m \times k}$.
Assume that we can find $\matS_2 \in \R^{m \times r_1}$, i.e.,
$\matR_1 = (\matA\transp \matS_2)\transp \in \R^{r \times n},$ with $r_1 = O(k)$ rows, such that
($\matE_2\transp = \matA\transp - \matA\transp \matZ_2\transp \matZ_2 
$): 
\eqan{
\FNormS{ \matA -  \matA \pinv{\matR}_1 \matR_1 } 
&\le& \FNormS{\matE_2} + \FNormS{\matE_2 \matS_2 \pinv{ (\matZ_2\transp \matS_2) }} \\
&=& \FNormS{\matA - \matZ_2 \matZ_2\transp \matA} + \FNormS{(\matA -  \matZ_2 \matZ_2\transp \matA) \matS_2 \pinv{ (\matZ_2\transp \matS_2) }} \\
&\le& O(1)  \FNormS{\matA -  \matZ_2 \matZ_2\transp \matA} \hspace{1.2in} (\text{\ref{eqn:crux3}}) \label{eqn:crux3}
}
The last inequality in Equation~\ref{eqn:crux3} also requires that 
$
\FNormS{\matE_2\matS_2 (\matZ_2\transp \matS_2)^{\dagger}} = O(1) \FNormS{\matE_2},
$
which is possible after constructing the matrix $\matS_2$ appropriatelly.

So far, we have $\matC$ and a subset of rows of $\matR$ in the optimal CUR that we would like to construct.
We need two additional steps:
(i) we use the adaptive sampling method for CUR that appeared in~\cite{WZ13} and further sample another $r_2 = O(k/\varepsilon)$ rows in $\matR_1$, i.e., construct 
$\matR \in \R^{r \times n}$ with
$r = O(k + k/\varepsilon)$ rows, and (ii) for $\matZ_2\matZ_2\transp \matA \pinv{\matR} \matR$, we replace $\matZ_2$ with $\matC \matTh,$ for an appropriate $\matTh$
(such a $\matTh$ always exists because $\matZ_2$ is in the span of $\matC$)  and take
$\matU = \matTh \matZ_2\transp \matA \pinv{\matR} \in \R^{c \times r}$. So overall,
$$ \matC \matU \matR = \matZ_2\matZ_2\transp \matA \pinv{\matR} \matR,$$ 
and
the bound is,
\eqan{
\FNormS{ \matA - \matZ_2 \matZ_2\transp \matA \pinv{\matR}\matR }
&\le& \FNormS{ \matA - \matZ_2 \matZ_2\transp \matA }  + \frac{\rank(\matZ_2\matZ_2\transp\matA)}{r_2}  \FNormS{\matA-\matA\pinv{\matR}_1\matR_1} \\ 
&=&   \FNormS{ \matA -  \matZ_2 \matZ_2\transp\matA } +O(\varepsilon) \FNormS{\matA-\matA\pinv{\matR}_1\matR_1}  \\                               
&\le& \left(1+O(\varepsilon) \right) \FNormS{ \matA -  \matZ_2 \matZ_2\transp \matA }   \\               							           
&\le& \left(1+O(\varepsilon) \right) \FNormS{\matA-\matC\matX_{opt} } 	\\														    
&\le& \left(1+O(\varepsilon) \right) \FNormS{\matA-\matA_k}                  															   
}
The first inequality is from the adaptive sampling argument (see also Lemma~\ref{lem:adaptiverows}).
The equality is from our choice of $r_2$.
The second inequality is from Eqn.~\ref{eqn:crux3}.
The third inequality is from Eqn.~\ref{eqn:crux2}. 
The last inequality is from Eqn.~\ref{eqn:crux15}. 

From all the above derivations, we now identify four
\emph{sufficient conditions} that give an optimal, relative-error, rank $k$ CUR. 
We call these conditions \emph{primitives}. Designing matrices $\matC, \matU,$ and
$\matR$ satisfying those basic primitives, an optimal, relative-error, rank $k$ CUR is secured. 

\subsubsection{CUR Primitives}
To construct an optimal, relative-error, and rank-$k$ CUR, we relied on four basic primitives:
\begin{enumerate}
\item  There is $\matC \in \R^{m \times c}$, $\matX_{opt} \in \R^{c \times n}$ with a relative-error bound to $\matA-\matA_k$ and $c = O(k / \varepsilon)$: Equation~\ref{eqn:crux15}.
\item  There is 
$\matZ_2 \in \R^{m \times k}$ ($\matZ_2 \in span(\matC)$ and $\matZ_2\transp \matZ_2 = \matI_k$)
with a relative-error bound to $\matA-\matC\matX_{opt}$: Eqn.~\ref{eqn:crux2}.
\item  There is $\matR_1 \in \R^{r_1 \times n}$ with a constant factor error to $\matA- \matZ_2 \matZ_2\transp\matA$ and $r_1 = O(k)$: Equation~\ref{eqn:crux3}.
\item  There is an adaptive sampling algorithm for CUR, i.e., an algorithm
that turns a constant factor CUR with $O(k/\varepsilon)$ columns and $O(k)$ rows
to a relative-error CUR by sampling only $O(k/\varepsilon)$ additional rows. 
\end{enumerate}
The user can pick $\matC, \matZ_2, \matR_1$ in the above conditions the way she likes. Below, 
we discuss various implementation choices which lead to desirable CUR algorithms. 

\subsubsection{CUR proto-algorithm}
To address primitive (1), we combine known ideas for
column subset selection including leverage-scores sampling~\cite{DMM08}, BSS sampling~\cite{BDM11a}
(i.e., deterministic sampling similar to the method of Batson, Spielman, and Srivastava~\cite{BSS09}) and adaptive sampling~\cite{DV06} (see Section~\ref{sec:sampling}).
To find $\matZ_1,$ we use techniques for approximating the SVD (see Section~\ref{sec:approxSVD}).
To address primitive (2)  we use methods to find low-rank approximations within a subspace (see Section~\ref{sec:bestF}).
To address primitive (3), again we employ leverage-score sampling~\cite{DMM08} and BSS sampling~\cite{BDM11a},
as in primitive (1). Finally, to address primitive (4),
we use an adaptive sampling method~\cite{WZ13} which turns a constant factor CUR to relative error by sampling an additional $O(k / \varepsilon)$ rows.
We summarize this in Algorithm~\ref{alg1}, which we call \emph{proto-algorithm} for an optimal, relative-error, rank $k$ CUR.
To obtain a deterministic or an input-sparsity-time CUR, we need to implement the corresponding steps in this proto-algorithm in the appropriate setting. We discuss those issues below.
\begin{algorithm*}[t]
\begin{framed}
\caption{A proto-algorithm for an optimal, relative-error, rank-$k$ CUR}
\label{alg1}
{\bf Input:} $\matA \in \R^{m \times n};$ rank parameter $k < \rank(\matA);$ accuracy parameter $0 < \varepsilon < 1$. \\
{\bf Output:} $\matC \in \R^{m \times c}$ with $c=O(k/\varepsilon)$;  $\matR \in \R^{r \times n}$  with $r=O(k/\varepsilon)$;  $\matU \in \R^{c \times r}$ with $\rank(\matU) = k$.\\
{\bf 1. Construct $\matC$ with $O(k + k / \varepsilon)$ columns}
\begin{algorithmic}[1]
\STATE \emph{Approximate SVD}: find $\matZ_1 \in \R^{n \times O(k)}$ that approximates  $\matV_k \in \R^{n \times k}$ from the SVD of $\matA$. \textcolor[rgb]{0,0,1}{Primitive 1}
\STATE \emph{Leverage-scores sampling}: Sample $O(k \log k)$ columns from $\matA$ with probabilities from $\matZ_1$. \textcolor[rgb]{0,0,1}{Primitive 1}
\STATE \emph{BSS sampling}: Downsample these columns to $O(k)$ columns. \textcolor[rgb]{0,0,1}{Primitive 1}
\STATE \emph{Adaptive sampling}: Sample $O(k/\varepsilon)$ additional columns. \textcolor[rgb]{0,0,1}{Primitive 1}
\end{algorithmic}
{\bf 2. Construct $\matR$ with $O(k + k / \varepsilon)$ rows}
\begin{algorithmic}[1]
\STATE \emph{Restricted low-rank approximation}: find $\matZ_2 \in \R^{m \times O(k)}$ with $\matZ_2 \in span(\matC)$. \textcolor[rgb]{0,0,1}{Primitive 2}
\STATE \emph{Leverage-scores sampling}: Sample $O(k \log k)$ rows from $\matA$ with probabilities from $\matZ_2$. \textcolor[rgb]{0,0,1}{Primitive 3}
\STATE \emph{BSS sampling}: Downsample these rows to $O(k)$ rows. \textcolor[rgb]{0,0,1}{Primitive 3}
\STATE \emph{Adaptive sampling}: Sample $O(k/\varepsilon)$ additional rows. \textcolor[rgb]{0,0,1}{Primitive 4}
\end{algorithmic}
{\bf 3. Construct $\matU$ of rank $k$}
\begin{algorithmic}[1]
\STATE Let $\matZ_2\matZ_2\transp \matA \pinv{\matR} \matR$; then, replace $\matZ_2 = \matC \matTh,$ for an appropriate $\matTh,$
and use 
$$\matU = \matTh \matZ_2\transp \matA \pinv{\matR}.$$
\end{algorithmic}
\end{framed}
\end{algorithm*}

\subsubsection{Input-sparsity-time CUR} To design an algorithm that runs in time proportional to the number of non-zero entries of $\matA,$ we need to implement all these
primitives in input-sparsity-time, or equivalently, all the steps in the proto-algorithm should be implemented in input-sparsity-time.
It is already known in the literature how to implement approximate SVD in input-sparsity-time (see Section~\ref{sec:alg3}).
The leverage-scores sampling step - given that one knows the probabilities - can be implemented fast as well~(see Lemma~\ref{lem:random}).
For the rest of the steps of the proto-algorithm
we develop new tools.
First, we design an input-sparsity-time version of the BSS sampling step (see Lemma~\ref{lem:dualsets}). To do that, we combine the method from~\cite{BDM11a} with ideas from the sparse subspace embedding literature~\cite{CW13}.
Second, we develop input-sparsity-time versions of the adaptive sampling algorithms of~\cite{DV06,WZ13} (see Lemma~\ref{lem:adaptivecolumnss} and Lemma~\ref{lem:adaptiverowss}). To do that, we combine existing ideas in~\cite{DV06,WZ13} with the Johnson-Lindenstrauss lemma (see Lemma~\ref{lem:jlt}).
Third, we develop an input-sparsity-time algorithm to find a relative-error (to the best possible) low-rank matrix within a given subspace (see Lemma~\ref{lem:KVW}). To do that, we combine ideas for subspace-restricted matrix approximations (see Section~\ref{sec:bestrankk}) with ideas from~\cite{KVW13,CW13}.
Fourth, we present a method to compute a rank $k$ matrix $\matU$ in input sparsity time. To do that, we took the original construction of $\matU$, which is a series of products of various matrices, and reduced the dimensions of some of those matrices using the sparse subspace embeddings in~\cite{CW13} (see Lemma~\ref{lem:sparseU}). The crux in this part is to ``view'' the computation of $\matU$ as the solution of a generalized matrix approximation problem (see Section~\ref{sec:Uopt}) and then apply sketching ideas to this problem.

\subsubsection{Deterministic CUR} To design a deterministic CUR, we need to implement the CUR proto-algorithm in a deterministic way.
All steps can be implemented deterministically with tools from prior work (Section~\ref{sec:alg2}).
The only piece missing is a deterministic version of adaptive sampling~\cite{DV06,WZ13}, which we obtain
by derandomizing the adaptive sampling algorithms in~\cite{DV06,WZ13}~(Section~\ref{sec:newDet}).


\section{Related work}\label{sec:prior}
Randomized algorithms for CUR~\cite{DK03,DKM06c,DMM08,DM09,GM13,WZ13} are an instance
of a large body of work on approximation algorithms for matrix computations~\cite{FKV98,DK01,DK03,DKM06a,DKM06b,DKM06c,DV06,Sar06,RT08,CW09,HMT,CW13,BG13,ZF13}.
Those algorithms provide a new paradigm and a complementary perspective in speeding up basic kernels in numerical linear algebra, such as matrix multiplication~\cite{DK01}, least-squares regression~\cite{Sar06},
and low-rank matrix approximation~\cite{BG13}. The application areas of those algorithms are broad, as discussed in the survey article~\cite{HMT}.

Before discussing the details of some of the available CUR algorithms in~\cite{DK03,DKM06c,DMM08,DM09,GM13,WZ13}, we briefly mention a similar problem
which constructs factorizations of the form $\matA= \matC \matX + \matE,$ where $\matC$ contains columns of $\matA$ and $\matX$ has rank at
most $k$. Unlike CUR, there are optimal algorithms for this problem~\cite{BDM11a,GS12}, in both the spectral and the Frobenius norm. Indeed, to
obtain a relative-error optimal CUR in this paper we use a sampling method from~\cite{BDM11a}, which allows to select $O(k)$ columns and rows.
For a more detailed discussion of this CX problem, which is also known as CSSP (Column Subset Selection Problem) see~\cite{BMD09a,BDM11a,GS12}.

Drineas and Kannan brought CUR factorizations to the theoretical computer science community in~\cite{DK03}.
Their main algorithm is randomized and samples columns and rows from $\matA$ with probabilities proportional
to their Euclidian length. The running time of this algorithm is linear in $m$ and $n$ and proportional to a
small-degree polynomial in $k$ and $1/\varepsilon,$ for some $\varepsilon >0$,
but the approximation bound is weak (see Theorem 3.1 in~\cite{DK03}): with $c = O(k/\varepsilon^2)$ columns and $r=O(k/\varepsilon)$ rows
the bound is 
$
\FNormS{\matA - \matC \matU \matR} \le \FNormS{\matA - \matA_k} + \varepsilon \FNormS{\matA}.
$
Drineas, Kannan, and Mahoney~\cite{DKM06c} built upon~\cite{DK03} but the error remained additive (see Theorem 5 in~\cite{DKM06c}):
with $c = O(k/\varepsilon^4)$ columns and $r = O(k/\varepsilon^2)$ rows the error is
$
\FNormS{\matA - \matC \matU \matR} \le \FNormS{\matA - \matA_k} +  \varepsilon  \FNormS{\matA}.
$
The first relative-error CUR algorithm appeared in~\cite{DMM08} (see Theorem 2 of \cite{DMM08}). The
algorithm of~\cite{DMM08} is based on subspace sampling and requires 
$c = O(k \log(k / \varepsilon^2)  \log \delta^{-1})$ columns and
$r=O(c \log (c / \varepsilon^2) \log \delta^{-1})$ rows to construct a relative-error CUR with failure probability $\delta$.
The running time of the method in~\cite{DMM08} is $O( m n \min\{m,n\} )$, since subspace sampling is based on sampling with probabilities
proportional to the so-called leverage scores, i.e., the row norms of the matrix $\matV_k$ from the SVD of $\matA$.
Mahoney and Drineas~\cite{DM09}, using again subspace sampling, improved slightly upon the number of columns and rows,
compared to~\cite{DMM08}, but achieved only a constant factor error (see Eqn.(5) in~\cite{DM09}).
Gittens and Mahoney~\cite{GM13} discuss CUR decompositions on SPSD matrices and present approximation bounds for Frobenius,
trace, and spectral norms (see Lemma 2 in~\cite{GM13}).
Finally, the current state-of-the-art, relative-error
CUR algorithm is in~\cite{WZ13}.
Using the near-optimal column subset selection methods in~\cite{BDM11a} along with a novel adaptive sampling technique, Wang and Zhang present a CUR
algorithm selecting
$c = (2k/\varepsilon)(1+o(1))$
columns and
$r = (2k/\varepsilon^2)(1+\varepsilon)(1+o(1))$
rows from $\matA$
(see Theorem 8 in~\cite{WZ13}).
The running time of this algorithm is
$$
O( mnk\varepsilon^{-1} + mk^3\varepsilon^{-\frac{2}{3}} + nk^3\varepsilon^{-\frac{2}{3}} + mk^2\varepsilon^{-2} + nk^2\varepsilon^{-4}).
$$
We summarize all these CUR algorithms as well as the three algorithms of our work in Table~\ref{table:summary}.

\begin{table*}
\begin{center}
  \begin{tabular}{| l | l | l | l | l | l| }
    \hline
     Reference           & $c$                          & $r$                          & $u$   & $\FNormS{\matA-\matC\matU\matR}\le$
                                                                                                                                                    & \text{Time, if $ m=\Theta(n)$} \\ \hline
     Th 3.1~\cite{DK03}   & $O(k/\varepsilon^2)$         & $O(k/\varepsilon)$          & $k$  & $\FNormS{\matA-\matA_k} + \varepsilon\FNormS{\matA}$ & $O(nnz(A)+poly(k,1/\varepsilon))$ \\ \hline
     Thm 5~\cite{DKM06c} & $O(k/\varepsilon^4)$         & $O(k/\varepsilon^2)$          & $k$  & $\FNormS{\matA-\matA_k} + \varepsilon\FNormS{\matA}$ &  $O(nnz(A)+poly(k,1/\varepsilon))$ \\ \hline
     Thm 2~\cite{DMM08}  & $O(\frac{k \log k}{\varepsilon^2})$ & $O(\frac{c \log c}{ \varepsilon^2})$ & $c$     & $(1+\varepsilon)\FNormS{\matA-\matA_k}$ & $O( n^3 )$ \\ \hline
     Eqn. 5~\cite{DM09}   & $O(\frac{k \log k}{\varepsilon^2})$ & $O(\frac{k \log k}{\varepsilon^2})$  & $c$    & $(2+\varepsilon)\FNormS{\matA-\matA_k}$ & $O( n^3 )$ \\ \hline
     Thm 8~\cite{WZ13}   & $O(k/\varepsilon)$           & $O(k/\varepsilon^2)$          & $c$ & $(1+\varepsilon)\FNormS{\matA-\matA_k}$ & $O( n^2k/\varepsilon + nk^3/\varepsilon^{\frac{2}{3}} + nk^2/\varepsilon^{4})$ \\ \hline
     Thm~\ref{thm:main1} & $O(k/\varepsilon)$ & $O(k/\varepsilon)$ & $k$ & $(1+\varepsilon)\FNormS{\matA-\matA_k}$ & $O( n^2 k/ \varepsilon  +n \cdot poly(k, \log(k/\varepsilon), 1/\varepsilon))$ \\ \hline
     Thm~\ref{thm:main2} & $O(k/\varepsilon)$ & $O(k/\varepsilon)$ & $k$ & $(1+\varepsilon)\FNormS{\matA-\matA_k}$ & $O( n^4 k/\varepsilon)$ \\ \hline
     Thm~\ref{thm:main3} & $O(k/\varepsilon)$ & $O(k/\varepsilon)$ & $k$ & $(1+\varepsilon)\FNormS{\matA-\matA_k}$ & $O( nnz(\matA)\log n + n \cdot poly(\log n, k,1/\varepsilon) )$ \\ \hline
  \end{tabular}
\end{center}
\caption{  CUR algorithms constructing $\matC$ with $c$ columns, $\matR$ with $r$ rows, and $\matU$ with rank $u$.}\label{table:summary}
\end{table*}

We now  explain why the two relative-error CUR algorithms in Table~\ref{table:summary}~\cite{DMM08,WZ13} require more columns and rows in $\matC$ and $\matR,$ and larger $\rank(\matU)$,
than our optimal CUR algorithms.
First, both algorithms~\cite{DMM08,WZ13} use Lemma~\ref{lem:structural0} twice and adaptively, as we do
(this is not explicitly stated in those articles but this claim can be validated after a careful comparison of their proofs with ours).
Drineas et al.~\cite{DMM08} implement the first step of the proto-algorithm in Algorithm~\ref{alg1} via leverage-scores sampling \emph{only}.
At that time, the BSS sampling was unknown and the adaptive sampling step wouldn't be particularly helpful. For the second step in the proto-algorithm,
they simply choose $\matZ_2$ as an orthonormal basis for $\matC,$ hence $r = O(c \log c / \varepsilon^2)$ rows are required, where
$c = O(k \log k / \varepsilon^2)$ are the columns selected in the first step. Given the above limitations, in the third step of the proto-algorithm,
$\matU$ unavoidably has rank
$O(k \log k / \varepsilon^2)$. Wang and Zhang on the other hand~\cite{DMM08}, motivated by the near-optimal algorithm of Boutsidis et al.~\cite{BDM11a},
take advantage of BSS  and adaptive sampling to sample only $c = O(k/\varepsilon)$ columns in the first step of the proto-algorithm. However, they also use $\matZ_2$ as an orthonormal basis for $\matC$;
hence $r = O(k/\varepsilon^2)$ and $\rank(\matU) = O(k/\varepsilon)$. The $r = O(k/\varepsilon^2)$ term here is because the row-selection adaptive sampling step requires $O(\rho / \varepsilon)$ rows, where $\rho = \rank(\matZ_2\matZ_2\transp\matA)$.

Finally, there are several interesting results on CUR developed within the numerical linear algebra community~\cite{Tyr96, Tyr00, GTZ97a, GTZ97b,HP97, Pan03, MG03, GM04,BPSS05,Ste99}.
For example,~\cite{Tyr96, Tyr00, GTZ97a, GTZ97b} discuss the so-called skeleton approximation, which focuses on the spectral norm
version of the CUR problem via selecting exactly $k$ columns and $k$ rows. The algorithms there are deterministic, run
in time proportional to the time to compute the rank $k$ SVD of $\matA$, and achieve bounds of the order,
$$\TNorm{\matA-\matC \matU\matR} \le O( \sqrt{k(n-k)} + \sqrt{k(m-k)} )\TNorm{\matA-\matA_k}.$$


\section{Column subset selection tools from existing literature}\label{sec:back1}
This section summarizes known techniques and related results from the literature that we use throughout this work.

\subsection{A perturbation bound for column-based matrix reconstruction}
We start with a restatement of Lemma~\ref{lem:structural0}.
Recall that $\matA$ is the matrix that we would like to approximate with a rank $k$ matrix $\matC \matX,$
where $\matC$ contains columns of $\matA$, or a rank $k$ matrix $\matC \matU \matR$ where $\matC$ contains
columns of $\matA$ and $\matR$ contains rows of $\matA$. For an appropriate \emph{sampling} matrix $\matS,$
we can write $\matC = \matA \matS$. The following lemma provides a general perturbation bound for such sampling matrices $\matS,$
under some rank assumption. We will use this lemma extensively when analyzing our $\matC \matU \matR$ algorithms.
\begin{lemma}[Lemma 3.1 in~\cite{BDM11a}]
\label{lem:structural}
Let $\matA = \matA \matZ \matZ\transp + \matE \in \R^{m \times n}$, with $\matZ \in \R^{n \times k},$
and $\matZ\transp\matZ=\matI_{k}$.
Let $\matS\in\R^{n\times r}$ be any matrix such that $rank(\matZ\transp \matS) =
rank(\matZ)=k.$
Let $\matC = \matA \matS \in \R^{m \times r}$. Then, 
\begin{equation}\label{eqn1_pert}
\FNormS{ \matA - \matC \pinv{\matC}\matA } \le  \FNormS{\matA - \Pi^{\mathrm{F}}_{\matC,k}(\matA)} \le
\FNormS{ \matA - \matC \pinv{(\matZ\transp\matS)}\matZ\transp } \le
\FNormS{\matE} + 
\FNormS{\matE\matS \pinv{(\matZ\transp \matS)}}.
%
%
%
\end{equation}
Here, 
$$\Pi^{\mathrm{F}}_{\matC,k}(\matA) = \matC \matX_{opt} \in \R^{m \times n},$$
 where $\matX_{opt}\in \R^{c \times n}$ has rank at most $k$, $\matC \matX_{opt}$ is the
best rank $k$ approximation to $\matA$ in $span(\matC)$, and $\pinv{(\matZ\transp \matS)}$ 
denotes the Moore-Penrose pseudoinverse of $\matZ\transp \matS$.
\end{lemma}

\subsection{SVD and the pseudo-inverse}\label{sec:svd}
The singular value decomposition (SVD) appears often throughout the paper.
The SVD of $\matA \in \R^{m \times n}$ of rank $\rho \le \min\{m,n\}$ is
\begin{eqnarray}\label{eqn:svd}
\label{svdA} \matA
         = \underbrace{\left(\begin{array}{cc}
             \matU_{k} & \matU_{\rho-k}
          \end{array}
    \right)}_{\matU_{\matA} \in \R^{m \times \rho}}
    \underbrace{\left(\begin{array}{cc}
             \matSig_{k} & \bf{0}\\
             \bf{0} & \matSig_{\rho - k}
          \end{array}
    \right)}_{\matSig_\matA \in \R^{\rho \times \rho}}
    \underbrace{\left(\begin{array}{c}
             \matV_{k}\transp\\
             \matV_{\rho-k}\transp
          \end{array}
    \right)}_{\matV_\matA\transp \in \R^{\rho \times n}},
\end{eqnarray}
with singular values \math{\sigma_1\left(\matA\right)\ge\ldots\geq \sigma_k\left(\matA\right)\geq\sigma_{k+1}\left(\matA\right)\ge\ldots\ge\sigma_\rho\left(\matA\right) > 0}.
The matrices
$\matU_k \in \R^{m \times k}$ and $\matU_{\rho-k} \in \R^{m \times (\rho-k)}$ contain the left singular vectors of~$\matA$; and, similarly, the matrices $\matV_k \in \R^{n \times k}$ and $\matV_{\rho-k} \in \R^{n \times (\rho-k)}$ contain the right singular vectors.
$\matSig_k \in \R^{k \times k}$ and $\matSig_{\rho-k} \in \R^{(\rho-k) \times (\rho-k)}$ contain the singular values of~$\matA$.
It is well-known that $\matA_k=\matU_k \matSig_k \matV_k\transp$ minimizes \math{\FNorm{\matA - \matX}} over all
matrices \math{\matX \in \R^{m \times n}} of rank at most $k \le \rho$. We use $\matA_{\rho-k} = \matA - \matA_k = \matU_{\rho-k}\matSig_{\rho-k}\matV_{\rho-k}\transp$. Also, $ \FNorm{\matA} =
\sqrt{ \sum_{i=1}^\rho\sigma_i^2(\matA) }$ and $\TNorm{\matA} = \sigma_1(\matA)$. The best rank $k$ approximation to $\matA$ satisfies:
$\TNorm{\matA-\matA_k} = \sigma_{k+1}(\matA)$;
$\FNorm{\matA-\matA_k} = \sqrt{\sum_{i=k+1}^{\rho}\sigma_{i}^2(\matA)}$.

\subsubsection{Pseudo-inverse}
$\pinv{\matA} = \matV_\matA \matSig_\matA^{-1} \matU_\matA\transp \in \R^{n \times m}$
denotes the so-called Moore-Penrose pseudo-inverse of $\matA \in \R^{m \times n}$ (here $\matSig_\matA^{-1}$ is the inverse of $\matSig_\matA$),
i.e., the unique $n \times m$ matrix satisfying all four properties:
$\matA = \matA \pinv{\matA} \matA$,
$\pinv{\matA} \matA \pinv{\matA} = \pinv{\matA}$,
$(\matA \pinv{\matA} )\transp = \matA \pinv{\matA}$, and
$(\pinv{\matA} \matA )\transp = \pinv{\matA}\matA$.
By the SVD of $\matA$ and $\pinv{\matA}$, it is easy to verify
that, for all $i=1,\dots,\rho = \rank(\matA) = \rank(\pinv{\matA})$:
$\sigma_i(\pinv{\matA}) = \frac{1}{\sigma_{\rho - i + 1}(\matA)}.$
Finally, for any $\matA \in \R^{m \times n}, \matB \in \R^{n \times \ell}$:
\math{\pinv{(\matA\matB)}=\pinv{\matB}\pinv{\matA}} if any one
of the following three properties hold:
(1) \math{\matA\transp\matA=\matI_n};
(2) \math{\matB\transp\matB=\matI_{\ell}};
or, (3) $\rank(\matA) = \rank(\matB) = n$.

\subsection{Fast approximate low-rank matrix approximations}\label{sec:approxSVD}
The SVD provides the best rank $k$ matrix $\matA_k$ to approximate $\matA;$ however, it is somewhat costly to compute.
In this work, to speedup our $\matC \matU \matR$ algorithms, which extensively make use of the SVD, we use low-rank matrix factorization algorithms
that are approximately as good as the SVD but can be implemented  considerably faster (e.g., in linear time or in input sparsity time). We describe three such algorithms below. Indeed, we omit the details of the algorithms per se, instead we
discuss the approximation bounds and the corresponding running times. 

\subsubsection{Deterministic approximate SVD}
Recent work in~\cite{GP13} describes a deterministic algorithm for relative-error approximate low-rank matrix factorization that requires sub-cubic arithmetic operations.
\begin{lemma}[Theorem 3.1 in~\cite{GP13}]\label{lem:approxSVDdet}
Given \math{\matA\in\R^{m\times n}} of rank $\rho$, a target rank $1 \leq k < \rho$, and $0 < \epsilon \le 1$, there exists a deterministic algorithm that  computes
$\matZ \in \R^{n \times k}$ with $\matZ\transp\matZ = \matI_k$  and
$$\FNormS{\matA - \matA \matZ \matZ\transp} \leq \left(1+{\epsilon}\right)\FNormS{\matA - \matA_k}.$$
The proposed algorithm requires $O\left(m n k^2 \epsilon^{-2}\right)$ arithmetic operations. We denote this procedure as
$$\matZ = DeterministicSVD(\matA, k, \varepsilon).$$
\end{lemma}
\begin{proof}
Theorem 4.1 in~\cite{GP13} describes an algorithm that given $\matA, k$ and $\varepsilon$ constructs $\matQ_k \in \R^{k \times n}$
such that 
$\FNormS{\matA - \matA\matQ_k\transp\matQ_k} \leq \left(1+{\epsilon}\right)\FNormS{\matA - \matA_k}.$
 To obtain the desired
factorization, we just need an additional step to orthonormalize the columns of $\matQ_k\transp,$ which takes $O(nk^2)$ time. So, assume
that $\matQ_k\transp = \matZ \matR$ is a qr factorization of $\matQ_k\transp$ with $\matZ \in \R^{n \times k}$ and $\matR \in \R^{k \times k}$.
Then,
\eqan{
\FNormS{\matA - \matA \matZ \matZ\transp}
&\le& \FNormS{\matA - \matA \matQ_k\transp  (\matR\transp) \matZ} \\
&=& \FNormS{\matA - \matA \matQ_k\transp \matR\transp (\matR\transp)^{-1} \matQ_k} \\
&=& \FNormS{\matA - \matA \matQ_k\transp  \matQ_k} \\
&\le& \left(1+{\epsilon}\right)\FNormS{\matA - \matA_k}.
}
\end{proof}

\subsubsection{Randomized linear-time approximate SVD}
The following result corresponds to a standard randomized algorithm for speeding up the SVD.
\begin{lemma}[Lemma 3.4 in~\cite{BDM11a}]
\label{lem:approxSVD}
Given \math{\matA\in\R^{m\times n}} of rank $\rho$, a target rank $2\leq k < \rho$, and $0 < \epsilon \le 1$, there exists a randomized algorithm that  computes
$\matZ \in \R^{n \times k}$ with $\matZ\transp\matZ = \matI_k$  and
$$\Expect{\FNormS{\matA - \matA\matZ\matZ\transp}} \leq \left(1+{\epsilon}\right)\FNormS{\matA - \matA_k}.$$
The proposed algorithm requires $O\left(mnk\epsilon^{-1}\right)$ arithmetic operations. We denote this procedure as
$$\matZ = RandomizedSVD(\matA, k, \varepsilon).$$
\end{lemma}

\subsubsection{Randomized input-sparsity-time approximate SVD}
Clarkson and Woodruff~\cite{CW13} described a randomized algorithm that runs in time proportional to the number of non-zero entries
in $\matA$ plus low-order terms.

\begin{lemma}[Theorem 47 in~\cite{CW13arxiv}]\label{lem:approxSVDsparse}
Given \math{\matA\in\R^{m\times n}} of rank $\rho$, a target rank $1 \leq k < \rho$, and $0 < \epsilon \le 1$, there exists a randomized algorithm that  computes
$\matZ \in \R^{n \times k}$ with $\matZ\transp\matZ = \matI_k$  and with probability at least $0.99$,
$$\FNormS{\matA - \matA \matZ \matZ\transp} \leq \left(1+{\epsilon}\right)\FNormS{\matA - \matA_k}.$$
The proposed algorithm requires 
$O\left(  \nnz(\matA) \right) + \tilde{O}\left( nk^2 \varepsilon^{-4} + k^3\varepsilon^{-5} \right)$
arithmetic operations. We denote this procedure as
$$\matZ = SparseSVD(\matA, k, \varepsilon).$$
\end{lemma}
\begin{proof}
Theorem 47 in~\cite{CW13arxiv} describes an algorithm that constructs $\matL \in \R^{m \times k}$ with orthonormal columns, diagonal $\matD \in \R^{k \times k}$
and $\matW \in \R^{n \times k}$ with orthonormal columns such that 
$
\FNormS{\matA - \matL \matD \matW\transp} \leq \left(1+{\epsilon}\right)\FNormS{\matA - \matA_k}.
$
We can just use $\matZ = \matW,$ because $\FNormS{\matA - \matA \matW \matW\transp} \leq \FNormS{\matA - \matL \matD \matW\transp} $.
\end{proof}

\subsection{Column subset selection techniques}\label{sec:sampling}
We now summarize the various tools from existing literature that we use in order to sample columns and/or rows from matrices.

\subsubsection{Deterministic BSS sampling}
The lemma below is a generalization of the spectral sparsification method of Batson, Spielman, and Strivastava (BSS)~\cite{BSS09};
it was the main ingredient of the near-optimal algorithm in~\cite{BDM11a}.
\begin{lemma}[Dual Set Spectral-Frobenius Sparsification. Lemma 3.6 in~\cite{BDM11a}]
\label{lem:dualset}
Let \math{\cl V=\{\v_1,\ldots,\v_n\}} be a decomposition of the identity, where \math{\v_i\in\R^{k}} ($k < n$) and
$\sum_{i=1}^n\v_i\v_i\transp=\matI_{k}$; let \math{\cl A=\{\a_1,\ldots,\a_n\}} be an arbitrary set
of vectors, where \math{\a_i\in\R^{\ell}}. Then,
given an integer \math{r} such that \math{k < r \le n}, there exists a set of weights \math{s_i\ge 0} ($i=1\ldots n$), {at most \math{r} of which are non-zero}, such that
\eqan{
\lambda_{k}\left(\sum_{i=1}^ns_i\v_i\v_i\transp\right)
&\ge&
\left(1 - \sqrt{\frac{k}{r}}\right)^2,
\qquad \\
\trace\left(\sum_{i=1}^n s_i\a_i\a_i\transp\right)
&\le&
\trace\left(\sum_{i=1}^n \a_i\a_i\transp\right)
=
\sum_{i=1}^n \TNormS{\a_i}.
}
Equivalently,
if $\matV \in \R^{n \times k}$ is a matrix whose rows are the vectors $\v_i\transp$,
$\matA \in \R^{n \times \ell}$ is a matrix whose rows are the vectors $\a_i\transp$, and
$\matS \in \R^{n \times r}$ is the sampling matrix containing the weights $s_i > 0, $ then:
\eqan{
\sigma_{k}\left(\matV\transp\matS\right)
\ge
1 - \sqrt{{k}/{r}},
\qquad
\FNormS{\matA\transp\matS}
\le \FNormS{\matA\transp}.
}
The weights $s_i$ can be computed in $O\left(rnk^2+n\ell\right)$ time. We denote this procedure as
$$\matS = BssSampling(\matV, \matA, r).$$
\end{lemma}

\subsubsection{Randomized sampling}\label{sec:prior2}
Next, we introduce a randomized method for sampling rows from tall-skinny matrices.
\begin{definition}[Random Sampling with Replacement~\cite{RV07}] \label{def:sampling}
Let $\matX \in \R^{n \times k}$ with $n > k$; and $\x_i\transp \in \R^{1 \times k}$
denote the $i$-th row of $\matX$ and $0 < \beta \leq 1$.
For $i=1,...,n,$ if $\beta=1$, then $p_i = (\x_i\transp \x_i) / \FNormS{\matX}$,
otherwise compute some $p_i \geq \beta (\x_i\transp \x_i) / \FNormS{\matX}$ with $ \sum_{i=1}^{n} p_i = 1$.
Let $r$ be an integer with $ 1 \le r \le n$.
Construct a sampling matrix $\matOmega \in \R^{n \times r}$ and a rescaling matrix $\matD \in \R^{r \times r}$ as follows. Initially, $\matOmega = \bm{0}_{n \times r}$ and $\matD=\bm{0}_{r \times r}$.
Then, for every column $j=1,...,r$ of $\matOmega$, $\matD$, independently, and with replacement, pick an index $i$ from the set $\{1,2,...,n\}$ with probability $p_i$ and set $\matOmega_{ij} = 1$ and $\matD_{jj} = 1/\sqrt{p_i r}$. To denote this
$O(nk + n + r\log(r))$ time procedure we write
$$[\matOmega, \matD] = RandSampling(\matX, r, \beta).$$
\end{definition}
\begin{lemma}[Originally proved in \cite{RV07}]\label{lem:random}
Let $\matV \in \R^{n \times k}$ with $n > k$ and $\matV\transp \matV = \matI_{k}$.
Let $0 < \beta \le1,$ $0 < \delta \le 1$ and $ 4 k \ln( 2 k / \delta) < r \leq n$.
Let $[\matOmega, \matD] = RandSampling(\matV,  r, \beta)$.
Then, for all $i=1,...,k$, and with probability  at least $1 - \delta,$
\begin{equation}\label{eqn46}
1 -   \sqrt{ \frac{ 4 k \ln(2 k / \delta )} {\beta r} }   \leq  \sigma_i^2(\matV\transp \matOmega \matD)
\leq 1 +  \sqrt{ \frac{ 4 k \ln(2 k / \delta )} {\beta r} } .
\end{equation}
\end{lemma}
\begin{proof}
This is Theorem 2 of~\cite{Mag10} with $\matS = \matI$, the identity matrix.
\end{proof}
\begin{lemma} \label{lem:fnorm}
For any $\beta, r$, $\matX \in \R^{n \times k}$, and $\matY \in \R^{m \times n}$, let
$[\matOmega, \matD] = RandSampling(\matX, r)$; then, with probability at least $0.9$:
$ \FNormS{ \matY \matOmega \matD } \leq 10 \FNormS{ \matY }. $
\end{lemma}
\begin{proof}
Eqn.~(36) in \cite{DMM06b} gives $\Expect{ \FNormS{ \matY \matOmega \matD} } = \FNormS{ \matY }$;
apply Markov's inequality to wrap up.
\end{proof}

\subsubsection{Adaptive sampling}
Adaptive sampling aims at sampling columns from the input matrix in an adaptive fashion based on information of the residual error from approximating the matrix with the already sampled columns.
The following lemma implements one round of adaptive sampling and will be used extensively in our analysis to improve constant factor column-based matrix approximations to relative-error ones.
\begin{lemma}[Theorem 2.1 of~\cite{DRVW06}]
\label{lem:adaptivecolumns}
Given $\matA \in \R^{m \times n}$ and $\matV \in \R^{m \times c_1}$ (with $c_1 \le n, m$),
define the residual
$$\matB = \matA - \matV \matV^{\dagger} \matA \in \R^{m \times n}.$$
For $i=1,\ldots,n$, and some fixed constant  $\alpha >0 ,$ let $p_i$ be a probability distribution such that for each $i:$
$$p_i \ge \alpha {\TNormS{\b_{i}}}/{\FNormS{\matB}},$$
where $\b_i$ is the $i$-th column of the matrix $\matB$. Sample
$c_2$ columns from $\matA$ in \math{c_2} i.i.d. trials, where in each trial the $i$-th column is chosen with probability $p_i$.
Let $\matC_2 \in \R^{m \times c_2}$ contain the $c_2$ sampled columns and let $\matC = [\matV\ \ \matC_2] \in \R^{m \times (c_1+c_2)}$
contain the columns of $\matV$ and $\matC_2$. 
Then, for any integer $k > 0$,
$$\Expect{ \FNormS{ \matA - \Pi_{\matC,k}^{\mathrm{F}}(\matA) } } \le \FNormS{ \matA - \matA_k } + \frac{k}{\alpha \cdot c_2} \FNormS{ \matA - \matV \matV^{\dagger} \matA}.$$
We denote this procedure as
$$\matC_2 = AdaptiveCols(\matA, \matV, \alpha, c_2).$$
Given $\matA$ and $\matV,$ the above algorithm requires $O( c_1 m n + c_2 \log c_2 )$ arithmetic operations to find $\matC_2$.
\end{lemma}

A recent result generalizes the above adaptive sampling lemma to sampling rows from a matrix. 
This new adaptive sampling lemma is particularly useful in designing CUR matrix decompositions. 
\begin{lemma}[Theorem 4 in~\cite{WZ13}]
\label{lem:adaptiverows}
Given $\matA \in \R^{m \times n}$ and $\matV \in \R^{m \times c}$ such that $$\rank(\matV) = \rank(\matV \pinv{\matV} \matA) = \rho,$$
with $\rho \le c \le n,$
we let $\matR_1 \in \R^{r_1 \times n}$ consist of $r_1$ rows of $\matA$  and define the residual $$\matB = \matA - \matA\pinv{\matR}_1\matR_1 \in \R^{m \times n}.$$
For $i=1,\ldots,n$ 
let $p_i$ be a probability distribution such that for each $i:$
$$p_i = {\TNormS{\b_{i}}}/{\FNormS{\matB}},$$
where $\b_i$ is the $i$-th column of  $\matB$. Sample
$r_2$ rows from $\matA$ in \math{r_2} i.i.d. trials, 
where in each trial the $i$-th row is chosen with probability $p_i$.
Let $\matR_2 \in \R^{r_2 \times n}$ contain the $r_2$ sampled columns and let $\matR = [\matR_1\transp, \matR_2\transp]\transp \in \R^{(r_1 + r_2) \times n}$.
Then,
$$\Expect{ \FNormS{ \matA - \matV\pinv{\matV}\matA \pinv{\matR}\matR } }
\le \FNormS{ \matA - \matV\pinv{\matV}\matA } + \frac{\rho}{r_2} \FNormS{\matA - \matA\pinv{\matR}_1\matR_1}.$$
We denote this procedure as
$$\matR_2 = AdaptiveRows(\matA, \matV, \matR_1, r_2).$$
Given $\matA,$ $\matV,$ $\matR_1,$
the above algorithm requires $O( r_1 m n + r_2 \log r_2  )$ arithmetic operations to find $\matR_2$.
\end{lemma}

We now discuss connections between Lemma~\ref{lem:adaptiverows} and Lemma~\ref{lem:adaptivecolumns}. As it is discussed in Remark 6 in~\cite{WZ13}, setting 
$\matV = \matA_k$ in Lemma~\ref{lem:adaptiverows} and
using the identity $\matA_k \pinv{\matA}_k\matA = \matA_k,$ gives
$$\Expect{ \FNormS{ \matA - \matA_k \pinv{\matR}\matR } }
\le \FNormS{ \matA - \matA_k } + \frac{k}{r_2} \FNormS{\matA - \matA\pinv{\matR}_1\matR_1}.$$
Applying this result to the transpose of $\matA$ and switching from rows $\matR_1 \in \R^{r_1 \times n},$ $\matR_2 \in \R^{r_2 \times n},$ and $\matR^{r \times n}$ to columns $\matC_1 \in \R^{m \times c_1},$ $\matC_2 \in \R^{m \times c_2},$ and $\matC \in \R^{m \times c},$ respectively, we have,
$
\Expect{ \FNormS{ \matA - \matC \pinv{\matC} \matA_k } }
\le 
\FNormS{ \matA - \matA_k } + (k / c_2) \FNormS{\matA - \pinv{\matC}_1\matC_1\matA}.
$
Since, 
$\FNormS{ \matA - \matC \pinv{\matC} \matA } \le \FNormS{ \matA - \matC \pinv{\matC} \matA_k },$
and
$\FNormS{ \matA - \matC \pinv{\matC} \matA } \le \FNormS{ \matA -  \Pi_{\matC,k}^{\mathrm{F}}(\matA) } ,$
and
$\FNormS{ \matA -  \Pi_{\matC,k}^{\mathrm{F}}(\matA) } \le \FNormS{ \matA - \matC \pinv{\matC} \matA_k },$
we obtain
$
\Expect{ \FNormS{ \matA - \matC \pinv{\matC} \matA } }
\le$
$$
\Expect{ \FNormS{ \matA - \Pi_{\matC,k}^{\mathrm{F}}(\matA) } } \le
\Expect{ \FNormS{ \matA - \matC \pinv{\matC} \matA_k } }
\le 
\FNormS{ \matA - \matA_k } + \frac{k}{c_2} \FNormS{\matA - \pinv{\matC}_1\matC_1\matA}.
$$
This result is identical to Lemma~\ref{lem:adaptivecolumns} with $\alpha=1$ and
$\matV = \matC_1$ containing columns of $\matA$.

\subsection{Low-rank approximations within a subspace}\label{sec:bestF}
In this section we discuss algorithms to find low-rank approximations to matrices
constrained to a given subspace. 

\subsubsection{The best rank \math{k} matrix $\Pi_{\matV,k}^\xi(\matA)$ within a subspace $\matV$} \label{sec:bestrankk}

Let $\matA \in \mathbb{R}^{m \times n}$, let $k < n$ be an integer, and let $\matV \in \mathbb{R}^{m \times c}$ with $k < c< n $.
$\Pi_{\matV,k}^\mathrm{F}(\matA) \in \mathbb{R}^{m \times n}$ is the best rank \math{k} approximation to \math{\matA} in the column span of \math{\matV}.
Equivalently, we can write $\Pi_{\matV, k}^\mathrm{F}(\matA) = \matV \matX_{opt},$ where
$$
\matX_{opt} = \argmin_{\matX \in {\R}^{c \times n}:\rank(\matX)\leq k}\FNormS{\matA-
\matV \matX}.
$$
In order to compute $\Pi_{\matV,k}^{\mathrm{F}}(\matA)$ given $\matA$,
$\matV$, and $k$, we will use the following algorithm:
\begin{center}
\begin{algorithmic}[1]
\STATE $\matV = \matY \matPsi$ is a $qr$ decomposition of $\matV$ with $\matY \in \R^{m \times c}$ and $\matPsi \in \R^{c \times c}$.
This step requires $O(m c^2)$ arithmetic operations.
\STATE
$\matXi = \matY\transp \matA \in \R^{c \times n}$. This step requires \math{O(mnc)} arithmetic operations.
\STATE
 $\matXi_k = \matDelta \tilde{\matSig}_k \tilde{\matV}_k\transp \in \R^{c \times n}$ is a rank $k$ SVD of $\matXi$
 with $\matDelta \in \R^{c \times k}, \tilde{\matSig}_k \in \R^{k \times k}, $ and $\tilde{\matV}_k \in \R^{n \times k}.$
This step requires \math{O( nc^2)} arithmetic operations.
\STATE Return $ \matY \matDelta \matDelta\transp \matY\transp  \in \mathbb{R}^{m \times n}$ of rank at most $k$.
\end{algorithmic}
\end{center}
\medskip
$  \matY \matDelta \matDelta\transp \matY\transp  \in \mathbb{R}^{m \times n}$ is a rank $k$ matrix that lies in the column span of $\matV$.
The next lemma is a simple corollary of Lemma 4.3 in~\cite{CW09}.
\begin{lemma}\label{lem:bestF}
Given $\matA \in {\R}^{m \times n}$, $\matV\in\R^{m\times r}$ and an integer $k$, $\matY \matDelta \matDelta\transp \matY\transp$ and  $ \matQ \tilde{\matU}_k \tilde{\matSig}_k \tilde{\matV}_k\transp$ satisfy:
$
\FNormS{\matA -  \matY \matDelta \matDelta\transp \matY\transp\matA } \le \FNormS{ \matA-  \matY \matDelta \tilde{\matSig}_k \tilde{\matV}_k\transp  } = \FNormS{\matA-\Pi_{\matV,k}^{\mathrm{F}}(\matA)}
$
The above algorithm requires $O(mnc + nc^2)$ arithmetic operations to construct $\matY, \matPsi,$ and $\matDelta$.
We will denote the above procedure as
$$ [ \matY, \matPsi, \matDelta ] = BestSubspaceSVD(\matA, \matV, r). $$
\end{lemma}
\begin{proof}
The equality was proven in Lemma 4.3 in~\cite{CW09}. To prove the inequality, notice that for any matrix $\matX:$
$\FNormS{\matA -  \matY \matDelta \pinv{(\matY \matDelta)}\matA } \le \FNormS{ \matA -  \matY \matDelta \matX  }.$
Also,
$\pinv{\left(\matY \matDelta\right)} = \pinv{\matDelta} \pinv{\matY} = \matDelta\transp \matY\transp,$
because both matrices are orthonormal.
\end{proof}

\subsubsection{An approximate rank $k$ matrix within a subspace}
The previous lemma provides a method for constructing the best rank $k$ matrix within a given subspace. This method,
however, is somewhat costly if one wants to design algorithms that run in time proportional to the number of non zeros entries of $\matA$.
To address this, we use the lemma below, which finds a rank $k$ matrix that is almost as good as the best rank $k$ matrix within $span(\matV)$.
\begin{lemma}\label{lem:KVW}
Let $\matA \in \R^{m \times n}$ and $\matV \in \R^{m \times c}$. We further assume that for some rank parameter $k < c$ and accuracy parameter $0 < \varepsilon <1,$
$$ \FNormS{\matA-\Pi_{\matV,k}^{\mathrm{F}}(\matA)} \le (1+\epsilon) \FNormS{\matA - \matA_k}.$$
In words, we assume that in the given subspace $\matV,$ there exists a rank $k$ matrix that approximates the best rank $k$ matrix from the SVD within a relative error.
Let $\matV = \matY \matPsi$ be a $qr$ decomposition of $\matV$
with $\matY \in \R^{m \times c}$ and $\matPsi \in \R^{c \times c}$.
Let $\matXi = \matY\transp \matA \matW\transp \in \R^{c \times \xi},$ where $\matW\transp \in \R^{n \times \xi}$ is a sparse subspace
embedding matrix with $\xi = O(c^2/\varepsilon^2)$~(see Definition~\ref{def:sse}).  
Let $\matDelta \in \R^{c \times k}$ contain the top $k$ left singular vectors of $\matXi$, i.e.,
$\matXi_k = \matDelta \tilde{\matSig}_k \tilde{\matV}_k\transp \in \R^{c \times n},$ is a rank $k$ SVD of $\matXi,$
with $\matDelta \in \R^{c \times k}, \tilde{\matSig}_k \in \R^{k \times k}, $ and $\tilde{\matV}_k \in \R^{n \times k}.$
Then, with probability at least $0.99$,
$$  \FNormS{ \matA - \matY \matDelta \matDelta\transp \matY\transp \matA } \le (1+\varepsilon) \FNormS{\matA - \matA_k}. $$
$\matY$, $\matPsi,$ and $\matDelta$ can be computed in $O(\nnz(\matA) + m c \xi )$ time. We denote the construction of $\matY, \matPsi$ and $\matDelta$ as
$$ [\matY, \matPsi, \matDelta] = ApproxSubspaceSVD(\matA, \matV, k, \varepsilon). $$
\end{lemma}
\begin{proof}
This result was proven inside the proof of Theorem 1.5 in~\cite{KVW13}. Specifically,
the error bound proven in~\cite{KVW13} is for the transpose of $\matA$ (also $\matY,\matDelta$ are
denoted with $U, V$ in~\cite{KVW13}). The only
requirement for the embedding matrix $\matW$ (denoted with $P$ in the proof of Theorem 1.5 in~\cite{KVW13})
is to be a subspace embedding for $\matY\transp\matA$,
in the sense that $\matW$ is a subspace embedding for $\matA$ in Lemma~\ref{lem:subspacesparse}.
Since our choice of $\matW$ with $\xi = O(c^2/\varepsilon^2)$ satisfies this requirement we omit the details of the proof. The running time
of the algorithm is $O(\nnz(\matA) + m c \xi )$ because one can compute (i) $\matY$ in $O(mc^2)$ time;
(ii) $\matXi$ in $O(\nnz(\matA) + mc\xi)$ time; and (iii) $\matDelta$ in $O(c \xi \min\{c,\xi\})$ time.
The failure probability $0.01$ is due to Lemma~\ref{lem:subspacesparse}.
\end{proof}

\subsection{Sparse subspace embeddings}\label{sec:CWT}
Next, we discuss the so-called sparse subspace embedding matrices of Clarskon and Woodruff~\cite{CW13arxiv}.
Those are linear transformations for dimensionality reduction that preserve both the sparsity of the input points as
well as their geometry. 

\begin{definition}\label{def:sse} [Sparse Subspace Embedding~\cite{CW13arxiv}]
We call $\matW \in \R^{\xi \times n}$ a sparse subspace embedding of dimension $\xi$ if it is constructed as follows,
$ \matW = \matPsi \matY, $
with
\begin{itemize}

\item $h: [n] \rightarrow [\xi]$ is a random map so that for each $i \in [n], h(i) = \xi',$ for $\xi' \in [\xi]$ w.p. $1/\xi$.

\item $\matPsi \in \R^{\xi \times n}$ is a binary
matrix with $\matPsi_{h(i),i} = 1,$ and all remaining entries $0$.

\item $\matY \in \R^{n \times n}$ is a random diagonal matrix, with each diagonal entry independently chosen to be $+1$ or $-1$, with equal probability.
\end{itemize}
For any matrix $\matA$ with $n$ rows, computing $\matW \matA$ requires $O( \nnz(\matA) )$ time.
\end{definition}
\begin{lemma}[\cite{CW13arxiv,MM13,JH13}] \label{lem:subspacesparse}
Let $\matA \in \R^{n \times d}$ have rank $\rho$ and let $\matW \in \R^{\xi \times n}$
be a randomly chosen sparse subspace embedding with dimension $\xi  = \Omega(\rho^2 \varepsilon^{-2})$, for some $0 < \varepsilon < 1.$
Then, with probability at least $0.99,$ and for all vectors $\y \in \R^d$ simultaneously,
$$  (1-\varepsilon) \TNormS{\matA \y} \le   \TNormS{ \matW \matA \y} \le (1+\varepsilon) \TNormS{\matA \y}.$$
\end{lemma}

\begin{lemma}[Lemma 40 in~\cite{CW13arxiv}] \label{lem:affinesparse}
Let $\matA \in \R^{n \times d}$ and let $\matW \in \R^{\xi \times n}$
be  a randomly chosen sparse subspace embedding with dimension $\xi= \Omega(\varepsilon^{-2})$, for some $0 < \varepsilon < 1.$
Then, with probability at least $0.99,$
$$  (1-\varepsilon) \FNormS{\matA} \le   \FNormS{ \matW \matA } \le (1+\varepsilon) \FNormS{\matA}.$$
\end{lemma}


\begin{lemma}[\cite{CW13arxiv}] \label{lem:affinesparse2}
Let $\matA \in \R^{n \times d}$ have rank $\rho$ and $\matB \in \R^{n \times \omega}$. Let $\matW \in \R^{\xi \times n}$ be a randomly chosen sparse subspace embedding
with $\xi  = \Omega(\rho^2 \varepsilon^{-2})$. Then, with probability at least $0.99,$
$
\FNormS{ \matA \tilde{\matX}_{opt} - \matB} \le \FNormS{ \matA \matX_{opt} - \matB }, 
$
where
$$ \tilde{\matX}_{opt} \in \argmin_{ \matX \in \R^{d \times \omega} } \FNormS{\matW\matA \matX - \matW\matB} $$
and,
$$ \matX_{opt} \in \argmin_{\matX \in \R^{d \times \omega}} \FNormS{\matA \matX - \matB}.$$
\end{lemma}
\begin{proof}
Theorem 36 in~\cite{CW13arxiv} shows this bound under the assumption that the event of Lemma 22 in~\cite{CW13arxiv} occurs.
In the proof of Lemma 22 in~\cite{CW13arxiv}, it is shown that a sparse subspace embedding defined with $\xi  = \Omega(\rho^2 \varepsilon^{-2})$
satisfies the bound, where the polylogarithmic factors in the dimension $\xi$ were removed in~\cite{MM13,JH13}.
\end{proof}

\subsection{Johnson Lindestrauss transform}
Here, we review the Johnson Lindestrauss transform, a standard method for dimension reduction that preserves
the geometry of a set of points. 
\begin{lemma}\label{lem:jlt} [Theorem 1 in~\cite{Ach03} for fixed $\varepsilon = 1/2$]
Let $\matB \in \R^{m \times n}$. Given $\epsilon, \beta > 0,$ let 
$$s = \frac{4 + 2 \beta}{(1/2)^2 - (1/2)^3 } \log n.$$
Construct a matrix $\matS \in \R^{s \times m},$ each element of which is a random variable which takes values $\pm 1/\sqrt{s}$ with
equal probability. Let $\tilde{\matB} = \matS \matB$. Then, if $\b_i$ and $\tilde{\b}_i$ denote the $i$th column of $\matB$ and $\tilde{\matB}$,
respectively, with probability at least $1 - n^{-\beta},$ and for all $i=1,...n,$
$$ (1-\frac12) \TNormS{\b_i} \le \TNormS{ \tilde{\b}_i } \le (1+\frac{1}{2}) \TNormS{\b_i}.  $$
Given $\matB,$ it takes $O(\nnz(\matB) \log n )$ arithmetic operations to construct $\tilde{\matB}$. We will denote this procedure as
$$ \tilde{\matB} = JLT(\matB, \beta). $$
\end{lemma}

\subsection{Generalized rank-constrained matrix approximations}\label{sec:Uopt}
Let $\matA \in \R^{m \times n}$,
$\matC \in \R^{m \times c}$,
$\matR \in \R^{r \times n}$,
and $k \le c,r$ be an integer.
Consider the following optimization problem,
$$ \matU_{opt}  \in \argmin_{ \matU \in \R^{c \times r}, \rank(\matU)\le k } \FNormS{ \matA - \matC \matU \matR }.  $$
Then, the solution  $\matU_{opt} \in \R^{c \times r}$ with $\rank(\matU_{opt}) \le k$ that has the minimum $\FNorm{\matU_{opt}}$ out of all possible feasible solutions
is given via the following formula,
$$ \matU_{opt}  =  \pinv{\matC}  \left( \matU_{\matC}\matU_{\matC}\transp \matA \matV_{\matR}\matV_{\matR}\transp \right)_k \pinv{\matR}.$$
$\left( \matU_{\matC}\matU_{\matC}\transp \matA \matV_{\matR}\matV_{\matR}\transp \right)_k \in \R^{m \times n}$ of rank at most $k$ denotes the best rank $k$ matrix to
$\matU_{\matC}\matU_{\matC}\transp \matA \matV_{\matR}\matV_{\matR}\transp \in \R^{m \times n}$.
This result was proven in~\cite{FT07} (see also~\cite{SR12} for the spectral norm version of the problem).

\section{New column subset selection tools}\label{sec:back2}
To design our CUR algorithms in Sections~\ref{sec:alg1}, ~\ref{sec:alg3}, and~~\ref{sec:alg2}
we combine the subset selection tools from the previous section with novel tools that we describe below. 
The results in the present section could be of independent interest.  

\subsection{New tools for deterministic $\matC \matU \matR$ decompositions}

\subsubsection{Deterministic adaptive sampling}\label{sec:newDet}

First, in the next lemma we derandomize the recent row/column adaptive sampling method of Wang and Zhang \cite{WZ13} (stated as Lemma~\ref{lem:adaptiverows} in our article). The argument uses pairwise-independence, and may be known, though we could not find a reference.
\begin{lemma}
\label{lem:adaptiverowsd}
Given $\matA \in \R^{m \times n}$ and $\matV \in \R^{m \times c}$ such that $$\rank(\matV) = \rank(\matV \pinv{\matV} \matA) = \rho,$$
with $\rho \le c \le n,$
we let $\matR_1 \in \R^{r_1 \times n}$ consist of $r_1$ rows of $\matA$.
There exists a deterministic algorithm to construct
$\matR_2 \in \R^{r_2 \times n}$ with $r_2$ rows such that for 
$$\matR = [\matR_1\transp, \matR_2\transp]\transp \in \R^{(r_1 + r_2) \times n},$$
we have
$$ \FNormS{ \matA - \matV\pinv{\matV}\matA \pinv{\matR}\matR } \le \FNormS{ \matA - \matV\pinv{\matV}\matA } + \frac{4\rho}{r_2} \FNormS{\matA - \matA\pinv{\matR}_1\matR_1}.$$
We denote this procedure as
$$\matR_2 = AdaptiveRowsD(\matA, \matV, \matR_1, r_2).$$
Given $\matA,$ $\matV,$ $\matR_1,$
it takes $O( m^2 ( mc^2 + nr^2 + mn(c+r)) )$ time to find $\matR_2$. Here $c = c_1+c_2, r=r_1+r_2$.
\end{lemma}
\begin{proof}
This lemma corresponds to a derandomization of Theorem 5 of
\cite{WZ13}. Note that the matrix $\matC$ in their proof corresponds to our matrix $\matV$, and despite their notation, as they state in their theorem statement
and prove, the matrix $\matC$ need not be a subset of columns of $\matA$.
To prove their theorem, they actually restate it as Theorem 15 in their paper, switching
the role of rows and columns (so they will sample a subset $\matC_2$ of columns, given a subset $\matC_1$ of columns and an arbitrary matrix $\matR$), which we will now derandomize. Our lemma then follows by taking transposes.

In that theorem the authors define a distribution $p$
on $n$ columns ${\bf b}_1, \ldots, {\bf b}_n$ where $\matB = \matA-\matC_1 \matC_1^{\dagger} \mat A$
for a matrix $\matC_1 \in \mathbb{R}^{m \times c_1}$ consisting of $c_1$ columns of $\matA$. In their
proof they show that for $p_i = \|{\bf b}_i\|_2^2/\|\matB\|_{\mathrm{F}}^2$, if one samples $c_2$
columns of $\matA$ in i.i.d. trials where in each trial the $i$-th column is chosen with probability $p_i$,
then if $\matC_2 \in \mathbb{R}^{m \times c_2}$ consists of the $c_2$ sampled columns, and $\matC = [\matC_1, \matC_2]
\in \mathbb{R}^{m \times (c_1+c_2)}$ then
$${\bf E}_{\matC_2}\|\matA -\matC\matC^{\dagger}\matA\matR^{\dagger}\matR\|_\mathrm{F}^2 \leq \|\matA-\matA\matR^{\dagger}\matR\|_\mathrm{F}^2
+ \frac{\rho}{c_2}\|\matA-\matC_1\matC_1^{\dagger}\matA\|_\mathrm{F}^2.$$

To derandomize this, we first discretize the distribution $p$, defining a new distribution $q$ (and along the way, a distribution $r$). 
Let $i^* \in [n]$ be such that $p_{i^*} \geq p_i$ for all $i \neq i^*$. Then, let 
$$r_{i^*} = p_{i^*} + \sum_{i \ne i^*} \frac{p_i}{2},$$
and $r_i = p_i/2$ for $i \neq i^*$. Hence, $r$ is a distribution.
Now round each $r_i$, $i \ne i^*$, up to  the nearest integer multiple of $1/(4n)$ by adding $\gamma_i$, and let $q_i$ be the resulting value.
For $q_{i^*}$, let $q_{i^*}$ = $r_{i^*} - \sum_{i \neq i^*} \gamma_i$. Then $\sum_i q_i = 1$ and $0 \leq q_i \leq 1$ for all $i \neq i^*$.
Now consider $q_{i^*}$. We have, 
$$0 \le \sum_{i \ne i^*} \gamma_i \leq n \cdot \frac{1}{4n} = \frac{1}{4}.$$ On the other hand 
$$r_{i^*} =  p_{i^*} + \sum_{i \ne i^*} \frac{p_i}{2}  \ge \sum_i \frac{p_i}{2} = \frac{1}{2}.$$
Hence $q_{i^*} = r_{i^*} - \sum_{i \neq i^*} \gamma_i \geq \frac{1}{4}$. Also $q_{i^*} \leq r_{i^*} \leq 1$. It follows that $q$ is a distribution. 

Notice that for all $i \ne i^*, q_i \ge r_i \ge p_i/2,$
while also, $q_{i^*} \ge p_{i^*}/4$ since $p_{i^*} \leq 1$.  
Hence, all $q_i$ satisfy $q_i \ge p_i/4$.

Next we follow the argument in the proof of Theorem 15 of \cite{WZ13} replacing distribution $p$ with $q$, pointing
out the differences. For $\ell \in [c_2]$, we define the vector random variable 
$${\bf x}_{j, (\ell)} = \frac{v_{i,j}}{q_i} {\bf b}_i,$$ 
where
$v_{i,j}$ is the $(i,j)$-th entry of $\matV_{\matA\matR^{\dagger}\matR, \rho} \in \mathbb{R}^{n \times \rho}$, where
$\matV_{\matA\matR^{\dagger}\matR, \rho} \in \mathbb{R}^{n \times \rho}$ denotes the matrix of the top $\rho$ right singular
vectors of $\matA\matR^{\dagger}\matR$. We have that
$${\bf E}[{\bf x}_{j, (\ell)}] = \sum_{i=1}^n q_i \frac{v_{i,j}}{q_i} {\bf b}_i = \matB v_j.$$ Moreover,
\begin{eqnarray}\label{eqn:new}
{\bf E}\|{\bf x}_{j,(\ell)}\|^2 = \sum_{i=1}^n q_i \frac{v_{i,j}^2}{q_i^2} \|{\bf b}_i\|^2
\leq \frac{v_{i^*, j}^2 \|{\bf b}_{i^*}\|^2}{\frac{\|{\bf b}_{i^*}\|^2}{4\|\matB\|_\mathrm{F}^2}}
+ \sum_{i \neq i^*} \frac{v_{i,j}^2 \|{\bf b}_i\|^2}{\frac{\|{\bf b}_i\|^2}{4\|\matB\|_\mathrm{F}^2}}
\leq 4\|\matB\|_\mathrm{F}^2,
\end{eqnarray}
using that for all $j$, $\sum_{i} v_{i,j}^2 = 1$. 
This is slightly weaker than the corresponding upper bound of ${\bf E}\|{\bf x}_{j,(\ell)}\|^2 \leq \|\matB\|_\mathrm{F}^2$ shown for distribution
$p$ in \cite{WZ13},
but the constant $4$ makes little difference for our purposes, while the fact that $q$ is discrete will help with our derandomization.

Now let 
$${\bf x}_j = \frac{1}{c_2} \sum_{\ell = 1}^{c_2} {\bf x}_{j, (\ell)}.$$ 
By linearity of expectation,
$${\bf E}[{\bf x}_j] = {\bf E}[{\bf x}_{j, (\ell)}] = \matB {\bf v}_j.$$
Next we bound ${\bf E}\|{\bf x}_j - \matB{\bf v}_j\|_2^2$, and here we observe that pairwise independence of the trials
is enough to bound this quantity:
$$
{\bf E}\|{\bf x}_j - \matB{\bf v}_j\|^2 
= {\bf E}\|{\bf x}_j - {\bf E}[{\bf x}_j]\|^2
= \frac{1}{c_2} {\bf E}\|{\bf x}_{j, (\ell)} - {\bf E}[{\bf x}_{j, (\ell)}]\|^2 = \frac{1}{c_2} {\bf E}\|{\bf x}_{j, (\ell)} - \mat B {\bf v}_j\|^2,
$$
where the second equality uses pairwise-independence so that the sum of the variances is equal to the variance of the sum. That is, 
the variance ${\bf E}\|{\bf x}_j - {\bf E}[{\bf x}_j]\|^2$ is $1/c_2$ times the variance 
${\bf E}\|{\bf x}_{j, (\ell)} - {\bf E}[{\bf x}_{j, (\ell)}]\|^2$ since ${\bf x}_j$ is the 
average of the ${\bf x}_{j, (\ell)}$. 

The remainder of the proof of Theorem 15 in \cite{WZ13} only uses their slightly stronger bound than (\ref{eqn:new}).
With our weaker bound they obtain
$${\bf E}_{\mat C_2}\|\matA-\matC\matC^{\dagger}\matA\matR^{\dagger}\matR\|_\mathrm{F}^2 \leq \|\matA-\matA\matR^{\dagger}\matR\|_\mathrm{F}^2 + \frac{4\rho}{c_2}\|\matB\|_{\mathrm{F}}^2,$$
whereas their bound would not have the factor of $4$. In particular there exists
a choice of $\mat C_2$ for which for the corresponding $\mat C$ we have
$$
\|\matA-\matC\matC^{\dagger}\matA\matR^{\dagger}\matR\|_\mathrm{F}^2 \leq \|\matA-\matA\matR^{\dagger}\matR\|_\mathrm{F}^2 + \frac{4\rho}{c_2}\|\matB\|_\mathrm{F}^2.
$$

Now, $q$ is a discrete distribution and the trials need only be pairwise independent.
Let $h:[c_2] \rightarrow [4n]$ be drawn from a pairwise-independent hash function family $\mathcal{F}$. Then
$\mathcal{F}$ need only have size $O(n^2)$ \cite{cw79}. For each trial $\ell \in [c_2]$, we compute $h(\ell)$. Since the
probabilities $q_i$, $i = 1, \ldots, n$ are integer multiples of $1/(4n)$ and $\sum_i q_i = 1$, we pick the largest $i \in [n]$
for which $h(\ell)/(4n) \leq \sum_{i' = 1}^i q_i$. The probability of picking $i$ in each trial is therefore $q_i$. For each
$h \in \mathcal{F}$, we compute the corresponding column samples $\matC_2$ and compute
$\|\matA-\matC\matC^{\dagger}\matA\matR^{\dagger}\matR\|_\mathrm{F}^2$. We choose the $\matC_2$ resulting in the smallest such value. Computing this
value can be done in $O( mc^2 + nr^2 + mnr + mnc + mn )$ time, and since $|\mathcal{F}| = O(n^2)$, the overall time is $O( n^2 ( mc^2 + nr^2 + mnr + mnc ) )$. Here $c = c_1+c_2, r=r_1+r_2$. 

Switching back to our notation (by taking transposes), the overall running time is
$O(m^2(mc^2 + nr^2 + mn(r+c))$. 
\end{proof}

A simple corollary to the previous lemma is a deterministic algorithm that corresponds to a derandomization of the adaptive sampling procedure of \cite{DRVW06} (stated as Lemma~\ref{lem:adaptivecolumns} in our work). In fact, we prove a slightly weaker version of the result in~\cite{DRVW06} since we require $\matV$ to have columns of $\matA;$ on the other hand, the result in~\cite{DRVW06} holds for any $\matV$. We remark that a direct derandomization of Theorem 2.1 in~\cite{DRVW06} is possible using the same argument, but we opt to omit it for simplicity. 
\begin{lemma}
\label{lem:adaptivecolumnsd}
Given $\matA \in \R^{m \times n}$ and $\matV \in \R^{m \times c_1}$ (with $c_1 \le n, m$) containing columns of $\matA$,
define the residual
$
\matB = \matA - \matV \matV^{\dagger} \matA \in \R^{m \times n}.
$
There exists a deterministic algorithm to construct
 $\matC_2 \in \R^{m \times c_2}$ containing $c_2$ columns  of $\matA$, such that  $\matC = [\matV\ \ \matC_2] \in \R^{m \times (c_1+c_2)},$ i.e., the matrix that
contains the columns of $\matV$ and $\matC_2$, satisfies: for any integer $k > 0$,
$$\FNormS{ \matA - \Pi_{\matC,k}^{\mathrm{F}}(\matA) }  \le \FNormS{ \matA - \matA_k } + \frac{4k}{ c_2} \FNormS{ \matA - \matV \matV^{\dagger} \matA}.$$
We denote this procedure as
$$\matC_2 = AdaptiveColsD(\matA, \matV, c_2).$$
Given $\matA$ and $\matV,$ the algorithm requires $O(mn^3c)$ arithmetic operations to find $\matC_2$. Here $c=c_1 + c_2$.
\end{lemma}
\begin{proof}
Let us denote with $\matV'$ the matrix $\matV$ in Lemma~\ref{lem:adaptiverowsd} and reserve $\matV$ for the matrix $\matV$ in this lemma. 
Setting $\matV' = \matA_k$ in Lemma~\ref{lem:adaptiverowsd} and
using  $\matA_k \pinv{\matA}_k\matA = \matA_k$:
$$ \FNormS{ \matA - \matA_k \pinv{\matR}\matR } 
\le \FNormS{ \matA - \matA_k } + \frac{4 k}{r_2} \FNormS{\matA - \matA\pinv{\matR}_1\matR_1}.$$
Applying this result to the transpose of $\matA$ and switching from rows $\matR_1 \in \R^{r_1 \times n},$ $\matR_2 \in \R^{r_2 \times n},$ and $\matR^{r \times n}$ to columns $\matV = \matC_1 \in \R^{m \times c_1},$ $\matC_2 \in \R^{m \times c_2},$ and $\matC \in \R^{m \times c},$ respectively, we have,
$
 \FNormS{ \matA - \matC \pinv{\matC} \matA_k } 
\le 
\FNormS{ \matA - \matA_k } + (4 k / c_2) \FNormS{\matA - \pinv{\matC}_1\matC_1\matA}.
$
Since, 
$\FNormS{ \matA - \matC \pinv{\matC} \matA } \le \FNormS{ \matA - \matC \pinv{\matC} \matA_k },$
and
$\FNormS{ \matA - \matC \pinv{\matC} \matA } \le \FNormS{ \matA -  \Pi_{\matC,k}^{\mathrm{F}}(\matA) } ,$
and
$\FNormS{ \matA -  \Pi_{\matC,k}^{\mathrm{F}}(\matA) } \le \FNormS{ \matA - \matC \pinv{\matC} \matA_k },$
we conclude the proof as follows:
$$
\FNormS{ \matA - \matC \pinv{\matC} \matA } 
\le
 \FNormS{ \matA - \Pi_{\matC,k}^{\mathrm{F}}(\matA)  } \le
 \FNormS{ \matA - \matC \pinv{\matC} \matA_k } 
\le 
\FNormS{ \matA - \matA_k } + \frac{4 k}{c_2} \FNormS{\matA - \pinv{\matC}_1\matC_1\matA}.
$$
To summarize: $\matC_2\transp = AdaptiveRowsD(\matA\transp, \matA_k\transp, \matC_1\transp, c_2)$, with $\matC_2 \in \R^{m \times c_2}$. %
\end{proof}

\subsection{New tools for input-sparsity-time $\matC \matU \matR$ decompositions}

\subsubsection{Input-sparsity time BSS sampling}
Next, we develop an ``input-sparsity-time" version of the so-called ``BSS sampling'' algorithm~\cite{BSS09}; to be precise, 
we develop an input-sparsity-time version of an extension of the original BSS algorithm that appeared in~\cite{BDM11a}. 

\begin{lemma}[Input-Sparsity-Time Dual-Set Spectral-Frobenius Sparsification.]
\label{lem:dualsets}
Let \math{\cl V=\{\v_1,\ldots,\v_n\}} be a decomposition of the identity, where \math{\v_i\in\R^{k}} ($k < n$) and
$\sum_{i=1}^n\v_i\v_i\transp=\matI_{k}$; let \math{\cl A=\{\a_1,\ldots,\a_n\}} be an arbitrary set
of vectors, where \math{\a_i\in\R^{\ell}}. Let $ \matW \in \R^{\xi \times \ell}$ be a randomly chosen
sparse subspace embedding with $\xi = O( n^2 / \varepsilon^2 ) < \ell$, for some $0 < \varepsilon < 1$ (see Definition~\ref{def:sse}).
Consider a new set of vectors  \math{\cl B = \{\matW \a_1,\ldots, \matW \a_n\}},
with \math{\matW \a_i\in\R^{\xi}}. Run the algorithm of Lemma~\ref{lem:dualset} with \math{\cl V=\{\v_1,\ldots,\v_n\}},
\math{\cl B = \{\matW \a_1,\ldots, \matW \a_n\}}, and some integer \math{r} such that \math{k < r \le n}. Let the output
of this be a set of weights \math{s_i\ge 0} ($i=1\ldots n$), at most \math{r} of which are non-zero. 
Then, with probability at least $0.98,$
\eqan{
\lambda_{k}\left(\sum_{i=1}^ns_i\v_i\v_i\transp\right)
&\ge&
\left(1 - \sqrt{\frac{k}{r}}\right)^2,
\qquad\\
\trace\left(\sum_{i=1}^n s_i\a_i\a_i\transp\right)
&\le&
\frac{1 + \varepsilon }{1 - \varepsilon} \cdot
\trace\left(\sum_{i=1}^n \a_i\a_i\transp\right) \\
&=&
\frac{1 + \varepsilon }{1 - \varepsilon} \cdot
\sum_{i=1}^n \TNormS{\a_i}.
}
Equivalently,
if $\matV \in \R^{n \times k}$ is a matrix whose rows are the vectors $\v_i\transp$,
$\matA \in \R^{n \times \ell}$ is a matrix whose rows are the vectors $\a_i\transp$,
$\matB = \matA \matW\transp \in \R^{n \times \xi}$ is a matrix whose rows are the vectors $\a_i\transp\matW\transp$,
and
$\matS \in \R^{n \times r}$ is the sampling matrix containing the weights $s_i > 0, $ then with probability at least $0.98,$
\eqan{
\sigma_{k}\left(\matV\transp\matS\right)
\ge
1 - \sqrt{{k}/{r}}
\qquad
\FNormS{\matA\transp\matS}
\le
\frac{1 + \varepsilon }{1 - \varepsilon} \cdot
\FNormS{\matA}.
}
The weights $s_i$ can be computed in $O\left( \nnz(\matA) + rnk^2+n \xi \right)$ time. We denote this procedure as
$$\matS = BssSamplingSparse(\matV, \matA, r, \varepsilon).$$
\end{lemma}
\begin{proof}
The algorithm constructs $\matS$ as follows,
$$ \matS = BssSampling(\matV, \matB, r). $$
The lower bound for the smallest singular value of $\matV$ is immediate from Lemma~\ref{lem:dualset}. That lemma also ensures,
$$
\FNormS{ \matB\transp \matS } \le \FNormS{\matB\transp}, 
$$ 
i.e.,
$$
\FNormS{ \matW \matA\transp \matS } \le \FNormS{\matW \matA\transp}.
$$
Since $\matW$ is a subspace embedding, from Lemma~\ref{lem:subspacesparse} we have that with probability at least $0.99$ and for all vectors $\y \in \R^{n}$ simultaneously,
$$
\left( 1 - \varepsilon \right) \TNormS{\mat\matA\transp \y} \le \TNormS{\matW \matA\transp \y}.
$$
Apply this $r$ times for $\y \in \R^n$ being columns from $\matS \in \R^{n \times r}$ and take a sum on the resulting inequalities,
$$\left( 1 - \varepsilon \right) \FNormS{\mat\matA\transp \matS} \le \FNormS{\matW \matA\transp \matS}.$$
Now, from Lemma~\ref{lem:affinesparse} we have that with probability at least $0.99,$
$$ \FNormS{\matW \matA\transp} \le  \left( 1 + \varepsilon \right)   \FNormS{\matA\transp}. $$
Combining all these inequalities together, we conclude that with probability at least $0.98,$
$$ \FNormS{\mat\matA\transp \matS}  \le \frac{1 + \varepsilon }{1 - \varepsilon} \cdot \FNormS{\matA\transp}. $$
\end{proof}

\subsubsection{Input-sparsity time adaptive sampling}
In this section, we develop ``input-sparsity-time'' versions of the adaptive sampling algorithms discussed in this paper.
The lemma below corresponds to such a fast version of the adaptive sampling algorithm of Lemma~\ref{lem:adaptivecolumns}.  
\begin{lemma}
\label{lem:adaptivecolumnss}
Given $\matA \in \R^{m \times n}$ and $\matV \in \R^{m \times c_1}$ (with $c_1 \le n, m$),
there exists a randomized algorithm to construct
 $\matC_2 \in \R^{m \times c_2}$ containing $c_2$ columns  of $\matA$, such that the matrix $\matC = [\matV\ \ \matC_2] \in \R^{m \times (c_1+c_2)}$
containing the columns of $\matV$ and $\matC_2$ satisfies: for any integer $k > 0$, and with probability $0.9-\frac{1}{n}$
$$\FNormS{ \matA - \Pi_{\matC,k}^{\mathrm{F}}(\matA) }  \le \FNormS{ \matA - \matA_k } +  \frac{30 k}{c_2} \FNormS{ \matA - \matV \matV^{\dagger} \matA}.$$
We denote this procedure
$$\matC_2 = AdaptiveColsSparse(\matA, \matV, c_2).$$
Given $\matA$ and $\matV,$ the  algorithm takes $O( \nnz(\matA) \log n +  m c_1 \log n +  m c_1^2)$ time to find $\matC_2$.
\end{lemma}
\begin{proof}
Define the residual $\matB = \matA - \matV \matV^{\dagger} \matA \in \R^{m \times n}.$ From Lemma~\ref{lem:jlt}, let
$$ \tilde{\matB} = JLT(\matB, 1). $$
The lemma gives us $\frac{3}{2} \TNormS{\b_i} \ge \TNormS{\tilde{\b}_i} \ge \frac{1}{2} \TNormS{\b_i}$ for all $i$. 
So from Lemma~\ref{lem:adaptivecolumns},
after using the following distribution for the sampling,
$$ p_i = \frac{ \TNormS{\tilde{\b}_i} }{\FNormS{\tilde{\matB}}} \ge  \frac{1}{2} \cdot  \frac{2}{3} \cdot \frac{ \TNormS{\b_i} }{ \FNormS{\matB} } =
 \frac{1}{3} \frac{  \TNormS{\b_i} }{ \FNormS{\matB} },$$
we obtain,
$$ \Expect{\FNormS{ \matA - \Pi_{\matC,k}^{\mathrm{F}}(\matA) }}  \le \FNormS{ \matA - \matA_k } +  \frac{3 k}{c_2} \FNormS{ \matA - \matV \matV^{\dagger} \matA}.$$
The expectation is taken with respect to the randomness in constructing $\matC_2,$ so
$$ \Expect{\FNormS{ \matA - \Pi_{\matC,k}^{\mathrm{F}}(\matA) } -  \FNormS{ \matA - \matA_k }}  \le  \frac{3 k}{c_2} \FNormS{ \matA - \matV \matV^{\dagger} \matA}.$$
The following relation is immediate, 
$$\FNormS{ \matA - \Pi_{\matC,k}^{\mathrm{F}}(\matA) } -  \FNormS{ \matA - \matA_k } > 0,$$
so, from Markov's inequality, with probability at least $0.9$
$$ \FNormS{ \matA - \Pi_{\matC,k}^{\mathrm{F}}(\matA) } -  \FNormS{ \matA - \matA_k }  \le  \frac{30 k}{c_2} \FNormS{ \matA - \matV \matV^{\dagger} \matA},$$
which implies,
$$ \FNormS{ \matA - \Pi_{\matC,k}^{\mathrm{F}}(\matA) }  \le \FNormS{ \matA - \matA_k }  + \frac{30 k}{c_2} \FNormS{ \matA - \matV \matV^{\dagger} \matA}.$$

The running time follows by a careful implementation of the random projection step inside the routine in Lemma~\ref{lem:jlt}.
First, we compute the matrices $\matS \matA$ and $\matS \matV$ in $O(\nnz(\matA) \log n)$ and $O( m c_1 \log n )$ arithmetic operations, respectively.
Computing $\pinv{\matV}$ requires $O(m c_1^2)$ operations, and $( \matS \matV ) \pinv{\matV}$ another $O( m c_1 \log n )$ operations.
Finally, computing $( \matS \matV  \pinv{\matV}) \matA$ takes $O (\nnz(\matA) \log n)$ operations and computing $\matS \matA - \matS \matV  \pinv{\matV} \matA$
 another $O(n \log n)$ operations. So, all these steps can be implemented in time
 $  O( \nnz(\matA) \log n +  m c_1 \log n +  m c_1^2) .$
\end{proof}

Next, we develop an input-sparsity-time version of the adaptive sampling algorithm of Lemma~\ref{lem:adaptiverows}.  
\begin{lemma}
\label{lem:adaptiverowss}
Given $\matA \in \R^{m \times n}$ and $\matV \in \R^{m \times c}$ such that 
$$\rank(\matV) = \rank(\matV \pinv{\matV} \matA) = \rho,$$
with $\rho \le c \le n,$
let $\matR_1 \in \R^{r_1 \times n}$ consist of $r_1$ rows of $\matA$.
There exists a randomized algorithm to construct
$\matR_2 \in \R^{r_2 \times n}$ with $r_2$ rows such that for $\matR = [\matR_1\transp, \matR_2\transp]\transp \in \R^{(r_1 + r_2) \times n}$,
we have that with probability at least $0.9 - 1/n,$
$$ \FNormS{ \matA - \matV\pinv{\matV}\matA \pinv{\matR}\matR } \le \FNormS{ \matA - \matV\pinv{\matV}\matA } + \frac{30 \rho}{r_2} \FNormS{\matA - \matA\pinv{\matR}_1\matR_1}.$$
We denote this procedure as
$$\matR_2 = AdaptiveRowsSparse(\matA, \matV, \matR_1, r_2).$$
Given $\matA,$ $\matV,$ $\matR_1,$
the algorithm takes 
$$O( \nnz(\matA) \log n +  n r_1 \log n +  n r_1^2)$$ arithmetic operations to find $\matR_2$.
\end{lemma}
\begin{proof}
First, there is an immediate generalization of Theorem 15 in~\cite{WZ13} to the case when the sampling probabilities satisfy $p_i \ge \alpha \frac{\TNormS{\b_i}}{\FNormS{\matB}}$, for some $\alpha < 1$,
rather than just for $\alpha = 1$ (we omit the details).
This leads to the following version of Lemma~\ref{lem:adaptiverows} described in our work: if the probabilities in that
lemma satisfy $p_i \ge \alpha \frac{\TNormS{\b_i}}{\FNormS{\matB}}$, for some $\alpha < 1,$ then the bound is
$$\Expect{ \FNormS{ \matA - \matV\pinv{\matV}\matA \pinv{\matR}\matR } } 
\le \FNormS{ \matA - \matV\pinv{\matV}\matA } + \frac{\rho}{\alpha r_2} \FNormS{\matA - \matA\pinv{\matR}_1\matR_1}.$$
Given this bound, the proof continues by repeating the ideas that we used in Lemma~\ref{lem:adaptivecolumnss}.
Although the proof is similar to that in Lemma~\ref{lem:adaptivecolumnss}, we include a proof for completeness.

Define the residual 
$$\matB = \matA - \matA\pinv{\matR}_1\matR_1 \in \R^{m \times n}.$$ 
From Lemma~\ref{lem:jlt}, let
$$ \tilde{\matB} = JLT(\matB, 1). $$
By doing this we have $\frac{3}{2} \TNormS{\b_i} \ge \TNormS{\tilde{\b}_i} \ge \frac{1}{2} \TNormS{\b_i}$. So,
after using the following distribution for the sampling,
$$ p_i = \frac{ \TNormS{\tilde{\b}_i} }{\FNormS{\tilde{\matB}}} \ge  \frac{1}{2} \cdot  \frac{2}{3} \cdot \frac{ \TNormS{\b_i} }{ \FNormS{\matB} } =
 \frac{1}{3} \frac{  \TNormS{\b_i} }{ \FNormS{\matB} },$$
we obtain,
$$\Expect{ \FNormS{ \matA - \matV\pinv{\matV}\matA \pinv{\matR}\matR } }\le \FNormS{ \matA - \matV\pinv{\matV}\matA } + \frac{3 \rho}{r_2} \FNormS{\matA - \matA\pinv{\matR}_1\matR_1}.$$
The expectation is taken with respect to the randomness in constructing $\matR_2,$ so
$$\Expect{ \FNormS{ \matA - \matV\pinv{\matV}\matA \pinv{\matR}\matR } - \FNormS{ \matA - \matV\pinv{\matV}\matA } } \le \frac{3 \rho}{r_2} \FNormS{\matA - \matA\pinv{\matR}_1\matR_1}.$$
We have 
$$\FNormS{ \matA - \matV\pinv{\matV}\matA \pinv{\matR}\matR } - \FNormS{ \matA - \matV\pinv{\matV}\matA }  > 0$$
since 
\eqan{
 \FNormS{ \matA - \matV\pinv{\matV}\matA \pinv{\matR}\matR } 
&=&  \FNormS{ \matA - \matV\pinv{\matV}\matA +\matV\pinv{\matV}\matA - \matV\pinv{\matV}\matA \pinv{\matR}\matR } \\
&=& \FNormS{ \matA - \matV\pinv{\matV}\matA } + \FNormS{\matV\pinv{\matV}\matA - \matV\pinv{\matV}\matA \pinv{\matR}\matR }. 
}
Hence, by Markov's inequality, with probability at least $0.9$
$$ \FNormS{ \matA - \matV\pinv{\matV}\matA \pinv{\matR}\matR } - \FNormS{ \matA - \matV\pinv{\matV}\matA }  \le \frac{30 \rho}{r_2} \FNormS{\matA - \matA\pinv{\matR}_1\matR_1},$$
which implies,
$$ \FNormS{ \ \matA - \matV\pinv{\matV}\matA \pinv{\matR}\matR }   \le \FNormS{ \matA - \matV\pinv{\matV}\matA }    + \frac{30 \rho}{r_2} \FNormS{\matA - \matA\pinv{\matR}_1\matR_1}.$$

The running time follows by a careful implementation of the random projection step inside the routine in Lemma~\ref{lem:jlt}.
First, we compute $\matS \matA$ in $O(\nnz(\matA) \log n)$ arithmetic operations.
Computing $\pinv{\matR}_1$ requires $O(n r_1^2)$ operations. Then, computing
 $ \left( \left( \matS \matA \right) \pinv{\matV}\right) \matR_1 $ requires another $O( n r_1 \log n )$ operations.
Finally, computing $\matS \matA - \matS \matA \pinv{\matR}_1 \matR_1$ requires $O(n \log n)$ operations. So, all these steps can be implemented in
$  O( \nnz(\matA) \log n +  n r_1 \log n +  n r_1^2)$ time. 
\end{proof}

\begin{algorithm}
\begin{framed}
\caption{A randomized, linear-time, optimal, relative-error, rank-$k$ CUR}
\label{algcur1}
{\bf Input:} $\matA \in \R^{m \times n};$ rank parameter $k < \rank(\matA);$ accuracy parameter $0 < \varepsilon < 1$. \\
{\bf Output:} $\matC \in \R^{m \times c}$ with $c=O(k/\varepsilon)$;  $\matR \in \R^{r \times n}$  with $r=O(k/\varepsilon)$;  $\matU \in \R^{c \times r}$ with $\rank(\matU) = k$. \\
{\bf 1. Construct $\matC$ with $O(k + k / \varepsilon)$ columns}
\begin{algorithmic}[1]
\STATE $\matZ_1 = RandomizedSVD(\matA, k, 1)$; $\matZ _1\in \R^{n \times k}$ ($\matZ_1\transp\matZ_1  = \matI_k$); $\matE_1 = \matA - \matA\matZ_1\matZ_1\transp \in \R^{m \times n}$.
\STATE $[\matOmega_1, \matD_1] = RandSampling(\matZ_1, h_1, 1);$ $h_1 = 16 k \ln( 20 k  );$ $\matOmega_1 \in \R^{n \times h_1};$ $\matD_1 \in \R^{h_1 \times h_1}$.\\
 $\matM_1 = \matZ_1\transp \matOmega_1 \matD_1 \in \R^{k \times h_1}$. $\matM_1 = \matU_{\matM_1} \matSig_{\matM_1} \matV_{\matM_1}\transp$ with $\rank(\matM_1)=k$ and
 $\matV_{\matM_1} \in \R^{h_1 \times k}$.
\STATE $\matS_1 = BssSampling(\matV_{\matM_1}, \left( \matE_1 \matOmega_1\matD_1 \right)\transp, c_1)$, with $c_1 = 4k$. $\matS_1 \in \R^{h_1 \times c_1}$.\\
 $\matC_1 = \matA \matOmega_1 \matD_1 \matS_1 \in \R^{m \times c_1}$ containing rescaled columns of $\matA$.
\STATE $\matC_2 = AdaptiveCols(\matA, \matC_1, 1, c_2)$, with $c_2=\frac{1620k}{\varepsilon}$ and $\matC_2 \in \R^{m \times c_2}$ with columns of $\matA$.\\
 $\matC = [\matC_1\ \ \matC_2] \in \R^{m \times c}$ containing $c = c_1 + c_2 = 4k + \frac{1620 k}{\varepsilon}$ rescaled columns of $\matA$.
\end{algorithmic}
{\bf 2. Construct $\matR$ with $O(k + k / \varepsilon)$ rows}
\begin{algorithmic}[1]
\STATE $[\matY, \matPsi, \matDelta] = BestSubspaceSVD(\matA, \matC, k); \matY \in \R^{m \times c}, \matPsi \in \R^{c \times c}, \matDelta \in \R^{c \times k};$ $\matB = \matY \matDelta$. \\
 $\matB = \matZ_2 \matD$ is a $qr$ of $\matB$ with $\matZ_2 \in \R^{m \times k}$ ($\matZ_2\transp\matZ_2=\matI_k$), $\matD \in \R^{k \times k}$, and $\matE_2 = \matA\transp - \matA\transp \matZ_2 \matZ_2\transp$.
\STATE $[\matOmega_2, \matD_2] = RandSampling(\matZ_2, h_2,1);$ $h_2 = 8 k \ln( 20 k  );$ $\matOmega_2 \in \R^{m \times h_2};$ $\matD_2 \in \R^{h_2 \times h_2}$.\\
 $\matM_2 = \matZ_2\transp \matOmega_2 \matD_2 \in \R^{k \times h_2}$. $\matM_2 = \matU_{\matM_2} \matSig_{\matM_2} \matV_{\matM_2}\transp$ with $\rank(\matM_2)=k$ and $\matV_{\matM_2} \in \R^{h_2 \times k}$.
\STATE $\matS_2 = BssSampling(\matV_{\matM_2}, \left(\matE_2 \matOmega_2\matD_2\right)\transp, r_1)$, with $r_1 = 4k$ and $\matS_2 \in \R^{h_2 \times r_1}$.\\
$\matR_1 = \left( \matA\transp \matOmega_2 \matD_2 \matS_2 \right)\transp \in \R^{r_1 \times n}$ containing rescaled rows from $\matA$.
\STATE $\matR_2 = AdaptiveRows(\matA, \matZ_2, \matR_1, r_2),$ with $r_2=\frac{1620k}{\varepsilon}$ and $\matR_2 \in \R^{r_2 \times n}$ with rows of $\matA$.\\
 $\matR = [\matR_1\transp, \matR_2\transp]\transp \in \R^{(r_1 + r_2) \times n}$ containing $r = 4k + \frac{1620k}{\varepsilon}$ rescaled rows of $\matA$.
\end{algorithmic}
{\bf 3. Construct $\matU$ of rank $k$}
\begin{algorithmic}[1]
\STATE 
Construct $\matU \in \R^{c \times r}$ with rank at most $k$ as follows (all those formulas are equivalent),
\eqan{
\matU
=  \matPsi^{-1} \matDelta \matD^{-1} \matZ_2\transp \matA \pinv{\matR}
&=&  \matPsi^{-1} \matDelta \matD^{-1}  \left( \matC \matPsi^{-1} \matDelta \matD^{-1} \right)\transp  \matA \pinv{\matR} \\
&=&  \matPsi^{-1} \matDelta \matD^{-1}  \left( \matC \matPsi^{-1} \matDelta \matD^{-1} \right)^{\dagger}  \matA \pinv{\matR}\\
&=&  \matPsi^{-1} \matDelta \matD^{-1}  \matD \matDelta\transp \matPsi \matC^{\dagger}  \matA \pinv{\matR}\\
&=&  \matPsi^{-1} \matDelta                      \matDelta\transp \matPsi \matC^{\dagger}  \matA \pinv{\matR}\\
}
\end{algorithmic}
\end{framed}
\end{algorithm}

\section{Linear-time randomized CUR}\label{sec:alg1}
In this section, we present and analyze a randomized CUR algorithm that runs in linear time
\footnote{``Linear time'' here we mean running time proportional to $m n k \varepsilon^{-1}$.}
\footnote{To be precise, the algorithm we analyze here constructs $\matC$ and $\matR$ with rescaled columns and rows from $\matA$. To
convert this to a truly CUR decomposition, keep the un-rescaled versions of $\matC$ and $\matR$ and introduce the scaling factors in $\matU$.
The analysis carries over to that CUR version unchanged.}.
The goal of this section is not to design the fastest possible CUR algorithm.
Instead, we focus on simplicity. 
A potentially faster randomized algorithm, especially for sparse matrices $\matA,$ 
is presented in Section~\ref{sec:alg3}. The algorithm of this section might be faster though 
for dense matrices, depending on the various parameters in the algorithm and the dimensions of the matrix. 
The analysis of the algorithm in this section serves
as a stepping stone for the input-sparsity-time and the deterministic CUR algorithms presented
later. Indeed, we provide detailed proofs for the results in this section; in later sections, to prove similar results, we often quote proofs from this section outlining the differences.  

We start with the algorithm description, which closely follows the CUR proto-algorithm in
Algorithm~\ref{alg1}. Then, we give a detailed analysis of the running time complexity of
the algorithm. Finally, in Theorem~\ref{thm:main1} we analyze the approximation error
$\FNormS{\matA - \matC \matU \matR}$.

\subsection{Algorithm description}
Algorithm~\ref{algcur1} takes as input an
$m \times n$ matrix $\matA$, rank parameter $k < \rank(\matA)$, and accuracy parameter $0< \varepsilon < 1$.
These are precisely the inputs of the CUR problem in Definition~\ref{def:cur}. It returns matrix $\matC \in \R^{m \times c}$
with $c=O(k/\varepsilon)$ columns of $\matA$, matrix $\matR \in \R^{r \times n}$
with $r=O(k/\varepsilon)$ rows of $\matA,$ as well as matrix $\matU \in \R^{c \times r}$ with rank at most $k$.
Algorithm~\ref{algcur1} follows closely the CUR proto-algorithm in Algorithm~\ref{alg1}. In more detail,
Algorithm~\ref{algcur1} makes specific choices for the various steps of the proto-algorithm that can be implemented
in linear time. Algorithm~\ref{algcur1} runs in three steps: (i) in the first step, an optimal number of columns are
selected in $\matC$; (ii) in the second step, an optimal number of rows are selected in $\matR$; and (iii) in the third step,
an intersection matrix with optimal rank is constructed and denoted by $\matU$.
The algorithm itself refers to several other algorithms, which we analyze in detail in different sections.
Specifically,
\emph{RandomizedSVD} is described in Lemma~\ref{lem:approxSVD};
\emph{RandSampling}  in Lemma~\ref{lem:random};
\emph{BssSampling}  in Lemma~\ref{lem:dualset};
\emph{AdaptiveCols} in Lemma~\ref{lem:adaptivecolumns}.
\emph{BestSubspaceSVD}  in Lemma~\ref{lem:bestF}; and
\emph{AdaptiveRows} in Lemma~\ref{lem:adaptiverows}.

\subsection{Running time analysis}
Next, we give a detailed analysis of the arithmetic operations of Algorithm~\ref{algcur1}.

\begin{enumerate}

\item  We need $O(m n k/ \varepsilon + k^4 \ln k + \frac{k}{\varepsilon} \log \frac{k}{\varepsilon})$ time to find $c = 4k + \frac{1620k}{\varepsilon}$ columns of $\matA$ in $\matC \in \R^{m \times c}$. 

\begin{enumerate}
\item We need $O(mnk)$ time to compute $\matZ_1$ (from Lemma~\ref{lem:approxSVD}), and $O(mnk)$ time to form $\matE_1.$
\item We need $O( kn + k \ln(  k  ) \log (k \ln( k  )) )$ time to construct $\matOmega_1, \matD_1$ (from Lemma~\ref{lem:random}).\\
 We need $O( k^3 \ln k )$ time to construct $\matV_{\matM_1}$.
\item We need $O( m k \ln k + k^4 \ln k )$ time to construct $\matS_1$ (from Lemma~\ref{lem:dualset}).\\
 We need $O(m + k)$ time to construct $\matC_1$.
\item We need $O(m n k/ \varepsilon + \frac{k}{\varepsilon} \log \frac{k}{\varepsilon})$ time to construct $\matC_2$ (from Lemma~\ref{lem:adaptivecolumns}).\\
  We need $O(m+k/\varepsilon)$ time to construct $\matC$.
\end{enumerate}

\item We need $O(m n k/ \varepsilon + k^4 \ln k + \frac{k}{\varepsilon} \log \frac{k}{\varepsilon})$ time to find $r = 4k + \frac{1620k}{\epsilon}$ rows of $\matA$ 
in  $\matR \in \R^{r_1 \times n}$.

\begin{enumerate}
\item We need $O(mnk)$ to construct $\matY$, $\matPsi$, and $\matDelta$  (from Lemma~\ref{lem:bestF}).\\
 We need $O(m k^2)$ time to construct $\matZ_2$, and $O(mnk)$ time to form $\matE_2$.
\item We need $O( km + k \ln( k  ) \log (k \ln( k  )) )$ time to construct $\matOmega_2, \matD_2$ (from Lemma~\ref{lem:random}).\\
 We need $O( k^3 \ln k )$ time to construct $\matV_{\matM_2}$.
\item We need $O( n k \ln k +  k^4 \ln k )$ time to construct $\matS_2$ (from Lemma~\ref{lem:dualset}).\\
We need $O(n + k)$ time to construct $\matR_1$.
\item We need $O(m n k/ \varepsilon + \frac{k}{\varepsilon} \log \frac{k}{\varepsilon})$ time construct $\matR_2$ (from Lemma~\ref{lem:adaptiverows}).\\ 
We need $O(n+k/\varepsilon)$ time to construct $\matR$.
\end{enumerate}


\item We need $O( m^2 k /\varepsilon +  n^2 k /\varepsilon  +m k^2 / \varepsilon^2 + k^3/\varepsilon^3)$ time to construct $\matU$. First, we compute $\pinv{\matC}, \pinv{\matR},$ and $\matPsi^{-1}$;
then, we compute $\matU$ as follows:
$$\matU =  \left(  \matPsi^{-1}  \left( \matDelta  \left( \matDelta\transp  \left( \matPsi  \pinv{\matC} \right)\right)\right)\right) \cdot (\matA \pinv{\matR}).$$
\end{enumerate}

The total asymptotic running time of the algorithm is 

$$ O( n^2 k/ \varepsilon +  m^2 k/ \varepsilon  +m k^2 / \varepsilon^2 + k^3/\varepsilon^3 + k^4 \ln k + \frac{k}{\varepsilon} \log \frac{k}{\varepsilon}).$$

\subsection{Error bounds}
The theorem below presents our main quality-of-approximation result regarding Algorithm~\ref{algcur1}. We prove the theorem in Section~\ref{sec:proof1}. 

\begin{theorem}\label{thm:main1}
The matrices $\matC, \matU$, and $\matR$ in Algorithm~\ref{algcur1} satisfy with probability at least $0.2,$
$$ \FNormS{ \matA - \matC \matU \matR}  \le \left(1+ 20 \varepsilon \right) \FNormS{\matA-\matA_k}.$$
\end{theorem}

\subsubsection{Intermediate results}
To prove Theorem~\ref{thm:main1} in Section~\ref{sec:proof1}, we need several intermediate results, 
some of which might be of independent interest. 

First, we argue that the sampling of columns implemented via the matrices $\matOmega_1, \matD_1,$ and $\matS_1$ ``preserves'' the rank of $\matZ_1\transp$. This is necessary in order to prove that 
$\matC_1 = \matA \matOmega_1 \matD_1 \matS_1$ gives a ``good'' column-based, low-rank approximation to 
$\matA$. We make this statement precise in Lemma~\ref{lem:pr1}. 
\begin{lemma}\label{lem:pr0}
The matrices $\matZ_1, \matOmega_1, \matD_1, \matS_1$ in Algorithm~\ref{algcur1} satisfy with probability at least $0.9$,
$$\rank(\matZ_1\transp \matOmega_1 \matD_1 \matS_1) = k.$$
\end{lemma}
\begin{proof}
It suffices to show that $\sigma_k\left( \matZ_1\transp \matOmega_1 \matD_1 \matS_1 \right) > 0$.
Recall some notation that was introduced in the algorithm:
$\matM_1 = \matZ_1\transp \matOmega_1 \matD_1 \in \R^{k \times h_1}$.
From Lemma~\ref{lem:random} with $\matV = \matZ_1$, $\matOmega = \matOmega_1,$ $\matD = \matD_1,$ and $r = h_1 = 16 k \ln( 20 k  )$, we obtain
that with probability at least $0.9$
\begin{equation}\label{eqn1}
\sigma_k^2(\matZ_1\transp \matOmega_1 \matD_1) \ge \frac{1}{2}.
\end{equation}
This implies that $\rank(\matM_1)=k;$ hence, the SVD of $\matM_1$ is
$\matM_1 = \matU_{\matM_1} \matSig_{\matM_1} \matV_{\matM_1}\transp$,
with $\matU_{\matM_1} \in \R^{k \times k}$, $\matSig_{\matM_1} \in \R^{k \times k},$ and $\matV_{\matM_1} \in \R^{h_1 \times k}$.
Now recall that in step 1-d in the algorithm we constructed the matrix $\matS_1 \in \R^{h_1 \times 2k}$ as follows,
$\matS_1 = BssSampling(\matV_{\matM_1}, \matE_1 \matOmega_1\matD_1, 2k)$.  From Lemma~\ref{lem:dualset}, we have that
\begin{equation}\label{eqn2}
\sigma_k\left(  \matV_{\matM_1}\transp \matS_1 \right) \ge \frac{1}{2},
\end{equation}
so $\rank( \matV_{\matM_1}\transp) = k$.
To conclude,
$$ \rank(\matZ_1\transp \matOmega_1 \matD_1 \matS_1) = \rank( \matU_{\matM_1} \matSig_{\matM_1} \matV_{\matM_1}\transp \matS_1)=
\rank( \matSig_{\matM_1 } \matV_{\matM_1}\transp \matS_1)
 = \rank( \matV_{\matM_1}\transp \matS_1) = k.$$
\end{proof}

Next, we argue that the matrix $\matC_1$ in the algorithm offers a  
constant-factor, column-based, low-rank approximation to $\matA$. 
Notice that $\matC_1$ contains $O(k)$ columns of $\matA$. 
\begin{lemma}\label{lem:pr1}
The matrix $\matC_1$ in Algorithm~\ref{algcur1} satisfies with probability at least $0.7,$
$$
\FNormS{\matA-\matC_1 \pinv{\matC}_1 \matA} \le 1620 \cdot \FNormS{\matA-\matA_k}. 
$$
\end{lemma}
\begin{proof}
We would like to apply Lemma~\ref{lem:structural} with $\matZ = \matZ_1 \in \R^{n \times k}$ and $\matS = \matOmega_1 \matD_1 \matS_1 \in \R^{n \times c_1}$.
First, we argue that the rank assumption of the lemma is satisfied for our specific choice of $\matS$.
In Lemma~\ref{lem:pr0} we proved that with probability at least $0.9$: $\rank(\matZ_1\transp \matOmega_1 \matD_1 \matS_1) = k.$ So, for $\matC_1 = \matA \matOmega_1 \matD_1 \matS_1$
($\matE_1 = \matA - \matA\matZ_1\matZ_1\transp \in \R^{m \times n}$),
\eqan{
\FNormS{ \matA - \matC_1 \pinv{\matC}_1\matA } 
&\le&  \FNormS{\matA - \Pi^{\xi}_{\matC_1,k}(\matA)}\\ 
&\le& \FNormS{ \matA - \matC_1 \pinv{(\matZ_1\matOmega_1 \matD_1 \matS_1)}\matZ_1\transp } \\
&\le& \FNormS{\matE_1} + \FNormS{\matE_1 \matOmega_1 \matD_1 \matS_1
\pinv{(\matZ_1\matOmega_1 \matD_1 \matS_1)}}.
}
We manipulate the second term above as follows,
\eqan{
\FNormS{\matE_1 \matOmega_1 \matD_1 \matS_1(\matZ_1\matOmega_1 \matD_1 \matS_1)^{\dagger}}
&\buildrel(a)\over\le& \FNormS{\matE_1 \matOmega_1 \matD_1 \matS_1} \TNormS{(\matZ_1\matOmega_1 \matD_1 \matS_1)^{\dagger}} \\
&\buildrel(b)\over=  & \FNormS{\matE_1 \matOmega_1 \matD_1 \matS_1} \TNormS{(\matU_{\matM_1} \matSig_{\matM_1} \matV_{\matM_1}\transp \matS_1)^{\dagger}} \\
&\buildrel(c)\over=  & \FNormS{\matE_1 \matOmega_1 \matD_1 \matS_1} \TNormS{ \left(\matV_{\matM_1}\transp \matS_1\right)^{\dagger} \left( \matU_{\matM_1} \matSig_{\matM_1} \right)^{\dagger}} \\
&\buildrel(d)\over\le& \FNormS{\matE_1 \matOmega_1 \matD_1 \matS_1} \TNormS{ \left(\matV_{\matM_1}\transp \matS_1\right)^{\dagger}} \TNormS{ \left( \matU_{\matM_1} \matSig_{\matM_1} \right)^{\dagger}}  \\
&\buildrel(e)\over=  & \FNormS{\matE_1 \matOmega_1 \matD_1 \matS_1}  \cdot \frac{1}{ \sigma_k^2\left(\matV_{\matM_1}\transp \matS_1\right)}  \cdot \frac{1}{ \sigma_k^2\left( \matU_{\matM_1} \matSig_{\matM_1} \right)} \\
&\buildrel(f)\over\le & \FNormS{\matE_1 \matOmega_1 \matD_1 \matS_1} \cdot 8 \\
&\buildrel(g)\over\le&  \FNormS{\matE_1 \matOmega_1 \matD_1} \cdot 8 \\
&\buildrel(h)\over\le& 80 \FNormS{\matE_1}
}
(a) follows by the strong spectral submultiplicativity property of matrix norms.
(b) follows by replacing $\matZ_1\matOmega_1 \matD_1 = \matM_1 = \matU_{\matM_1} \matSig_{\matM_1} \matV_{\matM_1}\transp$
(c) follows by the fact that $\matU_{\matM_1} \matSig_{\matM_1}$ is a full rank $k \times k$ matrix.
(d) follows by the spectral submultiplicativity property of matrix norms.
(e) follows by the connection of the spectral norm of the pseudo-inverse with the singular values of the matrix to be pseudo-inverted.
(f) follows by Equations~\ref{eqn1} and~\ref{eqn2} (here there is a $0.1$ failure probability - in the same probability event $\rank(\matZ_1\transp \matOmega_1 \matD_1 \matS_1) = k$).
(g) follows by Lemma~\ref{lem:dualset}.
(h) follows by Lemma~\ref{lem:fnorm} (here there is a $0.1$ failure probability).
So, overall with probability at least $0.8,$
$$ \FNormS{\matE_1 \matOmega \matD \matS_1(\matZ_1\matOmega \matD \matS_1)^{\dagger}} \le 80 \FNormS{\matE_1},$$
hence,
with the same probability,
$$\FNormS{ \matA - \matC_1 \pinv{\matC}_1\matA } \le\FNormS{\matE_1} +  80 \FNormS{\matE_1}.$$
From Lemma~\ref{lem:approxSVD} we obtain: $\Expect{\FNormS{\matE_1}} \leq 2 \FNormS{\matA - \matA_k}.$
This implies that with probability at least $0.9$: $ \FNormS{\matE_1} \le 20 \FNormS{\matA - \matA_k};$ hence,
with probability at least $0.7$,
$$ \FNormS{ \matA - \matC_1 \pinv{\matC}_1\matA } \le 1620 \FNormS{\matA - \matA_k}.$$
\end{proof}

Next, we argue that the matrix $\matC$ in the algorithm offers a 
relative-error, column-based, low-rank
 approximation to $\matA$. 
\begin{lemma}\label{lem:pr2}
The matrix $\matC$ in Algorithm~\ref{algcur1} satisfies  with probability at least $0.6,$
$$ 
\FNormS{\matA-\matC \pinv{\matC} \matA} \le \FNormS{\matA - \Pi_{\matC,k}^{\mathrm{F}}(\matA)} \le (1+10 \varepsilon) \cdot \FNormS{\matA-\matA_k}. 
$$
\end{lemma}
\begin{proof}
From Lemma~\ref{lem:adaptivecolumns},
$$\Expect{ \FNormS{ \matA - \Pi_{\matC,k}^{\mathrm{F}}(\matA) } } \le \FNormS{ \matA - \matA_k } + \frac{\varepsilon}{1620} \cdot\FNormS{\matA-\matC_1 \pinv{\matC}_1 \matA}.$$
Since $ \FNormS{ \matA - \matA_k } $ is considered a constant with respect to the expectation operator, we can write
$$\Expect{ \FNormS{ \matA - \Pi_{\matC,k}^{\mathrm{F}}(\matA) } -  \FNormS{ \matA - \matA_k } } \le \frac{\varepsilon}{1620} \cdot\FNormS{\matA-\matC_1 \pinv{\matC}_1 \matA}.$$
Notice that
$$\FNormS{ \matA - \Pi_{\matC,k}^{\mathrm{F}}(\matA) } -  \FNormS{ \matA - \matA_k } > 0,$$ 
so we can apply Markov's inequality and obtain that with probability at least $0.9$,
$$\FNormS{ \matA - \Pi_{\matC,k}^{\mathrm{F}}(\matA) } -  \FNormS{ \matA - \matA_k }  \le 10  \frac{\varepsilon}{1620} \cdot\FNormS{\matA-\matC_1 \pinv{\matC}_1 \matA},$$
hence, with probability at least $0.9,$
$$\FNormS{ \matA - \Pi_{\matC,k}^{\mathrm{F}}(\matA) }  \le\FNormS{ \matA - \matA_k }  +  10  \frac{\varepsilon}{1620} \cdot\FNormS{\matA-\matC_1 \pinv{\matC}_1 \matA}.$$
In Lemma~\ref{lem:pr1} we proved that
for the matrix $\matC_1$ in the main algorithm and with probability at least $0.7,$
$$ \FNormS{\matA-\matC_1 \pinv{\matC}_1 \matA} \le 1620 \cdot \FNormS{\matA-\matA_k}. $$
Combining the last two bounds, we are in a probability event that fails with probability at most $0.4$ and guarantees that
$$ 
\FNormS{\matA-\Pi_{\matC,k}^{\mathrm{F}}(\matA)} \le (1+10 \varepsilon) \cdot \FNormS{\matA-\matA_k}.
$$
Finally, by the definition of $\Pi_{\matC,k}^{\mathrm{F}}(\matA)$ we have that 
$$ \FNorm{\matA - \matC \pinv{\matC} \matA} \le \FNormS{\matA - \Pi_{\matC,k}^{\mathrm{F}}(\matA)}.$$
\end{proof}

Next, we show that there is a rank $k$ matrix in $span(\matC)$ that achieves a similar relative-error bound as the relative-error bound achieved by $\matC \pinv{\matC} \matA$ in the previous lemma.
\begin{lemma}\label{lem:pr3}
The matrices $\matY, \matDelta$ in Algorithm~\ref{algcur1} satisfy with probability at least $0.6$:
$$\FNormS{ \matA - \matY \matDelta \matDelta\transp \matY \matA } \le (1+ 10 \varepsilon) \FNormS{\matA-\matA_k}.$$
\end{lemma}
\begin{proof}
This result is immediate from Lemma~\ref{lem:bestF} and Lemma~\ref{lem:pr2}.
\end{proof}

The following two lemmas prove similar results to Lemmas~\ref{lem:pr0} and~\ref{lem:pr1} but for the matrix $\matR_1$.
\begin{lemma}\label{lem:pr0R}
The matrices $\matZ_2, \matOmega_2, \matD_2, \matS_2$ in Algorithm~\ref{algcur1} satisfy with probability at least $0.9$,
$$\rank(\matZ_2\transp \matOmega_2 \matD_2 \matS_2) = k.$$
\end{lemma}
\begin{proof}
The proof is identical to the proof of Lemma~\ref{lem:pr0}. 
\end{proof}
\begin{lemma}\label{lem:pr4}
The matrix $\matR_1$ in Algorithm~\ref{algcur1} satisfies with probability at least $0.7,$
$$\FNormS{\matA-\matA\pinv{\matR}_1\matR_1} \le 1620 \cdot \FNormS{\matA-\matA_k}. $$
\end{lemma}
\begin{proof}
The proof is identical to the proof of Lemma~\ref{lem:pr1} with $\matA, \matC_1$ replaced by $\matA\transp$ and $\matR_1\transp$
\end{proof}

\subsubsection{Proof of Theorem~\ref{thm:main1}}\label{sec:proof1}
We are now ready to prove Theorem~\ref{thm:main1}. 
Our construction of $\matC, \matU$, and $\matR$ implies that
$$ \matC \matU \matR =   \matZ_2 \matZ_2\transp \matA \pinv{\matR}\matR,$$
so below we analyze the error
$$ \FNormS{\matA -  \matZ_2 \matZ_2\transp \matA \pinv{\matR}\matR}.$$
From Lemma~\ref{lem:adaptiverows} (with $\matV=\matZ_2$) and our choice of $r_2 = 1620 k/\varepsilon$ in that Lemma,
$$
\Expect{ \FNormS{ \matA - \matZ_2 \matZ_2\transp \matA \pinv{\matR}\matR } } 
\le \FNormS{ \matA - \matZ_2 \matZ_2\transp \matA } +  \frac{\varepsilon}{1620} \FNormS{\matA-\matA\pinv{\matR}_1\matR_1}.
$$
The expectation is taken with respect to the construction of $\matR_2,$ so $\FNormS{ \matA - \matZ_2 \matZ_2\transp \matA }$ is a constant
with respect to the expectation operator. Hence, we can write
$$
\Expect{ \FNormS{ \matA - \matZ_2 \matZ_2\transp \matA \pinv{\matR}\matR } - \FNormS{ \matA - \matZ_2 \matZ_2\transp \matA } } 
\le  \frac{\varepsilon}{1620} \FNormS{\matA-\matA\pinv{\matR}_1\matR_1}.
$$
Notice that
$$\FNormS{ \matA - \matZ_2 \matZ_2\transp \matA \pinv{\matR}\matR } - \FNormS{ \matA - \matZ_2 \matZ_2\transp \matA } > 0,$$ 
so we can apply
Markov's inequality and obtain that with probability at least $0.9,$
$$
\FNormS{ \matA - \matZ_2 \matZ_2\transp \matA \pinv{\matR}\matR } - \FNormS{ \matA - \matZ_2 \matZ_2\transp \matA } \le \frac{10 \varepsilon}{1620} \FNormS{\matA-\matA\pinv{\matR}_1\matR_1}.
$$
Equivalently, with probability at least $0.9$
$$
\FNormS{ \matA - \matZ_2 \matZ_2\transp \matA \pinv{\matR}\matR } \le \FNormS{ \matA - \matZ_2 \matZ_2\transp \matA }  + \frac{10 \varepsilon}{1620} \FNormS{\matA-\matA\pinv{\matR}_1\matR_1}.
$$
We further manipulate this bound as follows:
\eqan{
\FNormS{ \matA - \matZ_2 \matZ_2\transp \matA \pinv{\matR}\matR }
&\buildrel(a)\over\le& \FNormS{ \matA - \matZ_2 \matZ_2\transp \matA }  + \frac{10 \varepsilon}{1620} \FNormS{\matA-\matA\pinv{\matR}_1\matR_1} \\
&\buildrel(b)\over=&  \FNormS{ \matA - \matB \pinv{\matB} \matA } + \frac{10 \varepsilon}{1620} \FNormS{\matA-\matA\pinv{\matR}_1\matR_1}\\
&\buildrel(c)\over\le&  \FNormS{ \matA - \matB \pinv{\matB} \matA }   + 10 \varepsilon \FNormS{\matA-\matA_k}\\
&\buildrel(d)\over=&  \FNormS{ \matA - \matY \matDelta \matDelta\transp \matY \matA }   + 10 \varepsilon  \FNormS{\matA-\matA_k}\\
&\buildrel(e)\over\le& \left(1+10 \varepsilon \right) \FNormS{\matA-\matA_k} + 10 \varepsilon \FNormS{\matA-\matA_k}
}
(b) follows by the fact that $\matZ_2 \matZ_2\transp = \matB \pinv{\matB}$
(to see this, $\matB \pinv{\matB} =  \matZ_2 \matD \pinv{( \matZ_2 \matD)} = \matZ_2 \matD \matD^{-1} \matZ_2 = \matZ_2 \matZ_2\transp $).
(c) follows by Lemma~\ref{lem:pr4} (there is a $0.3$ failure probability to this bound).
(d) follows by the fact that $\matB = \matY \matDelta$ and
$\pinv{\matB} = \pinv{\left(\matY \matDelta\right)} = \pinv{\matDelta} \pinv{\matY} = \matDelta\transp \matY\transp,$
because both matrices are orthonormal.
(e) follows by Lemma~\ref{lem:pr3} (there is a $0.4$ failure probability to this bound).
So, overall we obtain that with probability at least $0.2,$
$$ \FNormS{ \matA - \matZ_2 \matZ_2\transp \matA \pinv{\matR}\matR }  \le \FNormS{\matA-\matA_k} + 20 \varepsilon \FNormS{\matA-\matA_k}, $$
which shows that with the same probability,
$$ \FNormS{ \matA - \matC \matU \matR}  \le \FNormS{\matA-\matA_k} + 20 \varepsilon \FNormS{\matA-\matA_k}. $$

\section{Input-Sparsity-Time CUR}\label{sec:alg3}
In this section, we present and analyze a CUR algorithm that runs in input-sparsity time\footnote{To be precise, the algorithm we analyze here constructs $\matC$ and $\matR$ with rescaled columns and rows from $\matA$. To
convert this to a truly CUR decomposition, keep the un-rescaled versions of $\matC$ and $\matR$ and introduce the scaling factors in $\matU$.
The analysis carries over to that CUR version unchanged.}.
We start with the algorithm description, which closely follows the CUR proto-algorithm in
Algorithm~\ref{alg1}. Then, we give a detailed analysis of the running time complexity of
the algorithm. Finally, in Theorem~\ref{thm:main3} we analyze the approximation error $\FNormS{\matA - \matC \matU \matR}$.

\begin{algorithm*}
\begin{framed}
\caption{An input-sparsity-time, optimal, relative-error, rank-$k$ CUR}
\label{algcur3}
{\bf Input:} $\matA \in \R^{m \times n};$ rank parameter $k < \rank(\matA);$ and accuracy parameter $0 < \varepsilon < 1$. \\
{\bf Output:} $\matC \in \R^{m \times c}$ with $c=O(k/\varepsilon)$;  $\matR \in \R^{r \times n}$  with $r=O(k/\varepsilon)$;  $\matU \in \R^{c \times r}$ with $\rank(\matU) = k$.\\
{\bf 1. Construct $\matC$ with $O(k + k / \varepsilon)$ columns}
\begin{algorithmic}[1]
\STATE $\matZ_1 = SparseSVD(\matA, k, 1)$; $\matZ _1\in \R^{n \times k}$ ($\matZ_1\transp\matZ_1  = \matI_k$).
\STATE
$[\matOmega_1, \matD_1] = RandSampling(\matZ_1, h_1, 1);$ $h_1 = 16 k \ln( 20 k  );$ $\matOmega_1 \in \R^{n \times h_1};$ $\matD_1 \in \R^{h_1 \times h_1}$. \\
 $\matM_1 = \matZ_1\transp \matOmega_1 \matD_1 \in \R^{k \times h_1}$. $\matM_1 = \matU_{\matM_1} \matSig_{\matM_1} \matV_{\matM_1}\transp$ with $\rank(\matM_1)=k$ and
 $\matV_{\matM_1} \in \R^{h_1 \times k}$.
\STATE $\matS_1 = BssSamplingSparse(\matV_{\matM_1}, \left( (\matA - \matA\matZ_1\matZ_1\transp) \matOmega_1\matD_1 \right)\transp, c_1, 0.5)$, with $c_1 = 4k$. $\matS_1 \in \R^{h_1 \times c_1}$
(see the running time analysis section for a detailed implementation of this step). \\
 $\matC_1 = \matA \matOmega_1 \matD_1 \matS_1 \in \R^{m \times c_1}$ containing rescaled columns of $\matA$.
\STATE $\matC_2 = AdaptiveColsSparse(\matA, \matC_1, 1, c_2)$; $c_2=\frac{4820k}{\varepsilon}$; $\matC_2 \in \R^{m \times c_2}$ with columns of $\matA$. \\
$\matC = [\matC_1\ \ \matC_2] \in \R^{m \times c}$ containing $c = c_1 + c_2 = 4k + \frac{4820 k}{\varepsilon}$ rescaled columns of $\matA$.
\end{algorithmic}
{\bf 2. Construct $\matR$ with $O(k + k / \varepsilon)$ rows}
\begin{algorithmic}[1]
\STATE $[\matY, \matPsi, \matDelta] = ApproxSubspaceSVD(\matA, \matC, k); \matY \in \R^{m \times c}, \matPsi \in \R^{c \times c}, \matDelta \in \R^{c \times k};$ $\matB = \matY \matDelta.$\\
 $\matB = \matZ_2 \matD$ is a $qr$ decomposition of $\matB$ with $\matZ_2 \in \R^{m \times k}$ ($\matZ_2\transp\matZ_2=\matI_k$), $\matD \in \R^{k \times k}$.
\STATE $[\matOmega_2, \matD_2] = RandSampling(\matZ_2, h_2,1);$ $h_2 = 8 k \ln( 20 k  );$ $\matOmega_2 \in \R^{m \times h_2};$ $\matD_2 \in \R^{h_2 \times h_2}$. \\
$\matM_2 = \matZ_2\transp \matOmega_2 \matD_2 \in \R^{k \times h_2}$. $\matM_2 = \matU_{\matM_2} \matSig_{\matM_2} \matV_{\matM_2}\transp$ with $\rank(\matM_2)=k$ and $\matV_{\matM_2} \in \R^{h_2 \times k}$.
\STATE $\matS_2 = BssSamplingSparse(\matV_{\matM_2}, \left((\matA\transp - \matA\transp \matZ_2 \matZ_2\transp) \matOmega_2\matD_2\right)\transp, r_1, 0.5)$, with $r_1 = 4k.$ $\matS_2 \in \R^{h_2 \times r_1}$
(see the running time analysis section for a detailed implementation of this step).\\
 $\matR_1 = \left( \matA\transp \matOmega_2 \matD_2 \matS_2 \right)\transp \in \R^{r_1 \times n}$ containing rescaled rows from $\matA$.
\STATE $\matR_2 = AdaptiveRowsSparse(\matA, \matZ_2, \matR_1, r_2);$  $r_2=\frac{4820k}{\varepsilon};$ $\matR_2 \in \R^{r_2 \times n}$ with rows of $\matA$. \\
$\matR = [\matR_1\transp, \matR_2\transp]\transp \in \R^{(r_1 + r_2) \times n}$ containing $r = 4k + \frac{4820k}{\varepsilon}$ rescaled rows of $\matA$.
\end{algorithmic}
{\bf 3. Construct $\matU$ of rank $k$}
\begin{algorithmic}[1]
\STATE 
Let $\matW \in \R^{\xi \times m}$ be a randomly chosen sparse subspace embedding with $\xi = \Omega(k^2 \varepsilon^{-2})$. Then,
\eqan{
\matU
&=&  \matPsi^{-1} \matDelta \matD^{-1} \left(\matW \matC  \matPsi^{-1} \matDelta \matD^{-1} \right)^{\dagger} \matW \matA \matR^{\dagger}
=  \matPsi^{-1} \matDelta \matDelta\transp \left(\matW \matC  \ \right)^{\dagger} \matW \matA \matR^{\dagger}
}
\end{algorithmic}
\end{framed}
\end{algorithm*}

\subsection{Algorithm description}
Algorithm~\ref{algcur3} takes as input an
$m \times n$ matrix $\matA$, rank parameter $k < \rank(\matA)$, and accuracy parameter $0< \varepsilon < 1$.
These are precisely the inputs of the CUR problem in Definition~\ref{def:cur}. It returns matrix $\matC \in \R^{m \times c}$
with $c=O(k/\varepsilon)$ columns of $\matA$, matrix $\matR \in \R^{r \times n}$
with $r=O(k/\varepsilon)$ rows of $\matA,$ as well as matrix $\matU \in \R^{c \times r}$ with rank at most $k$.
Algorithm~\ref{algcur3} follows closely the CUR proto-algorithm in Algorithm~\ref{alg1}. In more detail,
Algorithm~\ref{algcur3} makes specific choices for the various steps of the proto-algorithm that can be implemented
in input-sparsity-time. Algorithm~\ref{algcur3} runs in three steps: (i) in the first step, an optimal number of columns are
selected in $\matC$; (ii) in the second step, an optimal number of rows are selected in $\matR$; and (iii) in the third step,
an intersection matrix with optimal rank is constructed and denoted as $\matU$.
The algorithm itself refers to several other algorithms, which we analyze in detail in different sections.
Specifically,
\emph{SparseSVD} is described in Lemma~\ref{lem:approxSVDsparse};
\emph{RandSampling}  in Lemma~\ref{lem:random};
\emph{BssSamplingSparse}  in Lemma~\ref{lem:dualsets};
\emph{AdaptiveColsSparse} in Lemma~\ref{lem:adaptivecolumnss}.
\emph{ApproxSubspaceSVD}  in Lemma~\ref{lem:KVW}; and
\emph{AdaptiveRowsSparse} in Lemma~\ref{lem:adaptiverowss}.

It is worth pointing out the differences of Algorithm~\ref{algcur3} with Algorithm~\ref{algcur1}.
In a nutshell, Algorithm~\ref{algcur3} replaces the steps in Algorithm~\ref{algcur1} that can 
not be implemented in input-sparsity-time with similar steps that enjoy this property. In some more detail, 
in step $1-1,$ Algorithm~\ref{algcur3} uses \emph{SparseSVD} instead of the standard SVD method, 
in steps $1-3$ and $2-3,$ Algorithm~\ref{algcur3} uses \emph{BssSamplingSparse} instead of the standard version of \emph{BssSampling}, 
in step $1-4$ Algorithm~\ref{algcur3} uses \emph{AdaptiveColsSparse}  instead of the standard \emph{AdaptiveCols} procedure, and in step $2-4$ Algorithm~\ref{algcur3} uses \emph{AdaptiveRowsSparse} instead of \emph{AdaptiveRows}.

\subsection{Running time analysis}
Next, we give a detailed analysis of the arithmetic operations of Algorithm~\ref{algcur3}.

\begin{enumerate}

\item  We need $O( nnz(\matA)\log n + m \cdot poly(\log n, k,1/\varepsilon) )$ time to find $c = 4k + \frac{4820k}{\varepsilon}$ columns of $\matA$ in $\matC \in \R^{m \times c}$

\begin{enumerate}
\item We need $O\left(  \nnz(\matA) \right) + \tilde{O}\left( nk^2 \varepsilon^{-4} + k^3\varepsilon^{-5} \right)$ time to compute $\matZ_1$ (from Lemma~\ref{lem:approxSVDsparse}).
\item We need $O( kn + k \ln( k  ) \log (k \ln( k  )) )$ time to construct $\matOmega_1, \matD_1$ (from Lemma~\ref{lem:random}).\\
 We need $O( k^3 \ln k )$ time to construct $\matV_{\matM_1}$.
\item We need $O( \nnz(\matA) + k^3 \log^2 (k) \varepsilon^{-2} m + k^3 \log^2 (k) \varepsilon^{-2} n +  k^4 \log (k) )$ time for $\matS_1$ (from Lemma~\ref{lem:dualsets})\footnote{
Here, we actually do not explicitly form $(\matA - \matA \matZ_1\matZ_1\transp)\matOmega \matD$.
We only form  $\matA\matOmega \matD$ and $\matZ_1\transp \matOmega \matD$ in $O(1)$ time. Then, the algorithm of Lemma~\ref{lem:dualsets} multiplies $\left(\matA - \matA \matZ_1\matZ_1\transp\right)\matOmega \matD$
from the left with a sparse subspace embedding matrix $\matW \in \R^{\xi \times m}$ with $\xi = O( k^2 \log^2 k \varepsilon^{-2} )$. Computing $\matW \matA$ takes $O\left(  \nnz(\matA) \right)$ time.
Then computing, $(\matW\matA) \matZ_1$ and $(\matW\matA\matZ_1)\matZ_1\transp$ takes another $O(\xi m k)$ + $O(\xi n k)$ time, respectively. Finally, the sampling algorithm on
$\matW (\matA - \matA \matZ_1\matZ_1\transp)\matOmega \matD$ is $O( k^4 \log k + m k \log k)$ time.}.\\
We need $O(m + k)$ time to construct $\matC_1$.
\item We need $O(\nnz(\matA) \log n + m k \log n + mk^2)$ time to construct $\matC_2$ (from Lemma~\ref{lem:adaptivecolumnss}).\\
We need $O(m + k/\varepsilon)$ time to construct $\matC$.
\end{enumerate}

\item We need $O( nnz(\matA)\log n + n\cdot poly(\log n, k,1/\varepsilon) )$ time to find $r = 4k + \frac{4820k}{\epsilon}$ rows of $\matA$ in $\matR \in \R^{r_1 \times n}$

\begin{enumerate}
\item We need $O(nnz(\matA) + mk^3/\varepsilon^3)$ time to construct $\matY$, $\matPsi$, and $\matDelta$  (from Lemma~\ref{lem:bestF}).\\
 We need $O(m k^2)$ time to construct $\matZ_2$.
\item We need $O( km + k \ln( k  ) \log (k \ln( k  )) )$ time to construct $\matOmega_2, \matD_2$ (from Lemma~\ref{lem:random}).\\
 We need $O( k^3 \ln k )$ time to construct $\matV_{\matM_2}$.
\item We need $O( \nnz(\matA) + k^3 \log^2 (k) \varepsilon^{-2} m + k^3 \log^2 (k) \varepsilon^{-2} n +  k^4 \log (k) )$ time to construct $\matS_2$ (Lemma~\ref{lem:dualsets}).\\
 We need $O(n+k)$ time to construct $\matR_1$.
\item We need $O( \nnz(\matA) \log n +  n k \log n +  n k^2)$ time to construct $\matR_2$ (from Lemma~\ref{lem:adaptiverowss}). \\
We need $O(n + k/\varepsilon)$ time to construct $\matR$.
\end{enumerate}

\item We need $O( nnz(\matA) +  n k^3 /\varepsilon^4  + k^4 / \varepsilon^6)$ time to construct $\matU$. First, we compute $\matW\matC$ and $\matW\matA$ in $O(nnz(\matA))$ time.
Then, we compute $\pinv{(\matW\matC)}, \pinv{\matR},$ and $\matPsi^{-1}$;
finally, we compute $\matU$ as follows:
$$\matU =  \left(  \matPsi^{-1}  \left( \matDelta  \left( \matDelta\transp  \left( \matPsi  \pinv{\left(\matW\matC\right)} \right)\right)\right)\right) \cdot \left( \left(\matW\matA\right) \pinv{\matR} \right).$$
\end{enumerate}

The total assymptotic running time of the algorithm is 
$$O( nnz(\matA)\log n + (m+n) \cdot poly(\log n, k,1/\varepsilon) ).$$

\subsection{Error bounds}
The theorem below presents our main quality-of-approximation result regarding Algorithm~\ref{algcur3}. We prove the theorem in Section~\ref{sec:proof3}. 
\begin{theorem}\label{thm:main3}
The matrices $\matC, \matU$, and $\matR$ in Algorithm~\ref{algcur3} satisfy with probability at least $0.16 - 2/n,$
$$ \FNormS{ \matA - \matC \matU \matR}  \le  \left( 1+\varepsilon \right) \left(1 + 60 \varepsilon \right) \FNormS{\matA-\matA_k}.$$
\end{theorem}

\subsubsection{Intermediate Results}
To prove Theorem~\ref{thm:main3} in Section~\ref{sec:proof3}, we need several intermediate results, 
some of which might be of independent interest. 

First, we argue that the sampling of columns implemented via the matrices $\matOmega_1, \matD_1,$ and $\matS_1$ ``preserves'' the rank of $\matZ_1\transp$. This is necessary in order to prove that 
$\matC_1 = \matA \matOmega_1 \matD_1 \matS_1$ gives a ``good'' column-based, low-rank approximation to 
$\matA$. We make this statement precise in Lemma~\ref{lem:pr1s}. 

\begin{lemma}\label{lem:pr0s}
The matrices $\matZ_1, \matOmega_1, \matD_1, \matS_1$ in Algorithm~\ref{algcur3} satisfy with probability at least $0.9$,
$$
\rank(\matZ_1\transp \matOmega_1 \matD_1 \matS_1) = k.
$$
\end{lemma}
\begin{proof}
The proof is identical to the proof of Lemma~\ref{lem:pr0}.
\end{proof}

Next, we argue that the matrix $\matC_1$ in the algorithm offers a  
constant-factor, column-based, low-rank approximation to $\matA$. 
Notice that $\matC_1$ contains $O(k)$ columns of $\matA$. \begin{lemma}\label{lem:pr1s}
The matrix $\matC_1$ in Algorithm~\ref{algcur3} satisfies with probability at least $0.69,$
$$\FNormS{\matA-\matC_1 \pinv{\matC}_1 \matA} \le 4820 \cdot \FNormS{\matA-\matA_k}. $$
\end{lemma}
\begin{proof}
The proof is very similar to the proof of Lemma~\ref{lem:pr1} so we only highlight the differences.
Indeed, the only difference is in step (g) in the manipulations of the term 
$$\FNormS{\matE_1 \matOmega_1 \matD_1 \matS_1(\matZ_1\matOmega_1 \matD_1 \matS_1)^{\dagger}}.$$
In that step, due to Lemma~\ref{lem:dualsets} and our choice of $\varepsilon = 0.5$ we will have an extra multiplicative term $3$ in the bound, so the final bound instead of 80 will be $240$.
Here, we  will also have an extra failure probability $0.01$. So the final bound for $ \FNormS{\matA-\matC_1 \pinv{\matC}_1\matA}$ will have a constant $4820,$ as claimed.
\end{proof}

Next, we argue that the matrix $\matC$ in the algorithm offers a  
relative-error, column-based, low-rank approximation to $\matA$. 
\begin{lemma}\label{lem:pr2s}
The matrix $\matC$ in Algorithm~\ref{algcur3} satisfies with probability at least $0.59 - 1/n$:
$$ 
\FNormS{\matA-\matC \pinv{\matC} \matA} \le \FNormS{\matA - \Pi_{\matC,k}^{\mathrm{F}}(\matA)} \le (1+30 \varepsilon) \cdot \FNormS{\matA-\matA_k}. 
$$
\end{lemma}
\begin{proof}
From Lemma~\ref{lem:adaptivecolumnss} and with probability at least $0.9 - \frac{1}{n}$,
$$
\FNormS{ \matA - \Pi_{\matC,k}^{\mathrm{F}}(\matA) }  \le\FNormS{ \matA - \matA_k }  +  10  \frac{3 \varepsilon}{4820} \cdot\FNormS{\matA-\matC_1 \pinv{\matC}_1 \matA}.
$$
In Lemma~\ref{lem:pr1s} we proved that
for the matrix $\matC_1$ in the algorithm and with probability at least $0.69,$
$$ 
\FNormS{\matA-\matC_1 \pinv{\matC}_1 \matA} \le 4820 \cdot \FNormS{\matA-\matA_k}. 
$$
Combining the last two bounds, we are in a probability event that fails with probability at most $0.41 + 1/n$ and guarantees that
$$ \FNormS{\matA-\Pi_{\matC,k}^{\mathrm{F}}(\matA)} \le (1+30 \varepsilon) \cdot \FNormS{\matA-\matA_k}. $$
Finally, by the definition of $\Pi_{\matC,k}^{\mathrm{F}}(\matA)$ we have that 
$$ \FNorm{\matA - \matC \pinv{\matC} \matA} \le \FNormS{\matA - \Pi_{\matC,k}^{\mathrm{F}}(\matA)}.$$
\end{proof}

Next, we show that there is a rank $k$ matrix in $span(\matC)$ that achieves a similar relative-error bound as the relative-error bound achieved by $\matC \pinv{\matC} \matA$ in the previous lemma.
\begin{lemma}\label{lem:pr3s}
The matrices  $\matY, \matDelta$  in Algorithm~\ref{algcur3} satisfy with probability at least $0.58 - 1/n$:
$$
\FNormS{ \matA - \matY \matDelta \matDelta\transp \matY \matA } \le (1+ 30 \varepsilon) \FNormS{\matA-\matA_k}.
$$
\end{lemma}
\begin{proof}
The result follows by combining Lemma~\ref{lem:KVW} and Lemma~\ref{lem:pr2s}.
\end{proof}

The following two lemmas prove similar results to Lemmas~\ref{lem:pr0s} and~\ref{lem:pr1s} 
but for the matrix $\matR_1$.
\begin{lemma}\label{lem:pr0Rs}
The matrices  $\matZ_2, \matOmega_2, \matD_2, \matS_2$ in Algorithm~\ref{algcur3} satisfy with probability at least $0.9$:
$$
\rank(\matZ_2\transp \matOmega_2 \matD_2 \matS_2) = k.
$$
\end{lemma}
\begin{proof}
The proof is identical to the proof of Lemma~\ref{lem:pr0s}.
\end{proof}
\begin{lemma}\label{lem:pr4s}
The matrix $\matR_1$ in Algorithm~\ref{algcur3} satisfies with probability at least  $0.69$:
$$
\FNormS{\matA-\matA\pinv{\matR}_1\matR_1} \le 4820 \cdot \FNormS{\matA-\matA_k}. 
$$
\end{lemma}
\begin{proof}
The proof is identical to the proof of Lemma~\ref{lem:pr1s} with $\matA, \matC_1$ replaced by $\matA\transp$ and $\matR_1\transp$
\end{proof}

\subsubsection{Proof of Theorem~\ref{thm:main3}}\label{sec:proof3}
We are now ready to prove Theorem~\ref{thm:main3}. 
Notice that we construct $\matU \in \R^{c \times r}$ of rank at most $k$ as follows,
\begin{equation}\label{eqn:Usparse1}
\matU =  \matPsi^{-1} \matDelta \matD^{-1} \matY_{opt},
\end{equation}
where $\matY_{opt} \in \R^{k \times r}$ is a matrix of rank at most $k$ constructed via the following formula,
\eqan{
\matY_{opt}
& = & \left(\matW \matC  \matPsi^{-1} \matDelta \matD^{-1} \right)^{\dagger} \cdot   \matU_{ \matW \matC  \matPsi^{-1} \matDelta \matD^{-1} } \matU_{ \matW \matC  \matPsi^{-1} \matDelta \matD^{-1} }\transp (\matW \matA\matV_{\matR}\matV_{\matR}\transp) \matV_{\matR}\matV_{\matR}\transp \cdot \matR^{\dagger} \\
& = & \left(\matW \matC  \matPsi^{-1} \matDelta \matD^{-1} \right)^{\dagger} \matW \matA \matV_{\matR}\matV_{\matR}\transp \matR^{\dagger} \\
& = &\left(\matW \matC  \matPsi^{-1} \matDelta \matD^{-1} \right)^{\dagger} \matW \matA \matR^{\dagger}.
}

Here, $\matW \in \R^{\xi \times m}$ is a randomly chosen sparse subspace embedding (see Section~\ref{sec:CWT}).
For this choice of $\matU,$ we will analyze the error
$ \FNormS{ \matA - \matC \matU \matR }.$
First, notice that (see Section~\ref{sec:Uopt}):
$$ 
\matY_{opt} \in \argmin_{\matY \in \R^{k \times r}} \FNorm{\matW \matA\matV_{\matR}\matV_{\matR}\transp-\matW \matC  \matPsi^{-1} \matDelta \matD^{-1} \matY \matR }. 
$$
Define
\begin{equation}\label{eqn:Usparse2}
\tilde{\matX}_{opt} = \matY_{opt} \matR \in \R^{k \times n},
\end{equation}
which is also equivalent to the following,
\begin{equation}\label{eqn:Usparse3}
\tilde{\matX}_{opt} =  \argmin_{\matX \in \R^{k \times n}} \FNorm{\matW \matA\matV_{\matR}\matV_{\matR}\transp-\matW \matC  \matPsi^{-1} \matDelta \matD^{-1}\matX } = (\matW \matC  \matPsi^{-1} \matDelta \matD^{-1})^{\dagger}\matW \matA\matV_{\matR}\matV_{\matR}\transp.
\end{equation}
Also, define $ \matX_{opt} \in \R^{k \times n}$ as follows
\begin{equation}\label{eqn:Usparse4}
\matX_{opt} =  \argmin_{\matX \in \R^{k \times n}} \FNorm{\matA\matV_{\matR}\matV_{\matR}\transp-\matC  \matPsi^{-1} \matDelta \matD^{-1}\matX } 
= (\matC  \matPsi^{-1} \matDelta \matD^{-1})^{\dagger}\matA\matV_{\matR}\matV_{\matR}\transp = \matZ_2\transp \matA\matV_{\matR}\matV_{\matR}\transp.
\end{equation}
The proof of the theorem is immediate after combining Lemma~\ref{lem:sparseU}
and Lemma~\ref{lem:twoB} below. 

\begin{lemma}\label{lem:sparseU}
The matrices $\matC, \matU,$ and $\matR$ in Algorithm~\ref{algcur3}  satisfy with probability at least $0.99$:
$$ 
\FNormS{\matA-\matC  \matU \matR } \le  (1+\varepsilon)  \FNormS{\matA -  \matZ_2 \matZ_2\transp \matA \pinv{\matR}\matR}.
$$
\end{lemma}
\begin{proof}
\begin{scriptsize}
\eqan{
\FNormS{\matA-\matC  \matU \matR } 
&\buildrel(\alpha)\over=& \FNormS{\matA-\matC  \matPsi^{-1} \matDelta \matD^{-1} \matY_{opt} \matR } \\
&\buildrel(\beta)\over=&  \FNormS{\matA-\matC  \matPsi^{-1} \matDelta \matD^{-1} \tilde{\matX}_{opt} } \\
&\buildrel(\gamma)\over=& \FNormS{\matA\matV_{\matR}\matV_{\matR}\transp-\matC  \matPsi^{-1} \matDelta \matD^{-1} \tilde{\matX}_{opt} + \matA - \matA\matV_{\matR}\matV_{\matR}\transp} \\
&\buildrel(\delta)\over=& \FNormS{\matA\matV_{\matR}\matV_{\matR}\transp-\matC  \matPsi^{-1} \matDelta \matD^{-1} \tilde{\matX}_{opt}} + \FNormS{ \matA - \matA\matV_{\matR}\matV_{\matR}\transp} \\
&\buildrel(\epsilon)\over\le& (1+\varepsilon) \FNormS{\matA\matV_{\matR}\matV_{\matR}\transp-\matC  \matPsi^{-1} \matDelta \matD^{-1} \matX_{opt} } + \FNormS{ \matA - \matA\matV_{\matR}\matV_{\matR}\transp} \\
&\buildrel(\zeta)\over=& \varepsilon \FNormS{\matA\matV_{\matR}\matV_{\matR}\transp-\matC  \matPsi^{-1} \matDelta \matD^{-1} \matX_{opt} } + \FNormS{\matA\matV_{\matR}\matV_{\matR}\transp-\matC  \matPsi^{-1} \matDelta \matD^{-1} \matX_{opt} } + \FNormS{ \matA - \matA\matV_{\matR}\matV_{\matR}\transp} \\
&\buildrel(\eta)\over=& \varepsilon \FNormS{\matA\matV_{\matR}\matV_{\matR}\transp-\matC  \matPsi^{-1} \matDelta \matD^{-1} \matX_{opt} } + \FNormS{\matA\matV_{\matR}\matV_{\matR}\transp-\matC  \matPsi^{-1} \matDelta \matD^{-1} ( \matZ_2\transp \matA\matR^{\dagger} \matR ) } + \FNormS{ \matA - \matA\matV_{\matR}\matV_{\matR}\transp} \\
&\buildrel(\theta)\over=& \varepsilon \FNormS{\matA\matV_{\matR}\matV_{\matR}\transp-\matC  \matPsi^{-1} \matDelta \matD^{-1} \matX_{opt} } + \FNormS{\matA-\matC  \matPsi^{-1} \matDelta \matD^{-1} ( \matZ_2\transp \matA\matR^{\dagger} \matR )} \\
&\buildrel(i)\over=& \varepsilon \FNormS{\matA\matV_{\matR}\matV_{\matR}\transp-\matC  \matPsi^{-1} \matDelta \matD^{-1} (\matC  \matPsi^{-1} \matDelta \matD^{-1})^{\dagger}\matA\matV_{\matR}\matV_{\matR}\transp } + \FNormS{\matA-\matC  \matPsi^{-1} \matDelta \matD^{-1} ( \matZ_2\transp \matA\matR^{\dagger} \matR ) } \\
&\buildrel(\kappa)\over\le& \varepsilon \FNormS{\matA-\matC  \matPsi^{-1} \matDelta \matD^{-1} (\matC  \matPsi^{-1} \matDelta \matD^{-1})^{\dagger}\matA} + \FNormS{\matA-\matC  \matPsi^{-1} \matDelta \matD^{-1} ( \matZ_2\transp \matA\matR^{\dagger} \matR )} \\
&\buildrel(\lambda)\over=&  \varepsilon \FNormS{\matA-\matZ_2  \matZ_2\transp\matA}  + \FNormS{\matA -  \matZ_2 \matZ_2\transp \matA \pinv{\matR}\matR}\\
&\buildrel(\mu)\over\le&  \varepsilon \FNormS{\matA -  \matZ_2 \matZ_2\transp \matA \pinv{\matR}\matR}  + \FNormS{\matA -  \matZ_2 \matZ_2\transp \matA \pinv{\matR}\matR} \\
&\buildrel(\nu) \over=&   (1+\varepsilon)  \FNormS{\matA -  \matZ_2 \matZ_2\transp \matA \pinv{\matR}\matR}
}
\end{scriptsize}
$(\alpha)$ follows from Eqn.~\ref{eqn:Usparse1};
$(\beta)$ follows from Eqn.~\ref{eqn:Usparse2};
$(\gamma)$ follows because 
$$
\matA\matV_{\matR}\matV_{\matR}\transp - \matA\matV_{\matR}\matV_{\matR}\transp = {\bf 0}_{m \times n};
$$
$(\delta)$ follows by the Pythagorean Theorem (to see this notice that from Eqn.~\ref{eqn:Usparse3} we can write $\tilde{\matX}_{opt} = \matM \matV_{\matR}\transp$ - for an appropriate $\matM$; also,
$\matA - \matA\matV_{\matR}\matV_{\matR}\transp = \matA\left(\matI_n - \matV_{\matR}\matV_{\matR}\transp\right)$);
$(\epsilon)$ follows by Lemma~\ref{lem:affinesparse2} (there is a failure probability $0.01$;
$(\eta)$ follows from Eqn.~\ref{eqn:Usparse4};
$(\theta)$ follows by the Pythagorean Theorem;
$(i)$ follows from Eqn.~\ref{eqn:Usparse4};
$(\kappa)$ follows because $\matV_{\matR}\matV_{\matR}\transp$ is a projector matrix and can be dropped without increasing the Frobenius norm;
$(\mu)$ follows by the optimality of $\matZ_2$.
\end{proof}

\begin{lemma}\label{lem:twoB}
The matrices $\matR$ and 
$\matZ_2$ in Algorithm~\ref{algcur3} satisfy with probability at least   $0.17 - \frac{2}{n},$
$$ 
\FNormS{ \matA - \matZ_2 \matZ_2\transp \matA \pinv{\matR}\matR }  \le \FNormS{\matA-\matA_k} + 60 \varepsilon \FNormS{\matA-\matA_k}. 
$$
\end{lemma}

\begin{proof}
From Lemma~\ref{lem:adaptiverowss} (with $\matV=\matZ_2$) and our choice of $r_2 = 4820 k/\varepsilon$ in that Lemma, with probability at least $0.9-1/n$
$$
\FNormS{ \matA - \matZ_2 \matZ_2\transp \matA \pinv{\matR}\matR }
\le \FNormS{ \matA - \matZ_2 \matZ_2\transp \matA } +  \frac{30 \varepsilon}{4820} \FNormS{\matA-\matA\pinv{\matR}_1\matR_1}
$$
Furthermore,
\eqan{
\FNormS{ \matA - \matZ_2 \matZ_2\transp \matA }  + \frac{30 \varepsilon}{4820} \FNormS{\matA-\matA\pinv{\matR}_1\matR_1}
&\buildrel(b)\over=&  \FNormS{ \matA - \matB \pinv{\matB} \matA } + \frac{30 \varepsilon}{4820} \FNormS{\matA-\matA\pinv{\matR}_1\matR_1}\\
&\buildrel(c)\over\le&  \FNormS{ \matA - \matB \pinv{\matB} \matA }   + 30 \varepsilon \FNormS{\matA-\matA_k}\\
&\buildrel(d)\over=&  \FNormS{ \matA - \matY \matDelta \matDelta\transp \matY \matA }   + 30 \varepsilon  \FNormS{\matA-\matA_k}\\
&\buildrel(e)\over\le& \left(1+30 \varepsilon \right) \FNormS{\matA-\matA_k} + 30 \varepsilon \FNormS{\matA-\matA_k}
}
(b) follows by the fact that $\matZ_2 \matZ_2\transp = \matB \pinv{\matB}$
(to see this,
$\matB \pinv{\matB} =  \matZ_2 \matD \pinv{( \matZ_2 \matD)} = \matZ_2 \matD \matD^{-1} \matZ_2 = \matZ_2 \matZ_2\transp, $
by our specific choice of those matrices).
(c) follows by Lemma~\ref{lem:pr4s} (there is a $0.31$ failure probability to this bound).
(d) follows by the fact that $\matB = \matY \matDelta$ and
$$
\pinv{\matB} = \pinv{\left(\matY \matDelta\right)} = \pinv{\matDelta} \pinv{\matY} = \matDelta\transp \matY\transp,$$
because both matrices are orthonormal.
(e) follows by Lemma~\ref{lem:pr3s} (there is a $0.42 + \frac{1}{n}$ failure probability to this bound).
So, overall we obtain that with probability at least $0.17 - 2/n,$
$$
\FNormS{ \matA - \matZ_2 \matZ_2\transp \matA \pinv{\matR}\matR }  \le \FNormS{\matA-\matA_k} + 60 \varepsilon \FNormS{\matA-\matA_k}. 
$$
\end{proof}

\section{Deterministic CUR}\label{sec:alg2}

In this section, we present and analyze a deterministic, polynomilal-time CUR algorithm\footnote{To be precise, the algorithm we analyze here constructs $\matC$ and $\matR$ with rescaled columns and rows from $\matA$. To
convert this to a truly CUR decomposition, keep the un-rescaled versions of $\matC$ and $\matR$ and introduce the scaling factors in $\matU$.
The analysis carries over to that CUR version unchanged.}.
We start with the algorithm description, which closely follows the CUR proto-algorithm in
Algorithm~\ref{alg1} (indeed, in this case we do not need the leverage-scores sampling step in the proto-algorithm).
Then, we give a detailed analysis of the running time complexity of
the algorithm. Finally, in Theorem~\ref{thm:main2} we analyze the approximation error
$\FNormS{\matA - \matC \matU \matR}$.

\subsection{Algorithm description}
Algorithm~\ref{algcur2} takes as input an
$m \times n$ matrix $\matA$, rank parameter $k < \rank(\matA)$, and accuracy parameter $0< \varepsilon < 1$.
These are precisely the inputs of the CUR problem in Definition~\ref{def:cur}. It returns matrix $\matC \in \R^{m \times c}$
with $c=O(k/\varepsilon)$ columns of $\matA$, matrix $\matR \in \R^{r \times n}$
with $r=O(k/\varepsilon)$ rows of $\matA,$ as well as matrix $\matU \in \R^{c \times r}$ with rank at most $k$.
Algorithm~\ref{algcur2} follows closely the CUR proto-algorithm in Algorithm~\ref{alg1}. In more details,
Algorithm~\ref{algcur2} makes specific choices for the various steps of the proto-algorithm that can be implemented
deterministically in polynomial time. Algorithm~\ref{algcur2} runs in three steps: (i) in the first step, an optimal number of columns are
selected in $\matC$; (ii) in the second step, an optimal number of rows are selected in $\matR$; and (iii) in the third step,
an intersection matrix with optimal rank is constructed and denoted as $\matU$.
The algorithm itself refers to several other algorithms, which we analyze in detail in different sections.
Specifically,
\emph{DeterministicSVD} is described in Lemma~\ref{lem:approxSVDdet};
\emph{BssSampling}  in Lemma~\ref{lem:dualset};
\emph{AdaptiveColsD} in Lemma~\ref{lem:adaptivecolumnsd}.
\emph{BestSubspaceSVD}  in Lemma~\ref{lem:bestF}; and
\emph{AdaptiveRowsD} in Lemma~\ref{lem:adaptiverowsd}.
All those algorithms are \emph{deterministic}. 

\begin{algorithm*}
\begin{framed}
\caption{A deterministic, poly-time, optimal, relative-error, rank-$k$ CUR}
\label{algcur2}
{\bf Input:} $\matA \in \R^{m \times n};$ rank parameter $k < \rank(\matA);$ and accuracy parameter $0 < \varepsilon < 1$. \\
{\bf Output:} $\matC \in \R^{m \times c}$ with $c=O(k/\varepsilon)$;  $\matR \in \R^{r \times n}$  with $r=O(k/\varepsilon)$;  $\matU \in \R^{c \times r}$ with $\rank(\matU) = k$.\\
{\bf 1. Construct $\matC$ with $O(k + k / \varepsilon)$ columns}
\begin{algorithmic}[1]
\STATE $\matZ_1 = DeterministicSVD(\matA, k, 1)$; $\matZ _1\in \R^{n \times k}$ ($\matZ_1\transp\matZ_1  = \matI_k$); $\matE_1 = \matA - \matA\matZ_1\matZ_1\transp \in \R^{m \times n}$.
\STATE $\matS_1 = BssSampling(\matV_{\matM_1}, \matE_1 \transp, c_1)$, with $c_1 = 4k$. $\matS_1 \in \R^{h_1 \times c_1}$.\\
 $\matC_1 = \matA \matS_1 \in \R^{m \times c_1}$ containing rescaled columns of $\matA$.
\STATE $\matC_2 = AdaptiveColsD(\matA, \matC_1, 1, c_2)$, with $c_2=\frac{10k}{\varepsilon}$ and $\matC_2 \in \R^{m \times c_2}$ with columns of $\matA$.\\
 $\matC = [\matC_1\ \ \matC_2] \in \R^{m \times c}$ containing $c = c_1 + c_2 = 4k + \frac{10 k}{\varepsilon}$ rescaled columns of $\matA$.
\end{algorithmic}
{\bf 2. Construct $\matR$ with $O(k + k / \varepsilon)$ rows}
\begin{algorithmic}[1]
\STATE $[\matY, \matPsi, \matDelta] = BestSubspaceSVD(\matA, \matC, k); \matY \in \R^{m \times c}, \matPsi \in \R^{c \times c}, \matDelta \in \R^{c \times k};$ $\matB = \matY \matDelta.$\\
 $\matB = \matZ_2 \matD$ is a $qr$  of $\matB$ with $\matZ_2 \in \R^{m \times k}$ ($\matZ_2\transp\matZ_2=\matI_k$), $\matD \in \R^{k \times k}$,
 and $\matE_2 = \matA\transp - \matA\transp \matZ_2 \matZ_2\transp$.
\STATE $\matS_2 = BssSampling(\matV_{\matM_2}, \left(\matE_2 \right)\transp, r_1)$, with $r_1 = 4k$ and $\matS_2 \in \R^{h_2 \times r_1}$.\\
$\matR_1 = \left( \matA\transp  \matS_2 \right)\transp \in \R^{r_1 \times n}$ containing rescaled rows from $\matA$.
\STATE $\matR_2 = AdaptiveRowsD(\matA, \matZ_2, \matR_1, r_2),$ with $r_2=\frac{10k}{\varepsilon}$ and $\matR_2 \in \R^{r_2 \times n}$ with rows of $\matA$.\\
 $\matR = [\matR_1\transp, \matR_2\transp]\transp \in \R^{(r_1 + r_2) \times n}$ containing $r = 4k + \frac{10k}{\varepsilon}$ rescaled rows of $\matA$.
\end{algorithmic}
{\bf 3. Construct $\matU$ of rank $k$}
\begin{algorithmic}[1]
\STATE 
Construct $\matU \in \R^{c \times r}$ with rank at most $k$ as follows (all those formulas are equivalent),
\eqan{
\matU
=  \matPsi^{-1} \matDelta \matD^{-1} \matZ_2\transp \matA \pinv{\matR}
&= & \matPsi^{-1} \matDelta \matD^{-1}  \left( \matC \matPsi^{-1} \matDelta \matD^{-1} \right)\transp  \matA \pinv{\matR}\\
&=&  \matPsi^{-1} \matDelta \matD^{-1}  \left( \matC \matPsi^{-1} \matDelta \matD^{-1} \right)^{\dagger}  \matA \pinv{\matR}\\
&=&  \matPsi^{-1} \matDelta \matD^{-1}  \matD \matDelta\transp \matPsi \matC^{\dagger}  \matA \pinv{\matR}\\
&=& \matPsi^{-1} \matDelta                      \matDelta\transp \matPsi \matC^{\dagger}  \matA \pinv{\matR}
}
\end{algorithmic}
\end{framed}
\end{algorithm*}

\subsection{Running time analysis}
Next, we give a detailed analysis of the arithmetic operations of Algorithm~\ref{algcur2}.

\begin{enumerate}

\item  We need $O(mn^3k/\varepsilon)$ time to find $c = 4k + \frac{10k}{\varepsilon}$ columns of $\matA$ in $\matC \in \R^{m \times c}$.

\begin{enumerate}
\item We need $O(mnk^2)$ time to compute $\matZ_1$ (from Lemma~\ref{lem:approxSVDdet}), and $O(mnk)$ time to form $\matE_1$.
\item We need $O( m n + n k^3 )$ time to construct $\matS_1$ (from Lemma~\ref{lem:dualset}).\\
 We need $O(m + k)$ time to construct $\matC_1$.
\item We need $O(mn^3k/\varepsilon)$ time construct $\matC_2$ (from Lemma~\ref{lem:adaptivecolumnsd}).\\
       We need $O(m + k/\varepsilon)$ time to construct $\matC$.
\end{enumerate}

\item We need $O( mn^3 k/\varepsilon)$ time to find $r = 4k + \frac{10k}{\epsilon}$ rows of $\matA$ in $\matR \in \R^{r_1 \times n}$.

\begin{enumerate}
\item We need $O(mnk/\varepsilon)$ to construct $\matY$, $\matPsi$, and $\matDelta$  (from Lemma~\ref{lem:bestF}).\\
 We need $O(m k^2)$ time to construct $\matZ_2$, and $O(mnk)$ time to form $\matE_2$.
\item We need $O( n m + m k^3 )$ time to construct $\matS_2$ (from Lemma~\ref{lem:dualset}).\\
 We need $O(n + k)$ time to construct $\matR_1$.
\item We need $O( mn^3 k/\varepsilon)$ time construct $\matR_2$ (from Lemma~\ref{lem:adaptiverowsd}).\\
 We need $O(n + k\varepsilon)$ time to construct $\matR$.
\end{enumerate}

\item We need $O( m^2 k /\varepsilon +  n^2 k /\varepsilon  +m k^2 / \varepsilon^2 + k^3/\varepsilon^3)$ time to construct $\matU$. First, we compute $\pinv{\matC}, \pinv{\matR},$ and $\matPsi^{-1}$;
then, we compute $\matU$ as follows: 
$$
\matU =  \left(  \matPsi^{-1}  \left( \matDelta  \left( \matDelta\transp  \left( \matPsi  \pinv{\matC} \right)\right)\right)\right) \cdot (\matA \pinv{\matR}). 
$$
\end{enumerate}
The total asymptotic running time of the algorithm is $O( mn^3 k/\varepsilon)$.

\subsection{Error bounds}
The theorem below presents our main quality-of-approximation result regarding Algorithm~\ref{algcur2}. We prove the theorem in Section~\ref{sec:proof2}. 
\begin{theorem}\label{thm:main2}
The matrices $\matC, \matU$, and $\matR$ in Algorithm~\ref{algcur2} satisfy,
$$\FNormS{ \matA - \matC \matU \matR}  \le \FNormS{\matA-\matA_k} + 8\varepsilon \FNormS{\matA-\matA_k}.$$
\end{theorem}

\subsubsection{Intermediate results}
To prove Theorem~\ref{thm:main2} in Section~\ref{sec:proof2}, we need several intermediate results, 
some of which might be of independent interest. 

First, we argue that the sampling of columns implemented via the matrix 
$\matS_1$ ``preserves'' the rank of $\matZ_1\transp$. This is necessary in order to prove that 
$\matC_1 = \matA \matS_1$ gives a ``good'' column-based, low-rank approximation to 
$\matA$. We make this statement precise in Lemma~\ref{lem:pr1d}. 
\begin{lemma}\label{lem:pr0d}
The matrices $\matZ_1, \matS_1$ in Algorithm~\ref{algcur2} satisfy,
$$\rank(\matZ_1\transp  \matS_1) = k.$$
\end{lemma}
\begin{proof}
From Lemma~\ref{lem:dualset} we obtain $\sigma_k\left(  \matZ_1\transp \matS_1 \right) \ge \frac{1}{2},$ which implies $\rank( \matZ_1\transp \matS_1) = k$:
\end{proof}

Next, we argue that the matrix $\matC_1$ in the algorithm offers a constant-factor, column-based, low-rank 
approximation to $\matA$. Notice that $\matC_1$ contains $O(k)$ columns of $\matA$. 
\begin{lemma}\label{lem:pr1d}
The matrix $\matC_1$ in Algorithm~\ref{algcur2} satisfies
$$ \FNormS{\matA-\matC_1 \pinv{\matC}_1 \matA} \le 10 \cdot \FNormS{\matA-\matA_k}. $$
\end{lemma}
\begin{proof}
We would like to apply Lemma~\ref{lem:structural} with $\matZ = \matZ_1 \in \R^{n \times k}$ and $\matS = \matS_1 \in \R^{n \times c_1}$.
First, we argue that the rank assumption of the lemma is satisfied for our specific choice of $\matS$.
In Lemma~\ref{lem:pr0d} we proved that $\rank(\matZ_1\transp  \matS_1) = k.$ So, for $\matC_1 = \matA  \matS_1$
(also recall $\matE_1 = \matA - \matA\matZ_1\matZ_1\transp \in \R^{m \times n}$),
$$
\FNormS{ \matA - \matC_1 \pinv{\matC}_1\matA } \le  \FNormS{\matA - \Pi^{\xi}_{\matC_1,k}(\matA)} \le \FNormS{ \matA - \matC_1 \pinv{(\matZ_1 \matS_1)}\matZ_1\transp } \le
\FNormS{\matE_1} + \FNormS{\matE_1  \matS_1(\matZ_1 \matS_1)^{\dagger}}.
$$
We manipulate the second term,
\eqan{
\FNormS{\matE_1 \matS_1(\matZ_1 \matS_1)^{\dagger}}
&\buildrel(a)\over\le& \FNormS{\matE_1  \matS_1} \TNormS{(\matZ_1 \matS_1)^{\dagger}} \\
&\buildrel(e)\over=&   \FNormS{\matE_1  \matS_1}  \cdot \sigma_k^{-2}\left(\matZ_1\transp \matS_1\right)\\
&\buildrel(f)\over\le& \FNormS{\matE_1 \matS_1} \cdot 4 \\
&\buildrel(g)\over\le & \FNormS{\matE_1 } \cdot 4 \\
}
(a) follows by the strong spectral submultiplicativity property of matrix norms.
(e) follows by the connection of the spectral norm of the pseudo-inverse with the singular values of the matrix to be pseudo-inverted.
(f) follows because $\sigma_k\left(  \matZ_1\transp \matS_1 \right) \ge \frac{1}{2}$. 
(g) follows by Lemma~\ref{lem:dualset}.
So,
$$
\FNormS{\matE_1  \matS_1(\matZ_1 \matS_1)^{\dagger}} \le 4 \FNormS{\matE_1},
$$
hence,
$$ \FNormS{ \matA - \matC_1 \pinv{\matC}_1\matA } \le\FNormS{\matE_1} +  4 \FNormS{\matE_1}.$$
From Lemma~\ref{lem:approxSVDdet}: $\FNormS{\matE_1} \leq 2 \FNormS{\matA - \matA_k};$
hence,
$$ 
\FNormS{ \matA - \matC_1 \pinv{\matC}_1\matA } \le 10 \FNormS{\matA - \matA_k}.
$$
\end{proof}

Next, we argue that the matrix $\matC$ in the algorithm offers a relative-error, column-based, low-rank 
approximation to $\matA$. 
\begin{lemma}\label{lem:pr2d}
The matrix $\matC$ in Algorithm~\ref{algcur2} satisfies,
$$ 
\FNormS{\matA-\matC \pinv{\matC} \matA} \le \FNormS{\matA - \Pi_{\matC,k}^{\mathrm{F}}(\matA)} \le (1+4\varepsilon) \cdot \FNormS{\matA-\matA_k}. 
$$
\end{lemma}
\begin{proof}
From Lemma~\ref{lem:adaptivecolumnsd},
$$\FNormS{ \matA - \Pi_{\matC,k}^{\mathrm{F}}(\matA) }  \le \FNormS{ \matA - \matA_k } + \frac{4\varepsilon}{10} \cdot\FNormS{\matA-\matC_1 \pinv{\matC}_1 \matA}.$$
Combine this with Lemma~\ref{lem:pr1d} to wrap up.
\end{proof}

Next, we show that there is a rank $k$ matrix in $span(\matC)$ that achieves a similar relative-error bound as the relative-error bound achieved by $\matC \pinv{\matC} \matA$ in the previous lemma.
\begin{lemma}\label{lem:pr3d}
The matrices $\matY, \matDelta$ in Algorithm~\ref{algcur2} satisfy, 
$$
\FNormS{ \matA - \matY \matDelta \matDelta\transp \matY \matA } \le (1+ 4\varepsilon) \FNormS{\matA-\matA_k}.
$$
\end{lemma}
\begin{proof}
This result is immediate from Lemma~\ref{lem:bestF} and Lemma~\ref{lem:pr2d}.
\end{proof}

The following two lemmas prove similar results to Lemmas~\ref{lem:pr0d} and~\ref{lem:pr1d} 
but for the matrix $\matR_1$.
\begin{lemma}\label{lem:pr0Rd}
The matrices $\matZ_2, \matS_2$ in Algorithm~\ref{algcur2} satisfy,
$$\rank(\matZ_2\transp  \matS_2) = k.$$
\end{lemma}
\begin{proof}
The proof is identical to the proof of Lemma~\ref{lem:pr0d}.
\end{proof}

\begin{lemma}\label{lem:pr4d}
The matrix $\matR_1$ in Algorithm~\ref{algcur2} satisfies, 
$$
\FNormS{\matA-\matA\pinv{\matR}_1\matR_1} \le 10 \cdot \FNormS{\matA-\matA_k}. 
$$
\end{lemma}
\begin{proof}
The proof is identical to the proof of Lemma~\ref{lem:pr1d} with $\matA, \matC_1$ replaced by $\matA\transp$ and $\matR_1\transp$.
\end{proof}

\subsubsection{Proof of Theorem~\ref{thm:main2}}\label{sec:proof2}
We are now ready to prove Theorem~\ref{thm:main2}. 
Our construction of $\matC, \matU$, and $\matR$ implies
$\matC \matU \matR =   \matZ_2 \matZ_2\transp \matA \pinv{\matR}\matR,$
so below we analyze the error
$ \FNormS{\matA -  \matZ_2 \matZ_2\transp \matA \pinv{\matR}\matR}.$
From Lemma~\ref{lem:adaptiverowsd} (with $\matV=\matZ_2$) and our choice of $r_2 = 10 k/\varepsilon$ in that Lemma,
$$
 \FNormS{ \matA - \matZ_2 \matZ_2\transp \matA \pinv{\matR}\matR }
\le \FNormS{ \matA - \matZ_2 \matZ_2\transp \matA } +  \frac{2\varepsilon}{10} \FNormS{\matA-\matA\pinv{\matR}_1\matR_1}
$$
We further manipulate this bound as follows:
\eqan{
\FNormS{ \matA - \matZ_2 \matZ_2\transp \matA \pinv{\matR}\matR }
&\buildrel(a)\over\le& \FNormS{ \matA - \matZ_2 \matZ_2\transp \matA }  + \frac{4\varepsilon}{10} \FNormS{\matA-\matA\pinv{\matR}_1\matR_1} \\
&\buildrel(b)\over=&  \FNormS{ \matA - \matB \pinv{\matB} \matA } + \frac{4\varepsilon}{10} \FNormS{\matA-\matA\pinv{\matR}_1\matR_1}\\
&\buildrel(c)\over\le&  \FNormS{ \matA - \matB \pinv{\matB} \matA }   + 4\varepsilon \FNormS{\matA-\matA_k}\\
&\buildrel(d)\over=&  \FNormS{ \matA - \matY \matDelta \matDelta\transp \matY \matA }   +4 \varepsilon  \FNormS{\matA-\matA_k}\\
&\buildrel(e)\over\le& \left(1+4\varepsilon \right) \FNormS{\matA-\matA_k} +  4\varepsilon \FNormS{\matA-\matA_k}
}
(b) follows by the fact that $\matZ_2 \matZ_2\transp = \matB \pinv{\matB}$
(to see this, $\matB \pinv{\matB} =  \matZ_2 \matD \pinv{( \matZ_2 \matD)} = \matZ_2 \matD \matD^{-1} \matZ_2 = \matZ_2 \matZ_2\transp $).
(c) follows by Lemma~\ref{lem:pr4d} 
(d) follows by the fact that $\matB = \matY \matDelta$ and
$\pinv{\matB} = \pinv{\left(\matY \matDelta\right)} = \pinv{\matDelta} \pinv{\matY} = \matDelta\transp \matY\transp,$
because both matrices are orthonormal.
(e) follows by Lemma~\ref{lem:pr3d}.
So, overall we obtain,
$$ \FNormS{ \matA - \matZ_2 \matZ_2\transp \matA \pinv{\matR}\matR }  \le \FNormS{\matA-\matA_k} + 8 \varepsilon \FNormS{\matA-\matA_k}, $$
which shows that,
$$ \FNormS{ \matA - \matC \matU \matR}  \le \FNormS{\matA-\matA_k} + 8 \varepsilon \FNormS{\matA-\matA_k}. $$

\section{Lower bound}\label{sec:lower}
In this section we will prove that a relative-error CUR is not possible unless $\matC$ has $\Omega(k/\varepsilon)$ columns of $\matA,$
$\matR$ has $\Omega(k/\varepsilon)$ rows of $\matA,$ and $\matU$ has rank $\Omega(k)$. We start with an outline of our approach.
In Lemma~\ref{lem:lower} below we prove that there is a symmetric matrix $\matA \in \R^{t \times t},$
such that, for any integer $k$ (the rank parameter) and $\varepsilon > 0$, no subset of $o(k/\varepsilon)$ columns of $\matA$ span a $(1+\varepsilon)$
approximation to $\matA$.
By symmetry (see Corollary~\ref{cor:lower}), this also implies that no subset of $o(k/\varepsilon)$ rows of $\matA$
span a $(1+\varepsilon)$ approximation to $\matA$.
Those two observations, along with an observation on the minimum possible rank for $\matU,$ give the lower bound.
The theorem below is the main result in this section. 
\begin{theorem}\label{thm:lower2}
Consider the matrix $\matA \in \R^{t \times t}$ in Lemma~\ref{lem:lower}.
Consider a factorization $\matC \matU \matR$, with $\matC \in \R^{t \times c}$ containing $c$ columns of $\matA$, $\matR \in \R^{r \times t}$ containing $r$ rows of $\matA$ and $\matU \in \R^{c \times r}$,
such that
$$
\FNormS{\matA - \matC \matU \matR} \le (1+\varepsilon) \FNormS{\matA -\matA_k}.
$$
Then, for any $\varepsilon < 1/3$ and any $k \ge 1$: 
$$c = \Omega(k/\varepsilon),$$
and 
$$r = \Omega(k/\varepsilon),$$ and 
$$\rank(\matU) \ge k/2.$$
\end{theorem}
\begin{proof}
Assume that $\matC, \matU, \matR$ are as described in the statement of the theorem, with
$$ \FNormS{\matA - \matC \matU \matR} \le (1+\varepsilon) \FNormS{\matA -\matA_k}.$$
Assume further that either $c = o(k/\varepsilon)$ or $r = o(k/\varepsilon)$ or $\rank(\matU)<k/2$.
We show that if any of these three assumptions is valid we obtain a contradiction;
hence, we conclude that to achieve a relative error bound $c = \Omega(k/\varepsilon)$ and $r = \Omega(k/\varepsilon)$ and $\rank(\matA) \ge k/2$.

If  $c=o(k/\varepsilon)$, then the columns of the matrix $\matC\matU\matR$ are in the column space of $\matC$, and so this implies that the columns of $\matC$ span a $(1+\varepsilon)$  approximation to $\matA$
(i.e., there is a matrix  $\matC \in \R^{t \times c}$ with $c=o(k/\varepsilon)$ columns of $\matA$ and $\matX = \matU \matR$ such that,
$ \FNormS{\matA - \matC \matX} \le (1+\varepsilon) \FNormS{\matA -\matA_k}$),
contradicting that no subset of $o(k/\varepsilon)$ columns of $\matA$ span a $(1+\varepsilon)$ approximation
(i.e., contradicting Lemma~\ref{lem:lower}).

If  $r=o(k/\varepsilon)$, then the rows of the matrix $\matC\matU\matR$ are in the row space of $\matR$, and so this implies that the rows of $\matR$ span a $(1+\varepsilon)$ approximation to $\matA$
(i.e., there is a matrix  $\matR \in \R^{r \times t}$ with $r=o(k/\varepsilon)$ rows of $\matA$ and $\matY = \matC \matU$ such that,
$ \FNormS{\matA - \matY \matR} \le (1+\varepsilon) \FNormS{\matA -\matA_k}$),
contradicting that no subset of $o(k/\varepsilon)$ rows of $\matA$ span a $(1+\varepsilon)$ approximation
(i.e., contradicting Corollary~\ref{cor:lower}).

To continue the proof we need the details for the specific construction of the adversarial matrix $\matA$. Those details
are given in the proof of Lemma~\ref{lem:lower} but we repeat them here for completeness.
For $\alpha > 0$ and integer $n>1$, consider the matrix
$$\matD = [\e_1+ \frac{\alpha}{\sqrt{k}} \e_2, \e_1+\frac{\alpha}{\sqrt{k}}\e_3,\ldots, \e_1+\frac{\alpha}{\sqrt{k}}\e_{n+1}] \in\R^{(n+1) \times n},$$
where, for $i=1:n+1,$ \math{\e_i\in\R^{n+1}} are the standard basis vectors.
$\matD$ looks like the following matrix,
$$ \matD =
\left(\begin{array}{ccccc}
             1 & 1 & 1  & \cdots & 1  \\
             \frac{\alpha}{\sqrt{k}} & & & & \\
              &  \frac{\alpha}{\sqrt{k}} & & & \\
              & & \frac{\alpha}{\sqrt{k}} & &  \\
              & &  & \ddots & \\
              &   & & & \frac{\alpha}{\sqrt{k}} \\
          \end{array}
    \right) \in \R^{(n+1) \times n}.
$$
Let $\matB \in \R^{m \times \ell}$ with $m=(n+1)k$ and $\ell = nk$ be constructed by repeating $\matD$ $k$ times along its
main diagonal,
$$ 
\matB =
\left(\begin{array}{ccc}
             \matD & &  \\

                    & \ddots &  \\
                   & &            \matD  \\
          \end{array}
    \right) \in \R^{k(n+1) \times kn}.
$$
Let $\matA \in \R^{t \times t}$ with $t = (2n+1)k$ be the following matrix,
$$ 
\matA =
\left(\begin{array}{cc}
              \matB &  \\
             & \matB\transp
          \end{array}
    \right) \in \R^{(2kn+k) \times (2kn+k)}.
$$
If $\rank(\matU) < k/2$, then $\rank(\matC \matU \matR) < k/2$, hence when we approximate $\matA$
with  $\matC \matU \matR$ there will be an error of at least $(3k/2)\FNormS{\matD}$, because
$\matA$ contains $2k$ blocks of $\matD$ and $\matD\transp$ along its main diagonal.
So,
$$
\frac{\FNormS{\matA-\matC\matU\matR}}{\FNormS{\matA-\matA_k} } \ge \frac{\frac{3k}{2}\FNormS{\matD}}{\ell(1+\frac{2 \alpha^2}{k})}=
\frac{\frac{3k}{2}(n + n \frac{\alpha^2}{k})}{\ell(1+\frac{2 \alpha^2}{k})} = \frac{3}{2} \cdot \frac{1 + \frac{\alpha^2}{k}}{1 + \frac{2\alpha^2}{k}}=
\frac{3}{2} \cdot \left( 1 -   \frac{\frac{ \alpha^2}{k}}{1 + \frac{2\alpha^2}{k}}  \right) =
\frac{3}{2} \cdot \left(\frac{k+ \alpha^2}{k + 2\alpha^2}  \right)
$$
In the above, we used 
$$\FNormS{\matA-\matA_k} = \ell(1+\frac{2 \alpha^2}{k}),$$ 
which we showed in the proof of Lemma~\ref{lem:lower}.
In the proof of Lemma~\ref{lem:lower} we also chose 
$$\alpha = 10^{-10}.$$ 
So, for example,
$$\left(\frac{k+ \alpha^2}{k + 2\alpha^2}  \right) \ge 8/9;$$ hence, the bound becomes,
$$\frac{\FNormS{\matA-\matC\matU\matR}}{\FNormS{\matA-\matA_k} } \ge 1 + \frac{1}{3},$$
contradicting that there is a relative error bound with $\varepsilon < 1/3$.
\end{proof}
\begin{lemma}\label{lem:lower}
There is a symmetric matrix $\matA \in \R^{t \times t}$ such that, for any $k$ and $\varepsilon>0,$
no subset of $c=o(k/\varepsilon)$ columns of $\matA$ span a $(1+\varepsilon)$ approximation to $\matA$,
i.e., there is no  $\matC \in \R^{t \times c}$ with $c = o(k/\varepsilon)$ columns of $\matA$ and  $\matX \in \R^{c \times t}$ such that
$$ 
\FNormS{\matA - \matC \matX} \le (1+\varepsilon) \FNormS{\matA -\matA_{k}}.
$$
\end{lemma}
\begin{proof}
For $\alpha > 0$ and integer $n>1$, consider the matrix
$$\matD = [\e_1+ \frac{\alpha}{\sqrt{k}} \e_2, \e_1+\frac{\alpha}{\sqrt{k}}\e_3,\ldots, \e_1+\frac{\alpha}{\sqrt{k}}\e_{n+1}] \in\R^{(n+1) \times n},$$
where, for $i=1:n+1,$ \math{\e_i\in\R^{n+1}} are the standard basis vectors. $\matD$ looks like the following matrix,
$$ \matD =
\left(\begin{array}{ccccc}
             1 & 1 & 1  & \cdots & 1  \\
             \frac{\alpha}{\sqrt{k}} & & & & \\
              &  \frac{\alpha}{\sqrt{k}} & & & \\
              & & \frac{\alpha}{\sqrt{k}} & &  \\
              & &  & \ddots & \\
              &   & & & \frac{\alpha}{\sqrt{k}} \\
          \end{array}
    \right) \in \R^{(n+1) \times n}.
$$
This matrix is popular in proving lower bounds for column-based low rank matrix factorizations~\cite{DR10,BDM11a}.
Let $\matB \in \R^{m \times \ell}$ with $m=(n+1)k$ and $\ell = nk$ be constructed by repeating $\matD$ $k$ times along its
main diagonal,
$$ \matB =
\left(\begin{array}{ccc}
             \matD & &  \\

                    & \ddots &  \\
                   & &            \matD  \\
          \end{array}
    \right) \in \R^{k(n+1) \times kn}.
$$
Let $\matA \in \R^{t \times t}$ with $t = (2n+1)k$ be the following matrix,
$$ \matA =
\left(\begin{array}{cc}
              \matB &  \\
             & \matB\transp
          \end{array}
    \right) \in \R^{(2kn+k) \times (2kn+k)}.
$$
This matrix $\matA$ is the hard instance for the lower bound (for some $\alpha$ that will be specified later).
Below, we are interested in estimating what is the best - smallest - error that can occur when
approximating $\matA$ with, let's say, $c$ columns from $\matA$. In particular, we will show that a relative
error approximation is not possible unless $c = \Omega(k/\varepsilon)$ columns of $\matA$ are selected.

Let $\matC \in \R^{t \times c}$
contain $c$ columns of $\matA$ and let it be choice of $c$ columns that result to the smallest possible
error $\FNormS{ \matA - \matC \pinv{\matC} \matA }.$ We prove a lower bound $\gamma$ of this form
$$  \frac{ \FNormS{ \matA - \matC \pinv{\matC} \matA } }{\FNormS{ \matA -\matA_{k} }} \ge \gamma. $$

\paragraph{Bound for $\FNormS{ \matA -\matA_{k} }$}
First, for the matrix $\matD$ described above (these observations are also made in~\cite{DR10,BDM11a}):
$$\matD\transp\matD=\bm{1}_n\bm{1}_n\transp+\alpha^2/k \matI_{n}, \qquad
\sigma_1^2(\matD)=n+\alpha^2/k, \qquad \mbox{and}  \qquad
\sigma_i^2(\matD)=\alpha^2/k \mbox{\ \ for\ \ } i>1.$$
Since $\matB$ is block diagonal containing $k$ blocks of $\matD$ we obtain:
$$
\sigma_i^2(\matB)=n+\alpha^2/k,  \mbox{\ \ for\ \ } i \le k;
\qquad$$
$$
\sigma_i^2(\matB)=\alpha^2/k \mbox{\ \ for\ \ } i>k.$$
Since $\matA$ is block diagonal containing $\matB$ and $\matB\transp$ along the main diagonal we obtain:
$$
\sigma_i^2(\matA)=n+\alpha^2/k,  \mbox{\ \ for\ \ } i \le 2k;
\qquad
$$
$$
\sigma_i^2(\matA)=\alpha^2/k \mbox{\ \ for\ \ } i> 2k.$$
Hence, for any $k \ge 1$ we obtain
$$
\FNormS{ \matA -\matA_{k}} = k ( n+\alpha^2/k )  + (2nk-k) (\alpha^2/k) = \ell + \alpha^2 + 2n\alpha^2 - \alpha^2= \ell + 2\frac{\ell}{k}\alpha^2.
$$
\paragraph{Bound for  $\FNormS{ \matA - \matC \pinv{\matC} \matA }$}
This ``optimum'' matrix $\matC \in \R^{t \times c}$ is a matrix of the following form,
$$ \matC =
\left(\begin{array}{cc}
              \matC_1 &\\
             & \matC_2
          \end{array}
    \right),
$$
where $\matC_1 \in \R^{m \times c_1}$ contains $c_1$ columns from $\matB$ and
$\matC_2 \in \R^{\ell \times c_2}$ contains $c_2$ columns from $\matB\transp$,
with
$ c = c_1 + c_2. $
Also,
\eqan{
\FNormS{
\matA - \matC \pinv{\matC} \matA}
 &=&
\FNormS{
\left(\begin{array}{cc}
              \matB &\\
             & \matB\transp
          \end{array}
    \right) -
    \left(\begin{array}{cc}
               \matC_1 &\\
             & \matC_2
          \end{array}
    \right)
 \left(\begin{array}{cc}
               \pinv{\matC}_1 \matB &\\
            &\pinv{\matC}_2\matB\transp
          \end{array}
    \right)
}\\
&=&
\FNormS{\matB - \matC_1 \pinv{\matC}_1 \matB} + \FNormS{\matB\transp - \matC_2 \pinv{\matC}_2 \matB\transp}
}
The last Equation in Section 9 in~\cite{BDM11a} shows precisely the following lower bound,
$$ \FNormS{\matB - \matC_1 \pinv{\matC}_1\matB} \ge
{\frac{\alpha^2}{k}}(\ell-c_1)
\left(1+\frac{k}{c_1+{\alpha}^2}\right).$$

Now, let $\matC_2 \in\R^{\ell \times c_2}$ contain any $c_2$ columns from $\matB\transp$. 
We now compute an upper bound for $ \FNormS{ \matB\transp - \matC_2 \pinv{\matC}_2 \matB\transp } $.
$\matC_2$ contains $c_2$ columns from $\matB\transp$, equivalently $\matC_2\transp$ contains $c_2$ rows from $\matB$.
Recall that $\matB$ contains $k$ copies of $\matD \in \R^{n \times n}$ along it's main diagonal. Let us denote those copies $\matD_i$, for $i=1:k$.
Let $c_2 = r_1 + r_2 + \cdots + r_k$, where $r_i$ is the number of rows selected from each $\matD_i$. Let also $\matR_i \in \R^{r_i \times n}$
contain these rows from the corresponding $\matD_i$.
Then,
$$ \FNormS{ \matB\transp - \matC_2 \pinv{\matC}_2 \matB\transp }  = \sum_{i=1}^k \FNormS{ \matD_i - \matD_i \pinv{\matR_i} \matR_i }.$$
We further manipulate this term as follows,
$$ \sum_{i=1}^k \FNormS{ \matD_i - \matD_i \pinv{\matR_i} \matR_1 } \ge \sum_{i=1}^k \frac{\alpha^2}{k} (n - r_i) =  \frac{\alpha^2}{k}\sum_{i=1}^k (n - r_i) = \frac{\alpha^2}{k} (kn - c_2) = \frac{\alpha^2}{k} (\ell - c_2).$$
We are now ready to prove the lower bound,
$$
\frac{ \FNormS{ \matA - \matC \pinv{\matC} \matA } }{\FNormS{ \matA -\matA_{k} }} \ge
\frac{ {\frac{\alpha^2}{k}}(\ell-c_1)
\left(1+\frac{k}{c_1+{\alpha}^2}\right) + \frac{\alpha^2}{k} (\ell - c_2) }{ (\frac{2 \ell}{k}) \alpha^2 + \ell} 
= \frac{\ell - c_1}{2 \ell }\left(1 + \frac{k}{c_1 + \alpha^2} \right) + \frac{\ell - c_2}{2 \ell}
$$
By using $\ell = nk \ge c_1 / \varepsilon$, $\ell = nk \ge c_2 / \varepsilon$
and $n = \omega(1/\varepsilon^2)$ we further manipulate the bound as follows,
$$
\frac{ \FNormS{ \matA - \matC \pinv{\matC} \matA } }{\FNormS{ \matA -\matA_{k} }} \ge
\frac{\ell - c_1}{2 \ell }\left(1 + \frac{k}{c_1 + \alpha^2} \right) + \frac{\ell - c_2}{2 \ell} \ge$$
$$
\frac{1}{2}\left(1-o(\varepsilon) \right)  \left(1 + \frac{k}{c_1+\alpha^2} \right) + \frac{1}{2}\left(1-o\left(\varepsilon\right) \right) =
1 + \frac{k}{c_1 + \alpha^2} - o(\varepsilon)
$$
Using, $c_1 < c$ we obtain,
$$
\frac{ \FNormS{ \matA - \matC \pinv{\matC} \matA } }{\FNormS{ \matA -\matA_{k} }} \ge
1 + \frac{k}{c + \alpha^2} - o(\varepsilon)
$$
The number of columns $c$ satisfies: $c > 1$. Now, choose $\alpha$ to be a sufficiently small positive constant, e.g.
$\alpha = 10^{-10}$. For this choice of $\alpha$: 
$$\frac{k}{c+\alpha^2} \ge \frac{k}{2c}.$$ Hence,
$$
\frac{ \FNormS{ \matA - \matC \pinv{\matC} \matA } }{\FNormS{ \matA -\matA_{k} }} \ge
1 + \frac{k}{c + \alpha^2} - o(\varepsilon) \ge
1 + \frac{k}{2c} - o(\varepsilon)
$$
So, to obtain a relative-error bound, we need at least $c = \Omega(k/\varepsilon)$ columns.
 \end{proof}

By symmetry, the following result is an immediate corollary to Lemma~\ref{lem:lower}.
\begin{corollary}\label{cor:lower}
There is a symmetric matrix $\matA \in \R^{t \times t}$ such that, for any $k$ and $\varepsilon>0,$
no subset of $o(k/\varepsilon)$ rows of $\matA$ span a $(1+\varepsilon)$ approximation to $\matA$,
i.e., there is no   $\matR \in \R^{c \times t }$ with $r = o(k/\varepsilon)$ rows of $\matA$
and $\matY \in \R^{t \times r}$ such that
$$ 
\FNormS{\matA - \matY \matR} \le (1+\varepsilon) \FNormS{\matA -\matA_{k}}.
$$
\end{corollary}

\paragraph{Remark} To prove Lemma~\ref{lem:lower}, we extend slightly the lower bound in~\cite{BDM11a}, which is stated
for a non-symmetric matrix. Here, we aim for a lower bound for a symmetric matrix. We should mention that a similar
lower bound for a symmetric matrix appeared in Lemma 6.2 in~\cite{GS12}. We chose not to use this result because
it was not clear to us how to address the lower bound for the rank of the intersection matrix $\matU$ in Theorem~\ref{thm:lower2}.

 \section*{Acknowledgement} 
David Woodruff would like to thank the Simons Institute at Berkeley where part of this work was done. He would also like to acknowledge
the XDATA program of the Defense Advanced Research Projects Agency (DARPA), administered through Air Force Research Laboratory contract 
FA8750-12-C0323, for supporting part of this work. 

\end{document}